\documentclass[a4paper,12pt]{article}
\usepackage{epsfig}
\usepackage{amssymb,latexsym,amsfonts,amsmath}
\usepackage{graphicx}
\usepackage{subfig}
\usepackage{hyperref}
\usepackage{color}

\catcode`\@=11

\@addtoreset{equation}{section}
\def\theequation{\thesection.\arabic{equation}}

\def\@normalsize{\@setsize\normalsize{15pt}\xiipt\@xiipt
\abovedisplayskip 14pt plus3pt minus3pt%
\belowdisplayskip \abovedisplayskip
\abovedisplayshortskip \z@ plus3pt%
\belowdisplayshortskip 7pt plus3.5pt minus0pt}

\def\small{\@setsize\small{13.6pt}\xipt\@xipt
\abovedisplayskip 13pt plus3pt minus3pt%
\belowdisplayskip \abovedisplayskip
\abovedisplayshortskip \z@ plus3pt%
\belowdisplayshortskip 7pt plus3.5pt minus0pt
\def\@listi{\parsep 4.5pt plus 2pt minus 1pt
      \itemsep \parsep
      \topsep 9pt plus 3pt minus 3pt}}

\relax

\catcode`@=12

\catcode`\@=11

\def\section{\@startsection{section}{1}{\z@}{3.5ex plus 1ex minus
    .2ex}{2.3ex plus .2ex}{\large\bf}}

\def\thesection{\arabic{section}}
\def\thesubsection{\arabic{section}.\arabic{subsection}}

\def\appendix{\setcounter{section}{0}
  \def\thesection{Appendix \Alph{section}}
  \def\thesubsection{\Alph{section}.\arabic{subsection}}
  \def\theequation{\Alph{section}.\arabic{equation}}}

\def\SymBoxes#1#2#3#4{\newdimen\un@t \un@t#3%
\raisebox{#1}{\rule{#2\un@t}{#4}\hskip-#2\un@t% lower horizontal
\@tempdimb\un@t \advance\@tempdimb by-#4\@tempcntb#2\relax%
\@whilenum{\@tempcntb>0}\do{%                         % #2 vertical lines
\rule{#4}{\un@t}\hskip\@tempdimb \advance\@tempcntb by\m@ne}%
\hskip-#2\un@t \rule[\un@t]{#2\un@t}{#4}%
\rule[\un@t]{#4}{#4}\hskip-#4%             % upper horizontal line
\rule{#4}{\un@t}}\hskip-#4}                % rightest vertical line

\newcommand{\beq}{\begin{equation}}
\newcommand{\eeq}{\end{equation}}
\newcommand{\bea}{\begin{eqnarray}}
\newcommand{\eea}{\end{eqnarray}}
\newcommand{\beas}{\begin{eqnarray*}}
\newcommand{\eeas}{\end{eqnarray*}}

\newcommand{\non}{\nonumber\\}
\newcommand{\bquo}{\begin{quote}}
\newcommand{\enqu}{\end{quote}}
\renewcommand{\(}{\begin{equation}}
\renewcommand{\)}{\end{equation}}
\def\p{\partial}
\def\Tr{ \hbox{\rm Tr}}

\def\Li{\hbox{\rm Li}}
\def\sign{\hbox{\rm sign}}

\newcommand{\arctanh}{{\mathop{\rm arctanh}}}

\begin{document}

\pagenumbering{Alph}
\begin{titlepage}
\def\thefootnote{\fnsymbol{footnote}}

\begin{center}
{\large {\bf Global two-monopoles}}
\end{center}

\bigskip
\begin{center}
{\large Sven Bjarke Gudnason\footnote{\texttt{bjarke(at)impcas.ac.cn}}
and Jarah Evslin\footnote{\texttt{jarah(at)impcas.ac.cn}}}
\end{center}

\renewcommand{\thefootnote}{\arabic{footnote}}

\begin{center}
{Institute of Modern Physics, Chinese Academy of Sciences,
  Lanzhou 730000, China
}
\end{center}

\bigskip

\noindent
\begin{center} {\bf Abstract} \end{center}

Global two-monopoles are unstable in their simplest formulation. We
construct a model with a metric-like prefactor which we show can
stabilize the global two-monopoles. The stabilizing construction is
realized with a Skyrme sector where the metric-like prefactor is a
function of the Skyrmion profile.

\noindent

\vfill

\begin{flushleft}
{\today}
\end{flushleft}
\end{titlepage}
\pagenumbering{arabic}

\hfill{}

\setcounter{footnote}{0}

\section{Introduction}

If the set of minima $\mathcal{M}$ of the potential energy of a field
is not 2-connected, for example if it is a 2-sphere $S^2$ or real
projective space $\mathbb{RP}^2$, then it admits solutions called
global monopoles.  In general the total energy of a global monopole
exhibits a long distance divergence (linear divergence). 

In the literature the global monopole has been considered mostly in a
cosmological context where the gravitational energy, their Goldstone
boson radiation and gravitational-wave emission and finally pair
annihilation sparked the primary interest
\cite{Kibble:1976sj,Barriola:1989hx,Linde:2005ht}.
Global monopoles may also be realized in nematic fluids
\cite{Poulin:1998}. 

The stability of a single global monopole has been debated in the
early literature. It was argued that the global monopole could itself
collapse to a zero-energy solution \cite{Goldhaber:1989na}, but it was
shortly after argued not to hold true
\cite{Rhie:1990kc}. Ref.~\cite{Perivolaropoulos:1991du} further argued
that the observation of Ref.~\cite{Goldhaber:1989na} should not be
interpreted as a collapse but rather as an acceleration of the entire
monopole. The false conclusion was thus reached by artificially
holding the origin of the monopole fixed. The bottom line is that the
single global monopole is indeed stable, both with respect to radial
and angular perturbations \cite{Perivolaropoulos:1991du}.

Ref.~\cite{Perivolaropoulos:1991du} made a further study of the mutual
forces between two different types of a global monopole and an 
anti-monopole, which we will call type B and type C.
We find -- in agreement with Ref.~\cite{Perivolaropoulos:1991du} --
that the type C monopole and anti-monopole attract each other and the
type B repel each other. We find, however, different values of the
forces than Ref.~\cite{Perivolaropoulos:1991du} does.
We further generalize this calculation to two monopole-monopole cases
which we will call type A and type B and they are both mutually
repulsive. 

The literature has nothing much to report about global multimonopole
configurations. This is because they are unstable. 
For completeness we carry out a numerical calculation of a perturbed
two-monopole configuration in \ref{app:mm} to confirm that they
are indeed unstable; that is they repel. 

We contemplate an interaction term which may be added to the global
monopole Lagrangian in order to stabilize said system. This is argued,
however, to be problematic due to the possibility that the monopoles
just to shrink upon such an intervention. We have also checked this
with numerical calculations. 

As a different approach on the path to stabilize the global
two-monopole configuration, we consider a modification of the global
monopole Lagrangian more similar to a nontrivial effective
metric. We choose to construct said effective metric being a function
of a Skyrmion field. The stabilization is thus dynamic but not
renormalizable. The Skyrme Lagrangian
\cite{Skyrme:1962vh,Skyrme:1961vq} itself is not renormalizable
either, so we do not regard that as a big problem but the model as an
effective field theory model generated by some fundamental theory.
A single gauged monopole inside a U(1) gauged single Skyrmion has been
studied in the literature already \cite{Brihaye:1998vr}. We are
studying the global analog of such a configuration and the
multimonopole instead of the single monopole. 
We show that the global two-monopole can indeed be stabilized inside a
Skyrmion for certain choices of parameters.

The paper is organized as follows. In the next section we review a
single global monopole to remind the reader and set the
notation. In Sec.~\ref{sec:puntosez} we review the
monopole-anti-monopoles interactions in the point-charge approximation
and make the same type of analysis for the two-monopole system.
After Sec.~\ref{sec:puntosez} we no longer use the point-charge
approximation and all the calculations and considerations are carried
out with the monopole profile functions and hence with true equations
of motion. 
In Sec.~\ref{sec:caging} we discuss the possibility of stabilizing the
global two-monopole using an interaction term.
In Sec.~\ref{sec:skyrmisez} we construct the two-monopole inside a
Skyrmion and dub it the Skyrmopole.
Finally, we conclude in Sec.~\ref{sec:conclusion}.
The \ref{app:mm} presents a numerical calculation showing the
instability of the normal global two-monopole.

\section{A single global monopole}

The theory consists of an adjoint-valued scalar field,
$\Phi=\Phi^a\sigma^a$, the gauge indices $a=1,2,3$ are summed over,
$\sigma^a$ are the Pauli matrices and the Lagrangian density is 
\beq
\mathcal{L} = -\frac{v^2}{2}\Tr(\p_\mu\Phi)^2
-\frac{\lambda v^4}{4}\left(1 - \Tr[\Phi^2]\right)^2,
\label{eq:Lmonopole}
\eeq
which has the equation of motion
\beq
\p_\mu\p^\mu \Phi = - \lambda v^2\left(1 - \Tr[\Phi^2]\right) \Phi.
\label{eq:eom}
\eeq
The Greek letters are used for spacetime indices, $\mu=0,1,2,3$ and we
use the mostly-positive metric. The potential breaks the global
SO(3)-symmetry down to SO(2) and the coset SO(3)/SO(2) $\sim S^2$ when
mapped from spatial infinity, which is also $S^2$, defines the
topological charge by means of the second homotopy group. 
The topological charge of the monopole is given by the winding number,
i.e.
\beq
Q = -\frac{1}{8\pi}\oint dS^{ij}\; \epsilon_{abc}
\Phi^a\partial_i\Phi^b\partial_j\Phi^c,
\label{eq:Q}
\eeq
where $a,b,c=1,2,3$ are gauge indices and $i,j=1,2,3$ are spatial
indices. The charge $Q$ is also called the monopole number. 
The mass of the monopole is the total energy in the rest frame
\beq
E[\Phi] = - \int d^3x \; \mathcal{L}[\Phi].
\eeq

Let us first review the standard hedgehog configuration which
describes a single global monopole in the theory at hand. 
Since $Q=1$, we can use a spherical Ansatz
\beq
\Phi = \frac{1}{r} h(r) x^a \sigma^a,
\eeq
for which the energy density reads 
\beq
\frac{1}{v^2}\mathcal{E} = \frac{1}{2} (h')^2 + \frac{1}{r^2}h^2
+ \frac{\lambda v^2}{4}(h^2-1)^2,
\eeq
where $'=\p_r$.
Defining a dimensionless coordinate, $\rho\equiv\sqrt{\lambda}v r$, we
can write the energy density as
\beq
\frac{1}{\lambda v^4}\mathcal{E} = \frac{1}{2} (h')^2 + \frac{1}{\rho^2}h^2
+ \frac{1}{4}(h^2-1)^2,
\eeq
where $h=h(\rho)$; now $'=\p_\rho$ and the equation of motion reads
\beq
h'' + \frac{2}{\rho}h' - \frac{2}{\rho^2}h - (h^2-1)h = 0.
\eeq
The boundary conditions for the monopole are: $h(0)=0$ and
$h(\infty)=1$. 
Expanding around $\rho=0$, we get
\beq
h = a \rho
- \frac{a}{10} \rho^3
+ \frac{a+10a^3}{280} \rho^5
+ \mathcal{O}(\rho^7),
\eeq
where $a>0$ is a constant (also called the shooting parameter). 
This tells us that the energy density around the center goes as
\beq
\frac{1}{\lambda v^4}\mathcal{E} = 
\left(\frac{1}{4}+\frac{3}{2}a^2\right)
-a^2\rho^2
+\left(\frac{9}{50}a^2+\frac{1}{2}a^4\right)\rho^4
+\mathcal{O}(\rho^6),
\eeq
i.e.~roughly constant for small values of $a$. 
Asymptotically, $h$ is very close to one, so linearizing about
$\beta=1-h$, we get the asymptotic equation of motion
\beq
\beta'' + \frac{2}{\rho}\beta' - \frac{2}{\rho^2}(\beta-1) - 2\beta =
0,
\eeq
which has the exact solution
\beq
h = 1-\beta = 1 - \frac{1}{\rho^2}.
\eeq
This in turn determines the asymptotic energy density as
\beq
\frac{1}{\lambda v^4}\mathcal{E} = \frac{1}{\rho^2}
- \frac{1}{\rho^4}
+ \frac{2}{\rho^6}
+ \frac{1}{4\rho^8},
\eeq
which for large $\rho$ goes like $1/\rho^2$ and hence the total energy
diverges linearly. 
Finally, the topological charge can easily be calculated using
Eq.~\eqref{eq:Q} as
\beq
Q = -\frac{1}{8\pi}\oint dS^{ij}\;\epsilon_{aij}h^3\frac{x^a}{r^3} =
1.
\eeq

\begin{figure}[!htp]
\begin{center}
\mbox{
\subfloat[]{\includegraphics[width=0.45\linewidth]{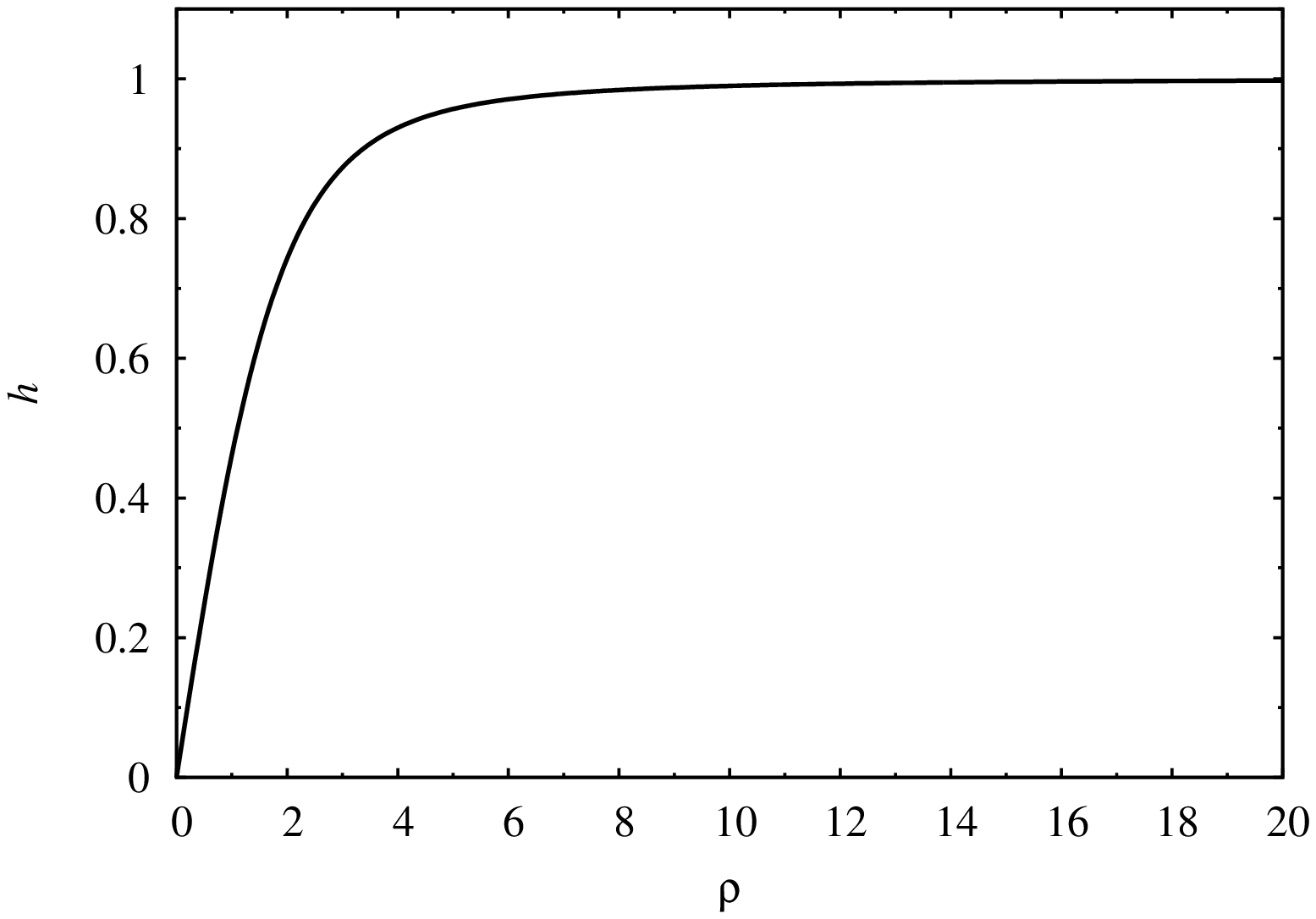}}\quad
\subfloat[]{\includegraphics[width=0.45\linewidth]{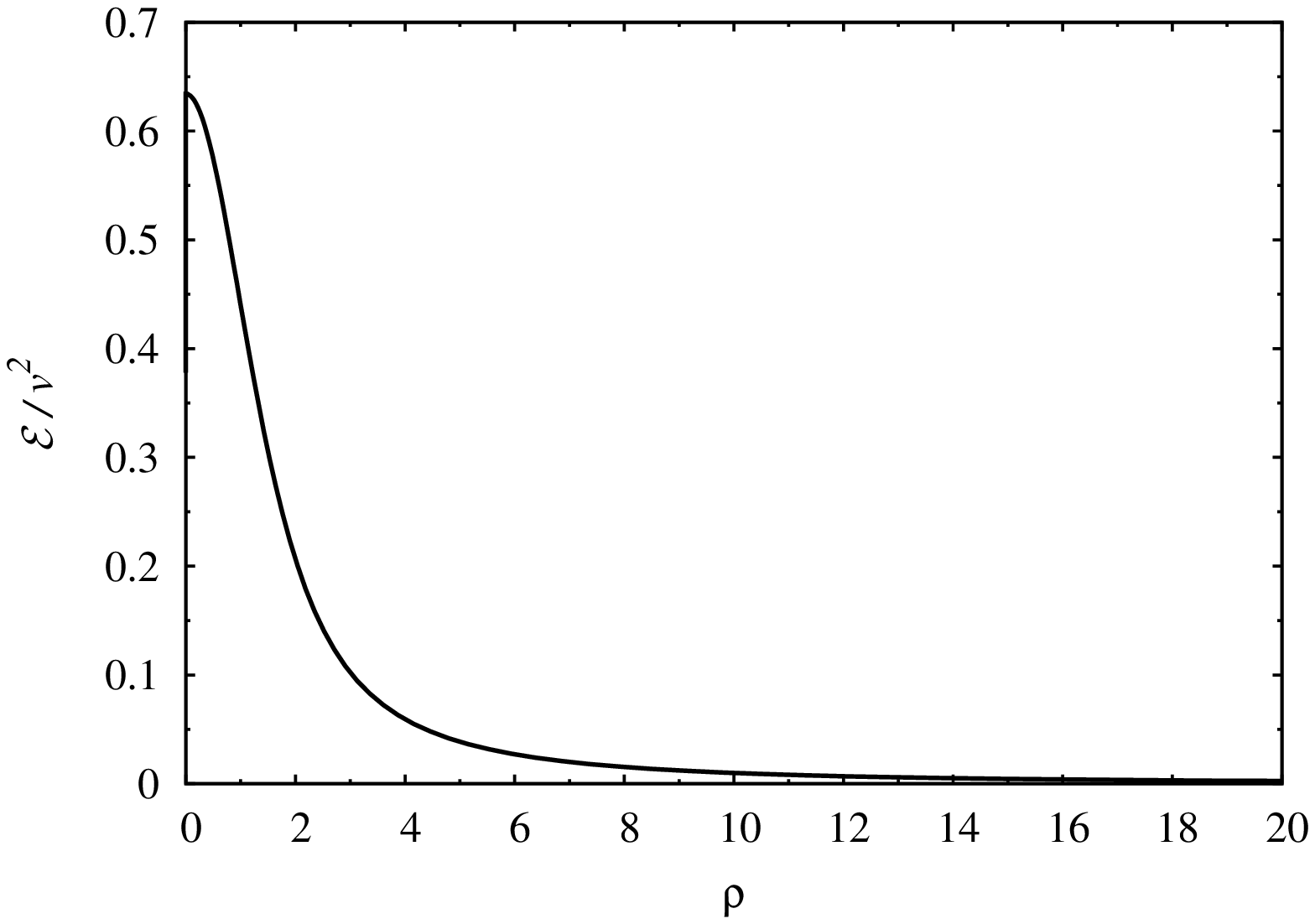}}}
\mbox{\subfloat[]{\includegraphics[width=0.45\linewidth]{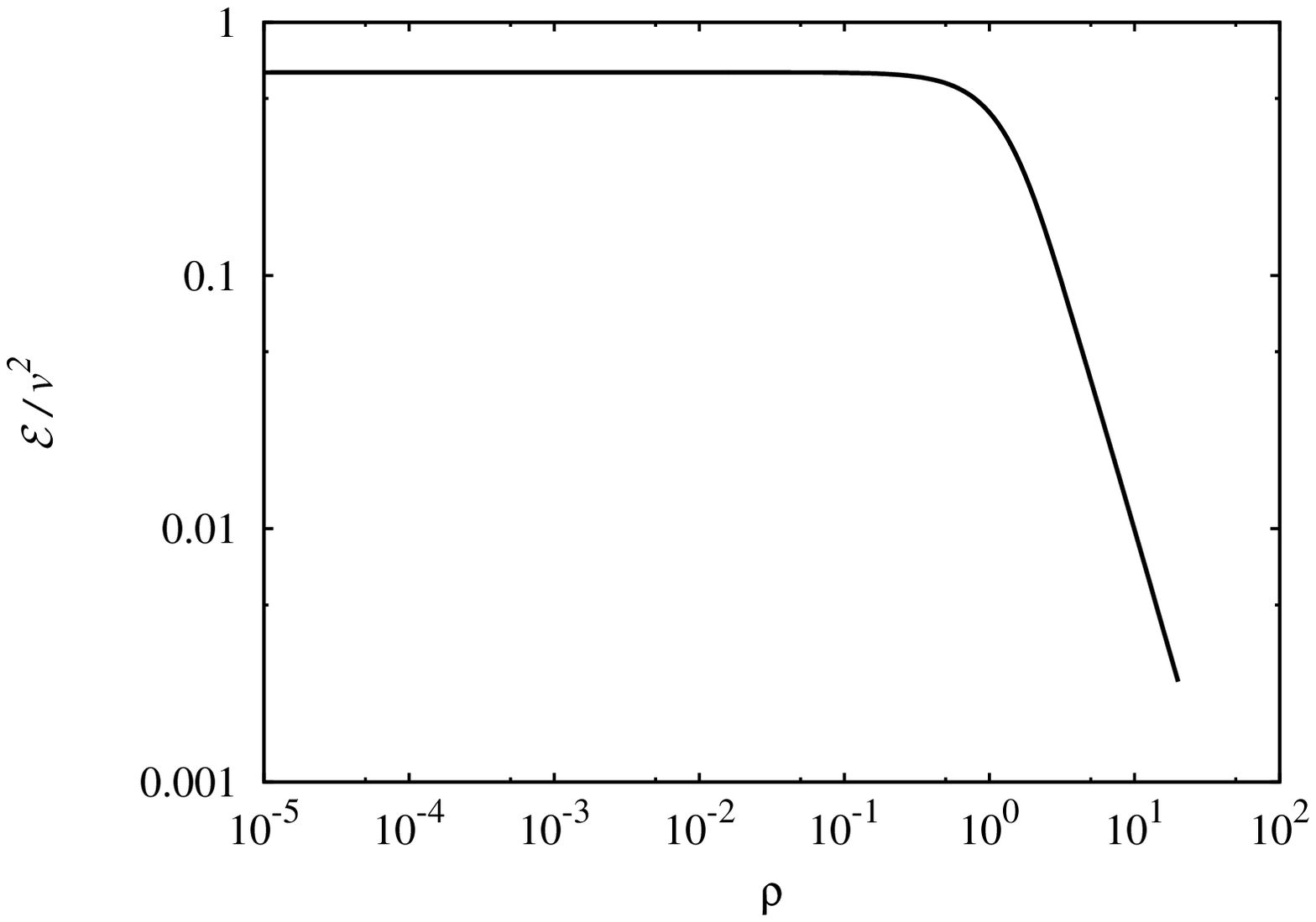}}}
\caption{(a) profile function $h$, (b) energy density and (c) energy
  density on loglog scale for the hedgehog ($Q=1$) monopole. }
\label{fig:Q1}
\end{center}
\end{figure}
In Fig.~\ref{fig:Q1} is shown the profile function and energy density
of the hedgehog configuration.

\section{Charge-two and charge-zero monopoles in the point-charge
  approximation}  \label{sec:puntosez}

Let us now turn to the double-winding global monopole configuration. 
The overall orientation of the monopole is not important, but the
relative orientation is important for the multi-monopole solution. 
In the following we will write the scalar field as
\beq
\Phi = h \left\{\sin g\cos k,\sin g \sin k,\cos g\right\},
\label{eq:ansatz}
\eeq
which describes a charge-two monopole if $k$ winds twice and $g$ winds
once.
Another possibility is to make $g$ wind two times and $k$ one time,
but it requires a twist (sign change) of for instance $k$, in order
not to create a monopole-anti-monopole pair. 
To see the charge explicitly, let us evaluate the integral
\eqref{eq:Q} for $g=m\theta$ and $k=n\phi$ at some asymptotic distance
where $h=1$:
\beq
Q = \frac{1}{8\pi}\oint dS^{ij}
\epsilon_{abc} \Phi^a\partial_i\Phi^b\partial_j\Phi^c
= n\sin^2\left(\frac{m\pi}{2}\right).
\label{eq:Q=2}
\eeq
We will study a number of cases in turn.

For physical and everywhere regular global monopoles, the profile
function $h$ cannot take the value 1 over all space, but should have
at least $Q$ zeros, counted with multiplicity. Regularity dictates
that the zeros of $h$ coincide with the origins of where the phases of
$g$ and $k$ wind about.
The approximation of setting $h:=1$ can physically be seen as the
limit of very small monopole sizes compared to the separation
distances.
Once we set $h:=1$, we introduce singularities at these origins or
positions of the charges. These singularities are dealt with by
subtracting the same singularity of a single singular global monopole
field. The difference is then a regular field.
This approximation captures the asymptotic interaction of global
monopoles, where the potential has no effect, but only the winding
phases interact at long distances via the kinetic term.
This interaction is important as it can never be neglected.

Independent of the local regularity of the global monopole, the total
energy of the configuration always diverges linearly. This divergence
is just an artifact of putting a global monopole in an infinite
space. In real world applications, such as liquid crystals, global
monopoles only live in a finite region of space. In this paper we will
consider the global monopoles on an infinite space, although the
numerical calculations will of course only be carried out on a finite
space, just like the physical global monopoles enjoy.

A comment is in order about the Ans\"atze that we will use for the
function $g$ and $k$ in the next subsections. They are not describing
\emph{solutions}, but describe the winding that necessarily will be
present on a given timeslice in a would-be solution.
Logically, there are two possible situations. If the two charges
attract, a solution will be close to a configuration where the charges
are coincident. By making an educated guess for the function $h$, the
full, time-dependent solution can then be found using the full
equations of motion.
The other possibility is that the charges repel each other. In that
case, no stable solution exists and this type of configuration is
never a static solution, but at best a snapshot of a time dependent
field configuration that evolves in time.
Real time dependent solutions to the equations of motion can be
constructed, but since we are not interested in unstable time
dependent configurations, we will simply disregard such cases as being 
what they are: unstable.

\subsection{Type A}

The first case is the charge-two monopole where $k$ winds twice and
$g$ once
\beq
\Phi_{\rm full} = h \left\{
\sin\theta\cos(\phi_1\pm\phi_2),
\sin\theta\sin(\phi_1\pm\phi_2),
\cos\theta
\right\}, \label{eq:tipoa}
\eeq
where we define the angles as
\beq
\theta = \arccos\frac{z}{r}, \quad
\phi_1 = \arctan\frac{y}{x-d}, \quad
\phi_2 = \arctan\frac{y}{x+d},
\eeq
and the two monopole charges are separated by a distance $2d$.
We will treat a case of a monopole-anti-monopole on the same footing
as the monopole-monopole by choosing the lower sign. 
The monopole charge of the configuration does not depend on $d$; hence
if we set $d=0$, we can see from Eq.~\eqref{eq:Q=2} that this
configuration has charge 2 for the upper sign and charge zero for the
lower sign.

In order to study the stability of this monopole, let us make a
point-charge approximation, where $h:=1$. 
This approximation corresponds to ignoring the potential and taking
into consideration the asymptotic effect of two charges affected only
by the kinetic term.
See also the discussion in the previous subsection.

The form \eqref{eq:tipoa} has been chosen for its simplicity.  However
the price of this simple form is that, when $d\neq 0$, the field
$\Phi_{\rm full}$ does not wind around the vacuum manifold when
restricted to a small sphere centered on the singular point at $x=\pm
d$.  Indeed, on such a small sphere the value of $\cos\theta$ is
always near zero.  The solution restricted to a 2-surface can only
have nontrivial winding on the vacuum manifold if at some point
$\sin\theta=0$, which means that the 2-surface must intersect the
$z$ axis and so extend a distance $d$ from the singular point.   In
fact, the kinetic energy diverges along the infinite lines extending
along the $z$ axis from the two singular points, which leads to a
short-distance logarithmic divergence in the energy.  

This divergence is simply an artifact of our simple functional form
\eqref{eq:tipoa}.
In \ref{app:mm} and in Sec.~\ref{sec:skyrmisez},
when we consider true solutions of the equations of motion, the field
configurations will be regular everywhere and so will the energy be.
However, in the present section we are only interested in the
interactions between monopoles.
What we will do is to subtract off a counter term, which cancels the
divergence. We will call this difference the interaction energy and it
is free from short-distance divergences.
A concern may be that the exact form of the counter term may change
the functional behavior of the interaction energy; however, to leading
order in $d$ it is independent of the counter term. 
To see this, we define the following interaction potential 
\beq
E_{\rm int} \equiv E[\Phi_{\rm full}] - E[\Phi_1] - E[\Phi_2],
\label{eq:Eint}
\eeq
where $\Phi_i$ is the $i$-th single monopole which we subtract off
\beq
\Phi_i = h \left\{
\sin\theta\cos\phi_i,
\sin\theta\sin\phi_i,
\cos\theta
\right\}.
\eeq
The single monopoles that are subtracted have the same two lines of
divergences at $x=\pm d,y=0$ and so the respective contributions of
their short-distance divergences to the interaction energy cancel.
The interaction energy is sufficient to determine the radial force
between two monopoles, which is simply its derivative with respect to
$d$.

The interaction energy is
\begin{align}
\frac{E_{\rm int}}{v^2} &= - \frac{1}{2} \int_{\rho\leq L,|z|<L} d^3x\;
\frac{d^4 - 2d^2\left[x^2-y^2\mp (y^2+x^2)\right] + (1\mp 2)(x^2+y^2)^2}
{\left[(d-x)^2+y^2\right]\left[(d+x)^2+y^2\right](x^2+y^2+z^2)} \\
&= -\int_0^L d\rho \int_0^{2\pi} d\phi \;
\frac{d^4\pm 2d^2\rho^2+(1\mp 2)\rho^4-2d^2\rho^2\cos 2\phi}
  {d^4+\rho^4-2d^2\rho^2\cos 2\phi} 
  \arctan\left(\frac{L}{\rho}\right) \non
&= -2\pi\int_0^L d\rho \;
\left(1\pm\sign(d-\rho)\frac{2\rho^2}{d^2+\rho^2}\right) 
\arctan\left(\frac{L}{\rho}\right) \non
&= 
-\frac{1\mp 2}{2}\pi(\pi+2\log 2)L
\mp 8\pi d \arctan\frac{L}{d}
\pm 2\pi^2 d \arctan\frac{2dL}{L^2-d^2} \non
&\phantom{=\ }
\mp 4\pi L\log\left(1+\frac{d^2}{L^2}\right) 
\mp 4\pi d\log 2\log \frac{L+d}{L-d} \non
&\phantom{=\ }
\pm\frac{1}{2}\pi d\Li_2\left[-\left(\frac{d-L}{d+L}\right)^2\right]
\mp\frac{1}{2}\pi d\Li_2\left[-\left(\frac{d+L}{d-L}\right)^2\right]
\pm\pi d\Li_2\frac{d+L}{d-L} \non
&\phantom{=\ }
\mp\pi d\Li_2\frac{d-L}{d+L}
\pm\pi d\Re\Li_2\frac{(d+L)(d+iL)}{(d-L)(d-iL)}
\mp\pi d\Re\Li_2\frac{(d-L)(d+iL)}{(d+L)(d-iL)} \non
&= -\frac{1\mp 2}{2}\pi(\pi+2\log 2)L \mp 4\pi^2d 
  + \mathcal{O}\left(\frac{d^2}{L}\right).
\nonumber
\end{align}
The domain of integration here is chosen, for convenience, to be a
cylinder (as opposed to a sphere). 
The first term in the linear expansion does depend on the geometry of
the integration region (cylinder versus sphere etc.), however the
second term, corresponding to the force does not. 
Again the upper sign is for the monopole-monopole case and the lower
sign is for the monopole-anti-monopole case. 

The mutual force between the monopole and the (anti-)monopole for
small $d$ is 
\beq
\frac{1}{v^2}F_{\rm int} \simeq \pm 4 \pi^2,
\eeq
and so the two monopoles (monopole-anti-monopole) repel (attract). 

\begin{figure}[!htp]
\begin{center}
\mbox{
\subfloat[monopole-monopole]{\includegraphics[width=0.49\linewidth]{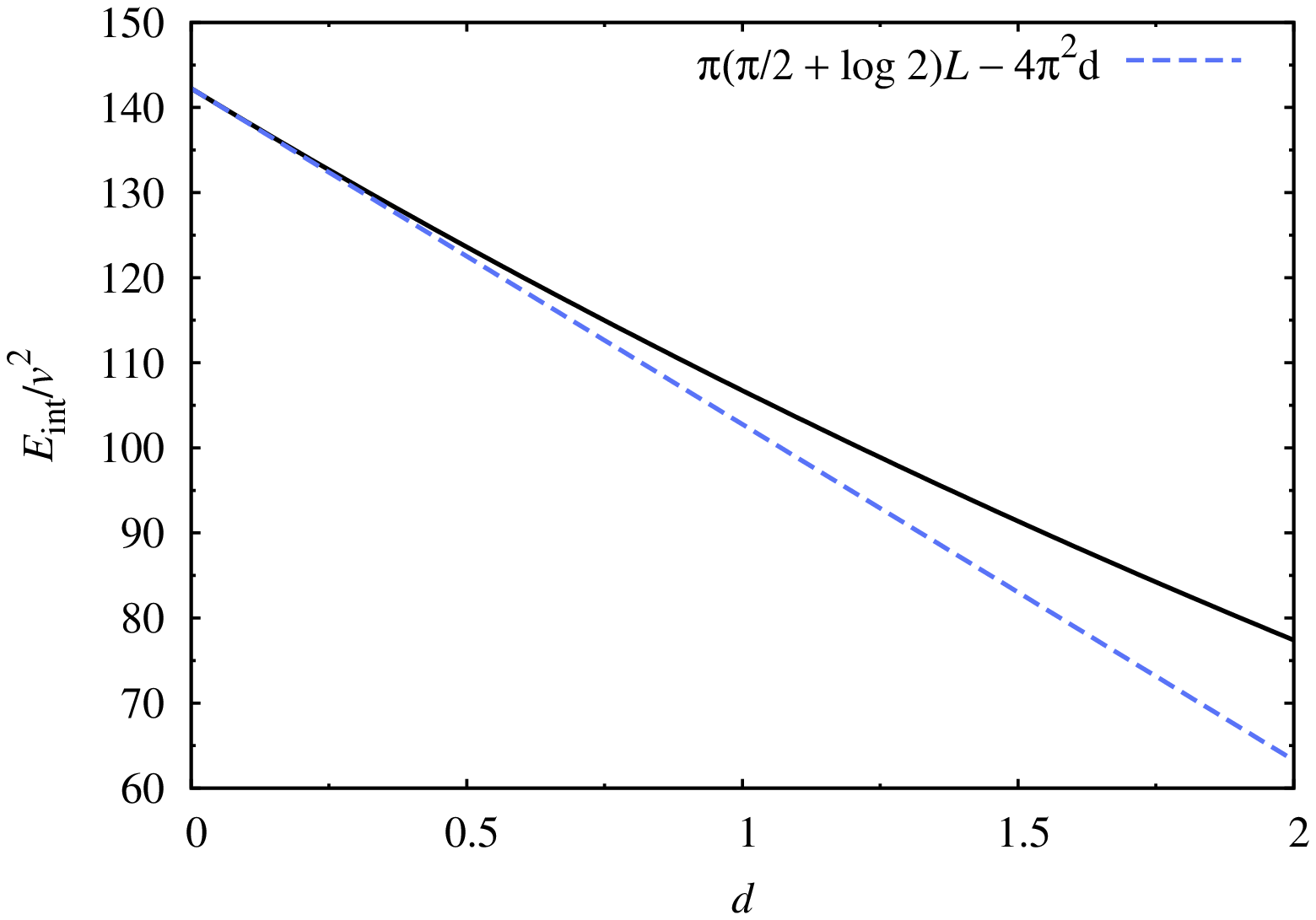}}
\subfloat[monopole-anti-monopole]{\includegraphics[width=0.49\linewidth]{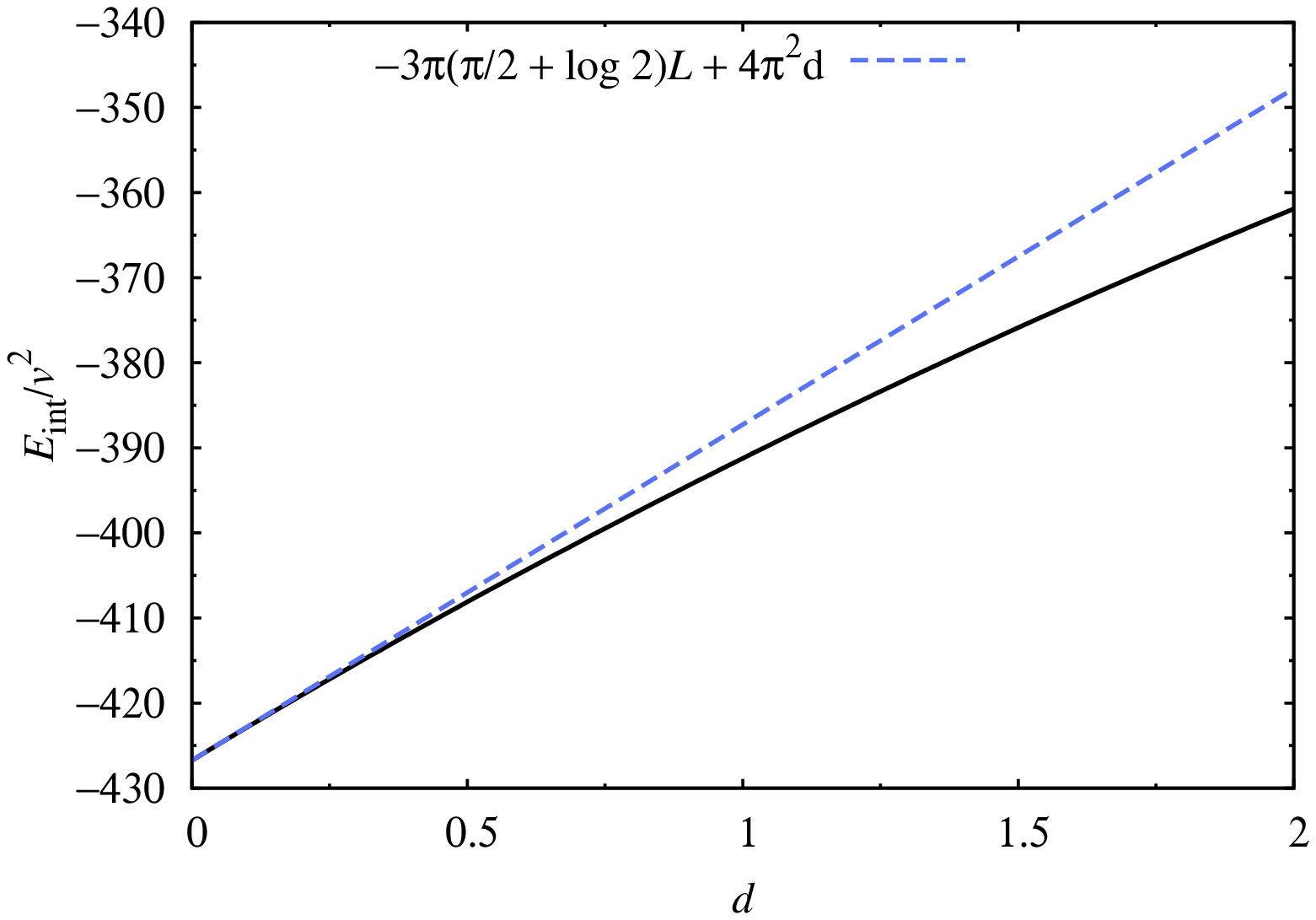}}}
\caption{Interaction energy of (a) the type A monopole-monopole pair
  and (b) the type A monopole-anti-monopole pair as a function of
  separation distance $2d$ for a cut-off value $L=20$. The blue dashed
  line shows linear expansions in $d$. } 
\label{fig:Eint_A}
\end{center}
\end{figure}

Fig.~\ref{fig:Eint_A} shows the interaction energies for both cases of
type A. 

\subsection{Type B}

The next case is a charge-two monopole where $k$ only winds once and
$g$ winds twice \cite{Perivolaropoulos:1991du}
\beq
\Phi_{\rm full} = h \left\{
\sin(\theta_1+\theta_2)\cos\phi,
\epsilon\sin(\theta_1+\theta_2)\sin\phi,
\cos(\theta_1+\theta_2)
\right\}, \label{eq:tipob}
\eeq
where we define the angles as
\begin{align}
\theta_1 = \arccos\frac{z-d}{\sqrt{x^2+y^2+(z-d)^2}}, \quad
\theta_2 = \arccos\frac{z+d}{\sqrt{x^2+y^2+(z+d)^2}}, \quad
\phi_1 = \arctan\frac{y}{x},
\label{eq:Bangles}
\end{align}
and the monopole charges are separated again by a distance $2d$.  
$\epsilon$ is the twist function; if $\epsilon=1$ the configuration
describes a monopole-anti-monopole pair, which was considered in
Ref.~\cite{Perivolaropoulos:1991du}.
In order to get a charge-two monopole configuration, we need to twist
the second cycle in $g$, i.e.,
\beq
\epsilon = (-1)^{\left\lfloor\frac{2\theta}{\pi}\right\rfloor},
\eeq
where $\lfloor x\rfloor$ is the floor function of $x$ and 
\beq
\theta = \arccos\frac{z}{\sqrt{x^2+y^2+z^2}},
\eeq
is $\theta$ measured from the origin which is the midpoint between the
two monopoles.

Let us now study the point-charge approximation, where $h:=1$. 
We again use the interaction potential \eqref{eq:Eint} and $\Phi_i$,
which is the $i$-th monopole, is now given by
\beq
\Phi_i = h \left\{
\sin\theta_i\cos\phi,
\sin\theta_i\sin\phi,
\cos\theta_i
\right\}.
\label{eq:Phi_i_theta}
\eeq
The interaction energy thus reads
\begin{align}
\frac{1}{v^2} E_{\rm int} &= -2 \int_{r\leq L} d^3x\;
\frac{d^2-z^2}{(d^2+r^2)^2 - 4d^2z^2} \non
&= \frac{\pi}{2d}\int_{-L}^L dz\;
\frac{d^2-z^2}{z}\left[\log\frac{(d-z)^2}{(d+z)^2}
  +\log\frac{d^2 + L^2 + 2 d z}{d^2 + L^2 - 2 d z}\right] \non
&= -\frac{\pi L}{2}\left(\frac{L^2}{d^2}-3+\frac{2\pi^2 d}{L}\right)
+\frac{\pi L}{2}\left(\frac{L^3}{d^3}+\frac{2L}{d}-\frac{3d}{L}\right)
\arctanh\left(\frac{d}{L}\right) \non
&\phantom{=\ }
+2\pi d \eta\left(\frac{d}{L}\right)
+\pi d \eta\left(\frac{2 d L}{d^2+L^2}\right) \non
&= \frac{8\pi L}{3} - d\pi^3 
+ \mathcal{O}\left(\frac{d^2}{L}\right),
\end{align}
where 
\beq
\eta(x) \equiv \Li_2(x) - \Li_2(-x),
\eeq
and $\Li_2$ is the Spence function (also called dilogarithm).
Note that the negative sign in front of the order $d$ term in the
expansion means that the system has negative binding energy and the
mutual force is repulsive
\beq
\frac{1}{v^2}F_{\rm int} \simeq \pi^3.
\eeq

Note that the interaction energy remains the same for both the case
with the twist function (the monopole-monopole case) and without the
twist function (the monopole-anti-monopole case). 

\begin{figure}[!htp]
\begin{center}
\includegraphics[width=0.49\linewidth]{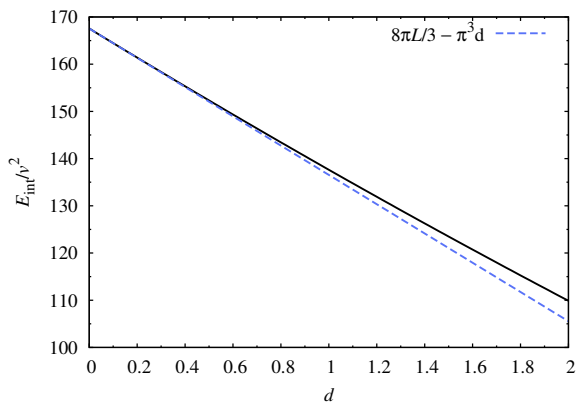}
\caption{Interaction energy of the type B monopole-monopole pair
  (which is the same as the type B monopole-anti-monopole) as a
  function of separation distance $2d$ for a cut-off value $L=20$. The
  blue dashed line shows the linear expansion in $d$. } 
\label{fig:Eint_B}
\end{center}
\end{figure}

In Fig.~\ref{fig:Eint_B} is shown the interaction energy of the
charge-two monopole (which is the same as the monopole-anti-monopole)
of type B.

Let us mention that this configuration has no lines of divergences, so
only the singular points at the origins of the charges are still
present in $\Phi_{\rm full}$ but $E_{\rm int}$ is again free from
these singularities. 

We will now consider the total energy, which is the same for both the
monopole-monopole and monopole-anti-monopole case,
\begin{align}
\frac{E_{\rm total}}{v^2} &= 2\int_{r\leq L} d^3x\; 
\frac{r^2+z^2}{(d^2+r^2)^2-4d^2z^2} \non
&= \int_{-L}^L \frac{dz}{z} \left[
(d^2-z^2-2dz)\log\frac{(d-z)^2}{d^2+L^2-2dz}
+(d^2-z^2+2dz)\log\frac{(d+z)^2}{d^2+L^2+2dz}\right] \non
&= -\frac{\pi L}{2}\left(\frac{L^2}{d^2}-11+\frac{2\pi^2d}{L}\right)
+\frac{\pi L}{2}\left(\frac{L^3}{d^3}+\frac{10L}{d}-\frac{11d}{L}\right)
\arctanh\left(\frac{d}{L}\right) \non
&\phantom{=\ }
+2\pi d\eta\left(\frac{d}{L}\right)
+\pi d\eta\left(\frac{2dL}{d^2+L^2}\right) \non
&=\frac{32\pi L}{3} - d\pi^3 + \mathcal{O}\left(\frac{d^2}{L}\right). 
\end{align}
Again the total energy $E[\Phi_{\rm full}]$ diverges linearly in $L$
for both configurations.

\subsection{Type C}

The last case, which we consider for completeness, is a
monopole-anti-monopole where $k$ winds once and $g$ winds and unwinds 
at spatially separate locations
\cite{Perivolaropoulos:1991du}
\beq
\Phi_{\rm full} = h \left\{
\sin(\theta_1-\theta_2)\cos\phi,
\sin(\theta_1-\theta_2)\sin\phi,
\cos(\theta_1-\theta_2)
\right\},
\eeq
where the angles are given in Eq.~\eqref{eq:Bangles} and the monopole
and anti-monopole charges are separated again by a distance $2d$.
This configuration was also considered in
Ref.~\cite{Perivolaropoulos:1991du}. 
We again use the point-charge approximation where $h:=1$. 
The interaction potential is given in Eq.~\eqref{eq:Eint} and the
single monopole fields are the same as in Eq.~\eqref{eq:Phi_i_theta}. 
The interaction energy thus is
\begin{align}
\frac{1}{v^2}E_{\rm int} &= 2\int_{r\leq L} d^3x\;
\frac{d^2-r^2}{(d^2+r^2)^2 - 4d^2z^2} \non
&= \pi\int_{-L}^L \frac{dz}{z}\;
\left[
(z-d)\log\frac{(z-d)^2}{d^2 + L^2 - 2 d z}
+(z+d)\log\frac{(z+d)^2}{d^2 + L^2 + 2 d z}
\right] \non
&= - \frac{\pi(L^2-d^2)}{d}\log\frac{(d+L)^2}{(d-L)^2}
- 4\pi L + 2\pi^3 d
- 4\pi d \eta\left(\frac{d}{L}\right)
- 2\pi d \eta\left(\frac{2 d L}{d^2+L^2}\right) \non
&= -8\pi L + 2 d\pi^3 + \mathcal{O}\left(\frac{d^2}{L}\right).
\end{align}
Note that the positive sign in front of the order $d$ term in the
expansion means that the system has a positive binding energy and the
mutual force is attractive
\beq
\frac{1}{v^2}F_{\rm int} \simeq -2\pi^3.
\eeq

\begin{figure}[!htp]
\begin{center}
\includegraphics[width=0.49\linewidth]{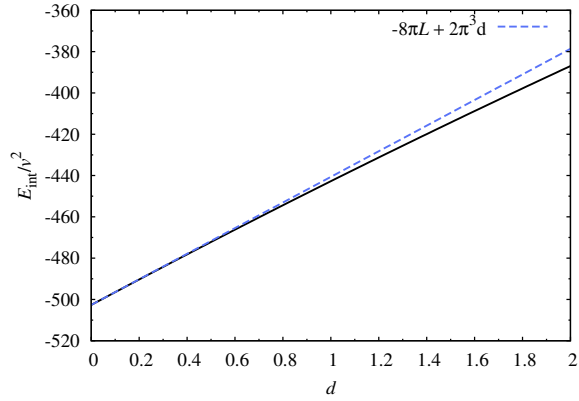}
\caption{Interaction energy of the type C monopole-anti-monopole as a 
  function of separation distance $2d$ for a cut-off value $L=20$. The
  blue dashed line shows the linear expansion in $d$. } 
\label{fig:Eint_C}
\end{center}
\end{figure}

In Fig.~\ref{fig:Eint_C} is shown the interaction energy of the
monopole-anti-monopole of type C.

Finally, we will evaluate the total energy for the type C
monopole-anti-monopole
\begin{align}
\frac{1}{v^2}E_{\rm total} &= 4d^2 \int_{r\leq L} d^3x\;
\frac{1}{(r^2+d^2)^2-4d^2z^2} \non
&= \pi d \int_{-L}^L \frac{dz}{z} \left[
\log\frac{(d+z)^2}{(d-z)^2}
+\log\frac{d^2+L^2-2dz}{d^2+L^2+2dz}\right] \non
&= 2\pi^3d - 4\pi d\eta\left(\frac{d}{L}\right)
-2\pi d\eta\left(\frac{2dL}{d^2+L^2}\right) \non
&= 2d\pi^3 + \mathcal{O}\left(\frac{d^2}{L}\right).
\end{align}
Notice that this and only this case (of those considered here) has a
finite total energy.

Let us finally mention that the interaction energy is regular in this
case but has a linear divergence in the integral (the constant),
whereas the total energy has 2 singular points at the monopole charges
but has a finite total energy.

\subsection{Summary}

The conclusion we can draw from the point-charge approximation is that
the charge-two monopole is unstable given both orientations we have 
constructed. 
The mutual repulsion could be slightly altered by including the
profile function $h$ of the configuration, which in turn includes the
effect of the potential. 
Asymptotically, however, the instability of the global charge-two
monopole remains, as we have demonstrated above.
For concreteness we carry out a numerical calculation of a perturbed 
two-monopole configuration in \ref{app:mm} to confirm that they
are indeed unstable.
In the next section we will begin to contemplate how to stabilize the
global two-monopole.

\section{Caging global monopoles}\label{sec:caging}

In this section and in the remainder of this paper, we consider the
monopole profile function and hence the full equations of motion. 
The first attempt at stabilizing the global monopoles is to modify the 
potential in such a way that it is energetically favorable to remain
near the spatial origin of a certain coordinate frame. This origin
could then be determined dynamically by a solution to a different
sector, which we here for concreteness take to be a one-Skyrmion
sector.

In this section, we would like to advocate that it is not as trivial
to realize this idea as one might expect.
We consider for concreteness the following theory
\begin{align}
-\mathcal{L} &= \frac{v^2}{2}\Tr(\p_\mu\Phi)^2
+ \frac{\widetilde{\lambda}v^4}{4}(1+n_4)^\alpha
\left(1-\Tr[\Phi^2]\right)^2,
\end{align}
where the field $n_4$ is a component of the Skyrmion sector which
obeys the boundary conditions $n_4(0)=-1$ (at the origin of the
coordinate frame) and $n_4(\infty)\to 1$.

The idea is that at the origin, the potential vanishes and hence the
contribution to the energy from the potential is minimal when the
monopole is situated at the origin. 
If the monopole tried to get out, it should start to feel the potential
growing and thus it would be energetically favorable to return to the 
origin. The hope is then that by tuning the potential parameters, the
two monopoles still prefer to stay at the origin although they are
mutually repulsive. There is however a monkey wrench around the
corner. Hence, instead of presenting a numerical calculation, let us
examine the expectations with a very crude calculation. 

Let us assume that the Skyrmion has the size $L_{\rm sk}$, which for all
practical purposes here just means that
\beq
n_4(r) = -1 + \frac{2r}{L_{\rm sk}}, \qquad
\textrm{for} \ r<L_{\rm sk},
\eeq
and $r$ is the radial coordinate measured from the origin.
Let us further assume that a single monopole has the profile shape
$\Phi^a=\sigma^a r/L_{\rm m}$ for $r<L_{\rm m}$. We consider the
case $L_{\rm sk}\gg L_{\rm m}$, which is what we would prefer for
physical reasons.
The repulsive force comes from the kinetic term and was estimated in
Sec.~\ref{sec:puntosez} to be
\beq
F_{\rm repulsive} = -4\pi^2 v^2,
\eeq 
and the attractive force due to the above potential can be estimated
as
\begin{align}
F_{\rm attractive} &= \p_d \int d^3x\;
\frac{\widetilde{\lambda}v^4}{4}[1+n_4(d)]^\alpha
\left(1-\Tr[\Phi^2]\right)^2 \non
&= \frac{2\alpha\widetilde{\lambda}v^4}{L_{\rm sk}}\int dr \;r^2
\left(\frac{2d}{L_{\rm sk}}\right)^{\alpha-1}
\left(1-\frac{r^2}{L_{\rm m}}\right)^2.
\end{align}
Since we assumed that the monopole size is much smaller than the size
of the Skyrmion ($L_{\rm m}\ll L_{\rm sk}$), we can crudely pull the 
factor of $(2d/L_{\rm sk})^{\alpha-1}$ out of the integral and we get
\begin{align}
F_{\rm attractive} =
\frac{16\alpha\pi\widetilde{\lambda}v^4L_{\rm m}^3}{105L_{\rm sk}}
\left(\frac{2d}{L_{\rm sk}}\right)^{\alpha-1}.
\end{align}
Now, the distance from the center of the origin $d$, has to be small
for the potential to be effective (in the sense of exerting a force on
the monopole); that is, $d<L_{\rm sk}$.\footnote{One may naively think
  that the factor is helping at making the force bigger when
  $d>L_{\rm sk}/2$, but in the real Skyrmion solution, the tail is
  much flatter at such distances from the origin and hence the
  derivative becomes small.}
Let us now compare the forces, i.e.~the attractive one versus the
repulsive one
\beq
\frac{4\alpha\widetilde{\lambda}v^2L_{\rm m}^3}{105L_{\rm sk}}
\left(\frac{2d}{L_{\rm sk}}\right)^{\alpha-1} > \pi.
\eeq
The inequality shows that a larger Skyrmion provides a smaller
attractive force (this is also expected as the derivative of the field
profile becomes smaller). The real kicker is that by trying to
increase the potential factor, i.e.~$\widetilde{\lambda}v^2$, the
monopole size decreases by the same factor, namely
$L_{\rm m}^{-2}\sim \widetilde{\lambda}v^2$. This means that the only
parameter we can dial to stabilize the monopole(s) is
$\alpha$. However, a large value of $\alpha$ also makes the factor
$(2d/L_{\rm sk})^{\alpha-1}$ small.
In other words, once we try to squeeze in the monopole(s) by the
potential, they shrink and can escape as easily as before. 

This is not supposed to be a proof of no-go, but merely a
justification of not constructing the theory with just a modified
potential. We have tried several numerical attempts and have observed
exactly this behavior, namely that the monopoles shrink and are able
to escape. We do not consider those attempts as a proof of no-go
either and will not present this study here. 
In the next section we will pursue a different strategy.

\section{Skyrmopole}\label{sec:skyrmisez}

In this section we will place a global two-monopole inside a
potential so that each monopole will not run away despite mutual
repulsion.
In view of the estimates of the last section, we consider a coupling
between a one-Skyrmion and the two-monopole via a metric-like
prefactor of the monopole part of the Lagrangian density. The aim is
that the monopole constituents prefer to stay in the center of the 
``potential'', so much so that the repulsion is negligible. 
If a monopole constituent is to exit the Skyrmion, it will enjoy the 
instabilities described in the Sec.~\ref{sec:puntosez}. 

Let us first contemplate the construction in order to know what to
expect. We start with a Skyrmion which has a kinetic term, a Skyrme
term and a mass term. The size of the Skyrmion is given by the balance
of these three terms. 
The global monopole, on the other hand, has a size squared given by
the inverse coefficient of the self-interacting potential term.
In front of the monopole sector we multiply by a factor depending on
the profile of the Skyrmion sector. 

Putting the pieces together gives us the Lagrangian density
\beq
\mathcal{L} = \mathcal{L}_{\rm sk}
+ \mathcal{G}(n_4)\mathcal{L}_{\rm m}
\label{eq:Lskm}
\eeq
where the Skyrmion Lagrangian is simply
\begin{align}
-\mathcal{L}_{\rm sk} &=
\frac{c_2}{2} (\p_\mu\mathbf{n}\cdot\p^\mu\mathbf{n})
+ \frac{c_4}{4} \left[(\p_\mu\mathbf{n}\cdot\p^\mu\mathbf{n})^2 -
  (\p_\mu\mathbf{n}\cdot\p_\nu\mathbf{n})^2\right]
+ m^2(1-n_4),
\label{eq:Lsk}
\end{align}
the monopole Lagrangian reads
\beq
-\mathcal{L}_{\rm m} =
\frac{v^2}{2}\Tr(\p_\mu\Phi\p^\mu\Phi)
+\frac{\lambda v^4}{4}\left(1-\Tr[\Phi^2]\right)^2,
\label{eq:Lm}
\eeq
and the coupling is chosen to be
\beq
\mathcal{G}(n_4) =
\left(\frac{1+b+n_4}{2+b}\right)^\alpha,
\label{eq:G}
\eeq
where the fields $n_p$, with $p=1,\ldots,4$ are unitless O(4) fields 
describing the Skyrmion part of the configuration and
$\Phi=\Phi^a\sigma^a$, $a=1,2,3$ are SU(2) adjoint unitless monopole
fields.
The coefficients in the Lagrangians are $c_2$ and $m$ with units
of mass squared, $v$ has units of mass, and finally $c_4$, $\lambda$,
$b$ and $\alpha$ are unitless parameters.
$c_2$, $c_4$ and $m$ determine the size of the Skyrmion and
$\sqrt{\lambda}v$ determines the size of the monopole(s). $b$ is
included in order for the monopole energy not to vanish at the center
of the Skyrmion and larger values of $\alpha$ translate to a larger
attractive force among the monopoles.
One may interpret the coupling, $\mathcal{G}$, as an effective metric
felt by the monopole sector of the theory (although here it is just
incarnated as a non-renormalizable field theory coupling). 

If we compare to the considerations made in the last section, the
difference is that now the size of the monopoles and the size of the
energy barrier are independent parameters of the theory. 

The static equations of motion are
\begin{align}
c_2 \p^2 n_p 
+ c_4 \left(\p_i\mathbf{n}\cdot\p_i\mathbf{n}\right) \p^2 n_p
+ c_4 \left(\p_i\p_j\mathbf{n}\cdot\p_j\mathbf{n}\right) \p_i n_p
- c_4 \left(\p^2\mathbf{n}\cdot\p_i\mathbf{n}\right)\p_i n_p 
\phantom{\ = 0,}\non
\mathop- \frac{\alpha v^2}{2+b}\left(\frac{1+b+n_4}{2+b}\right)^{\alpha-1}
\left[\frac{1}{2}\Tr(\p_\mu\Phi)^2 \delta^{p4}
  + \frac{\lambda v^2}{4}\left(1-\Tr[\Phi^2]\right)^2\right]
+ m^2 \delta^{p4} = 0, \label{eq:eom_sk_mono1}\\
\p^2 \Phi
+ \frac{\alpha}{1+b+n_4}\p_\mu n_4\p^\mu\Phi^a
+ \lambda v^2 \left(1-\Tr[\Phi^2]\right) \Phi = 0. 
\label{eq:eom_sk_mono2}
\end{align}
The vacuum equations are
\beq
\Tr[\Phi^2] = 1, \qquad
n_4 = 1.
\eeq
Note that even without the mass term for $\mathbf{n}$, there is no
explicit breaking of O(4) symmetry, the symmetry is still broken
spontaneously down to O(3) for any finite energy state. 
This breaking, O(4) $\to$ O(3), supports the Skyrmion(s). 
The spontaneous breaking of SU(2) down to U(1) by the potential
supports the monopoles.

There is a subtlety due to the coupling of the monopole theory to the
Skyrmion sector in this way, namely that the second last term in
Eq.~\eqref{eq:eom_sk_mono1} cannot exceed the value of the mass term
in order for the field, $n_4$, to be able to return to its vacuum
expectation value. When this is not fulfilled, so-called pion
condensation occurs and the vacuum equations are completely different,
see e.g.~Ref.~\cite{Gudnason:2015nxa}. This leads to the following
(strong) condition:
\beq
\frac{\alpha v^2}{2+b}
\left[\frac{1}{2}\Tr(\p_\mu\Phi)^2 
  + \frac{\lambda v^2}{4}\left(1-\Tr[\Phi^2]\right)^2\right]
< m^2,
\label{eq:flipconstraint}
\eeq
where we have neglected the factor of $((1+b+n_4)/(2+b))^{\alpha-1}$,
as it is always smaller or equal to unity (by construction).
This means that the region in parameter space we are interested in is
where $\alpha$ is big (to get a sizable attractive force) and $v$ is
small (to avoid breaking the vacuum conditions).\footnote{If the
  boundary conditions are switched so that $n_4=1$ at the origin and
  $n_4=-1$ at spatial infinity, the same condition holds with
  $n_4\to-n_4$, i.e.~it does not ameliorate the problem. }

The Skyrmion charge is also a topological charge and can be
calculated as
\beq
B = -\frac{1}{12 \pi^2} \int d^3x \; \epsilon^{pqrs} \epsilon^{ijk} 
\partial_i n_p \partial_j n_q \partial_k n_r n_s.
\eeq

In order to get a feeling for the parameters in the theory, let us
consider a 1-Skyrmion with a single monopole inside, i.e.~$Q=B=1$. 
For this we can choose the following Ans\"atze 
\begin{align}
\mathbf{n} &= \left\{\hat{x}^1\sin f(r),\hat{x}^2\sin
f(r),\hat{x}^3\sin f(r),\cos f(r)\right\}, \label{eq:nhedgehog}\\
\Phi &= h(r) \hat{x}^a \sigma^a, \label{eq:fhedgehog}
\end{align}
where $\hat{x}^a=x^a/r$ is the spatial unit vector and $r=|x|$ is the
radial coordinate. 
Inserting the above into the Lagrangians \eqref{eq:Lsk} and
\eqref{eq:Lm}, we get 
\begin{align}
-\mathcal{L}_{\rm sk} &= 
\frac{c_2}{2}f_r^2
+\frac{c_2}{r^2}\sin^2f
+\frac{c_4}{r^2}\sin^2(f)f_r^2
+\frac{c_4}{2r^4}\sin^4f
+ m^2(1-\cos f), \\
-\mathcal{G}(n_4)\mathcal{L}_{\rm m} &=
\left(\frac{1+b+\cos f}{2+b}\right)^\alpha
\left[\frac{v^2}{2}h_r^2
+\frac{v^2}{r^2}h^2
+\frac{\lambda v^4}{4}(1-h^2)^2\right],
\end{align}
which give rise to the following equations of motion
\begin{align}
c_2\left(f_{rr} + \frac{2}{r}f_{r} - \frac{1}{r^2}\sin 2f\right)
+c_4\left(\frac{2}{r^2}\sin^2(f) f_{rr} + \frac{1}{r^2}\sin(2f)f_r^2 -
\frac{1}{r^4}\sin^2 f\sin 2f\right) \non
\mathop- m^2 \sin f
+ \frac{\alpha v^2}{2+b}\left(\frac{1+b+\cos f}{2+b}\right)^{\alpha-1}
\left[\frac{1}{2}h_r^2
+\frac{1}{r^2}h^2
+\frac{\lambda v^2}{4}(1-h^2)^2\right] \sin f = 0, \\
h_{rr} + \frac{2}{r} h_r - \frac{2}{r^2} h
- \frac{\alpha\sin f}{1+b+\cos f} f_r h_r
+ \lambda v^2 (1-h^2)h = 0,
\end{align}
where $f_r\equiv \p_r f$. 
Let us now expand the equations of motion around $r=0$ as
\begin{align}
f &= \pi + f_1 r + \frac{1}{2} f_2 r^2 + \cdots, \qquad
h = h_1 r + \frac{1}{2} h_2 r^2 + \cdots.
\end{align}
The equations of motion to leading order in $r$ determine
$f_2=h_2=0$. 
We can thus calculate the energy density around the origin as
\beq
\mathcal{E} =
\frac{3}{2} c_2 f_1^2
+\frac{3}{2} c_4 f_1^4
+2m^2
+\frac{3}{2} v^2 \left(\frac{b}{2+b}\right)^\alpha
  \left[h_1^2 + \frac{1}{6}\lambda v^2\right]
+ \mathcal{O}(r^2).
\eeq
Using instead a scaling argument due to Derrick \cite{Derrick:1964ww}, 
we can estimate the size of the Skyrmion as follows. Neglecting the
monopole fields (as we assume that $v^2\ll(c_2,m)$), we denote by
$e_{2,4}$ the kinetic energy terms with 2 and 4 derivatives and $u$ is
the unitless potential
\beq
E = c_2 e_2 + c_4 e_4 + m^2 u. 
\eeq
Rescaling now the spatial coordinates $x^i\to\mu x^i$, we get
\beq
E \to \frac{1}{\mu} c_2 e_2 + \mu c_4 e_4 + \frac{1}{\mu^3} m^2 u, 
\eeq
which has an equilibrium when 
\beq
\mu^4 c_4 e_4 - \mu^2 c_2 e_2 - 3 m^2 u = 0.
\eeq
If we neglect the kinetic term, assuming that the mass is big and the
size of the Skyrmion is also big, then the characteristic length scale
and hence the Skyrmion size is roughly 
\beq
L_{\rm sk} \sim \sqrt[4]{\frac{c_4}{3m^2}}.
\label{eq:size_sk}
\eeq
The final step is to identify $f_1$ in the expansion of the solution
as an order-one number times the inverse length scale: 
$f_1\sim 1/L_{\rm sk}$. 
Now we can write the energy density of the Skyrmion part as
\beq
\mathcal{E}_{\rm sk} \sim \frac{3\sqrt{3}}{2} \frac{c_2 m}{\sqrt{c_4}}
+ 5m^2,
\eeq
whereas the energy density of the monopole part is roughly
\beq
\mathcal{E}_{\rm m} \sim 
\frac{1}{2} \lambda v^4 \left(\frac{b}{2+b}\right)^\alpha.
\eeq

\subsection{A single Skyrmopole}

\begin{figure}[!tpb]
\begin{center}
\includegraphics[width=\linewidth]{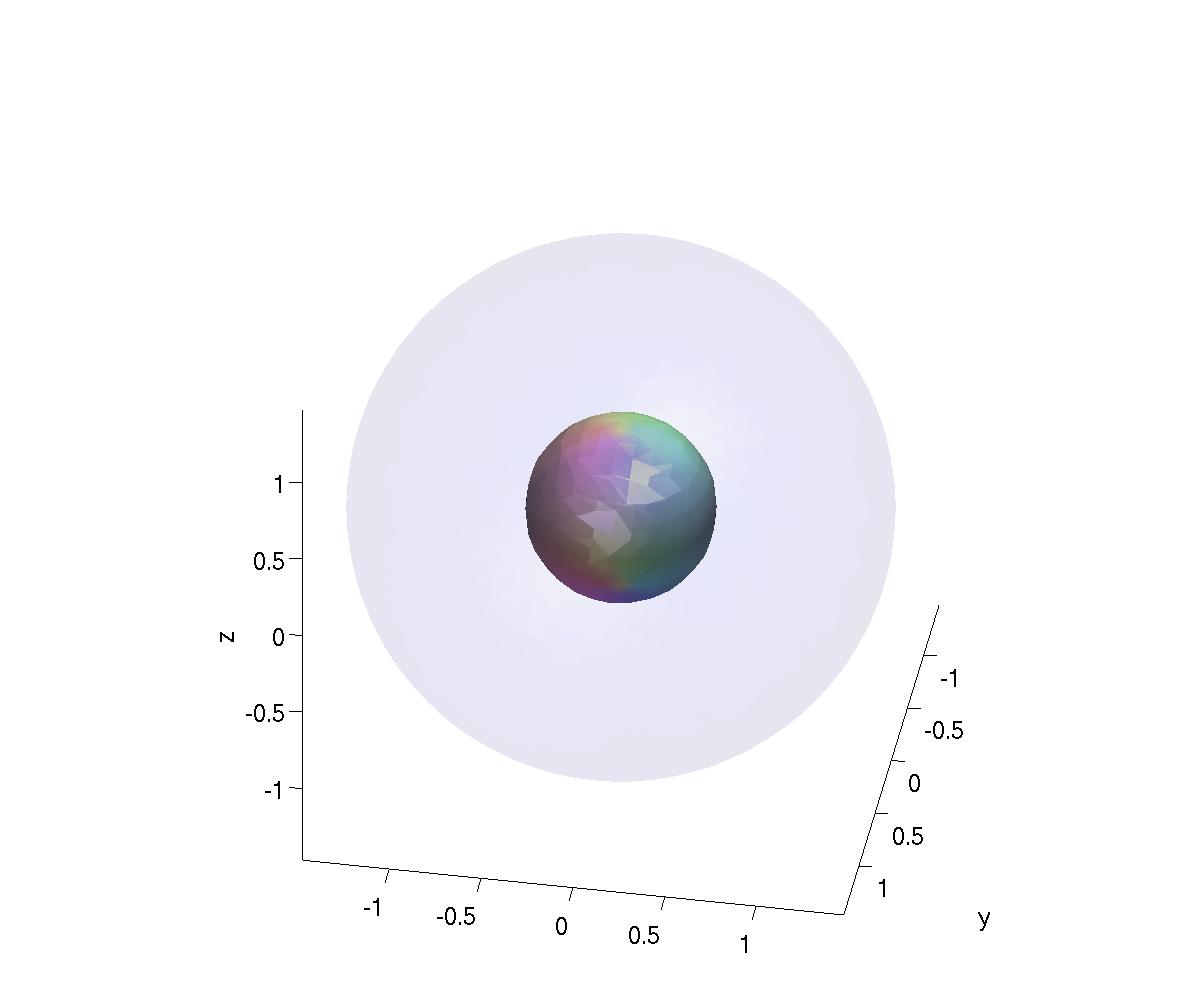}
\caption{A global monopole inside a Skyrmion. The figure shows
  isosurfaces of the monopole charge and Skyrmion charge densities, 
  respectively, at half their respective maximum values. 
  The gray cloud is the Skyrmion charge isosurface and the colored
  isosurface is the monopole charge.
  The coloring is made using an HSL (hue-saturation-lightness) map
  from the monopole field such that ${\rm arg}(\Phi_1/\Phi_2)$ is
  mapped to the hue and $|\Phi_3|$ determines the lightness. 
  The parameters are chosen as: $c_2=1$, $c_4=4$, $m=1$, $v=1/4$,
  $\lambda=128$, $b=1/2$ and $\alpha=4$, and the numerically
  integrated charges are $B=0.99993$ and $Q=0.96707$. The calculation 
  is made on a $201^3$ cubic lattice. } 
\label{fig:skm11}
\end{center}
\end{figure}

Let us try to solve the coupled equations numerically.
We use the finite-difference method on a cubic lattice of sizes
$121^3$ and $201^3$ and evolve a linear time operator, which is
standard in the relaxation method.\footnote{Clearly, the single
  Skyrmopole case, $Q=B=1$, can be solved using just the ordinary
  differential equations and not the three-dimensional partial
  differential equations (PDEs). In fact we solved this case also using an
  ODE solver and compared the solution to the one shown here found by
  solving the PDE as a check on our code. } After the equations of
motion are satisfied locally better than the one-permille level we
stop the relaxation algorithm.
We use the Ans\"atze \eqref{eq:nhedgehog}-\eqref{eq:fhedgehog} with
appropriate radial profile functions as the initial configuration for
the relaxation algorithm. 
We have chosen the parameters as $c_2=1$, $c_4=4$, $m=1$, $v=1/4$,
$\lambda=128$, $b=1/2$ and $\alpha=4$ in order to satisfy the
constraint \eqref{eq:flipconstraint}.
In Fig.~\ref{fig:skm11} is shown a single global monopole ($Q=1$)
inside a single Skyrmion ($B=1$). The figure shows the 3-dimensional
isosurfaces of the Skyrmion charge and monopole charge densities,
respectively, each at their respective half-maximum values. 

The size and density (lattice spacing) is the same as that used for
the two-monopole inside a Skyrmion and we have optimized said
dimensions for the two-monopole. This is why we do not capture the
$Q=1$ monopole charge better than to about three and a half
percent on this lattice. As we will see shortly, the two-monopole
charge is far better contained.

\begin{figure}[!tpb]
\begin{center}
\mbox{\subfloat[]{\includegraphics[width=0.49\linewidth]{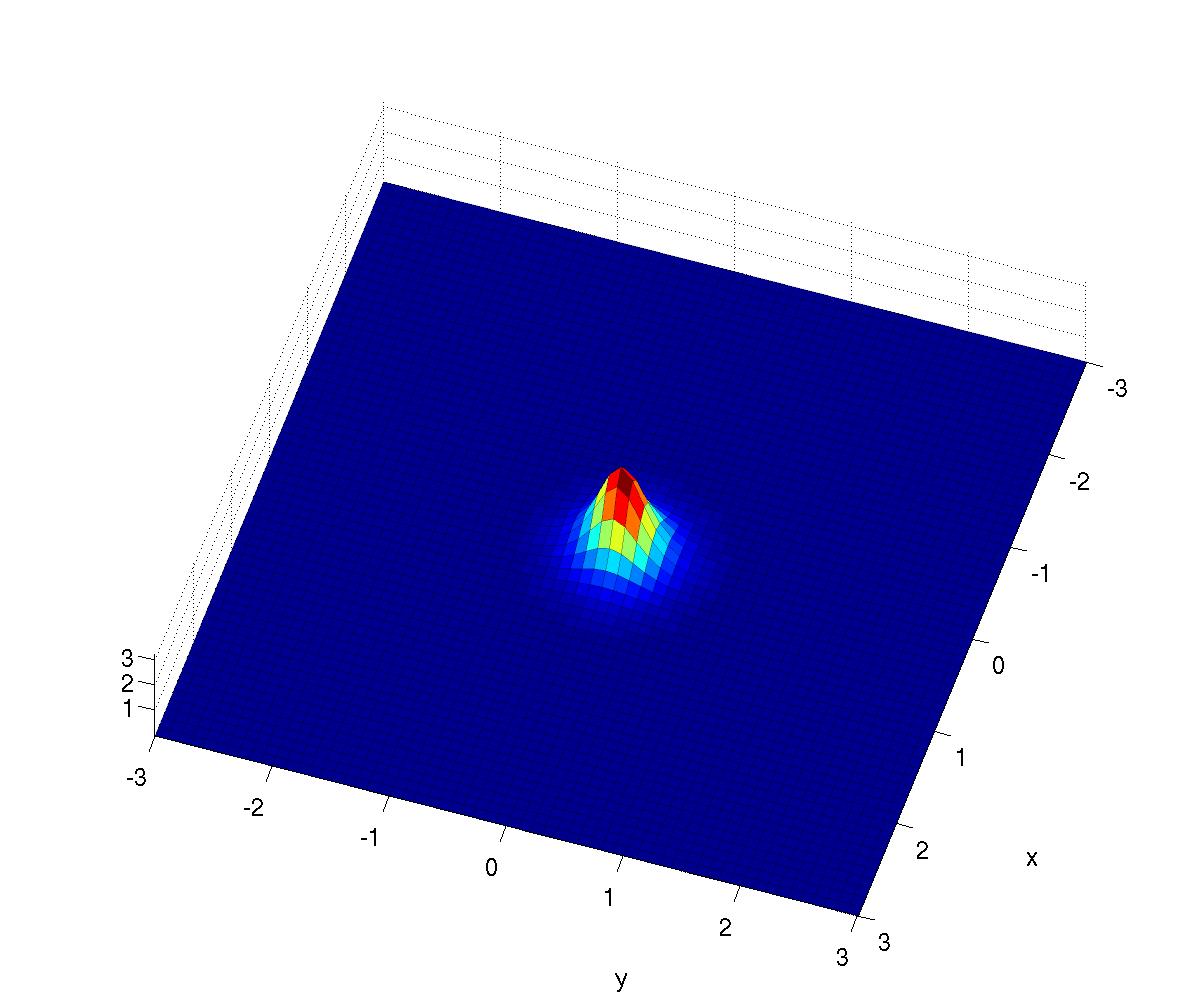}}
\subfloat[]{\includegraphics[width=0.49\linewidth]{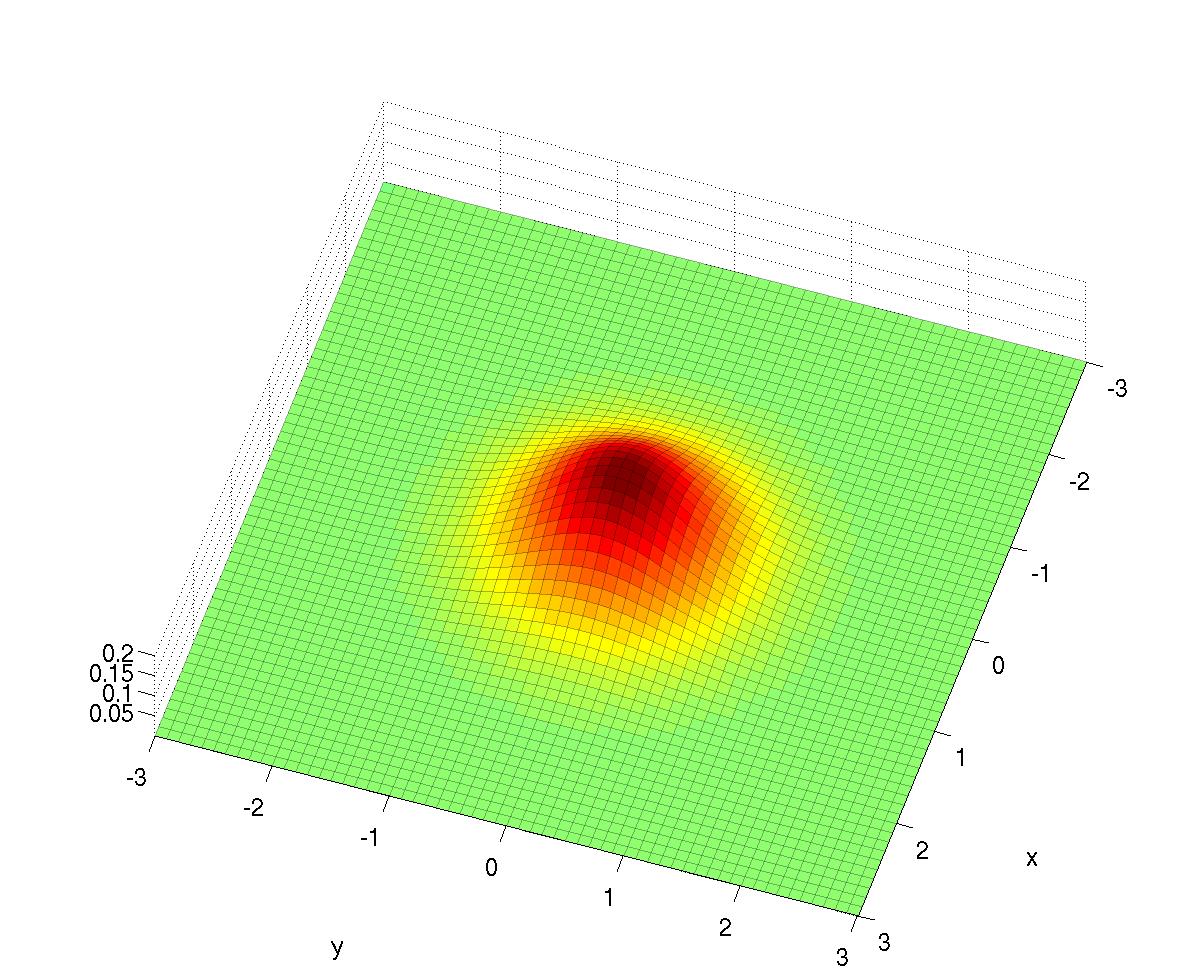}}}
\caption{Topological charges for the global monopole inside a
  Skyrmion; (a) is the monopole charge and (b) is the Skyrmion
  charge. The figures show slices in the $xy$ plane for $z=0$. The 
  numerically integrated charges are $B=0.99993$ and $Q=0.96707$. } 
\label{fig:skm11charges}
\end{center}
\end{figure}

In Fig.~\ref{fig:skm11charges} are shown $xy$ slices at $z=0$ of the
monopole charge and Skyrmion charge, respectively. Both have the shape
of solid balls and the monopole has roughly half the diameter of that
of the Skyrmion. 

\begin{figure}[!tpb]
\begin{center}
\mbox{\subfloat[]{\includegraphics[width=0.33\linewidth]{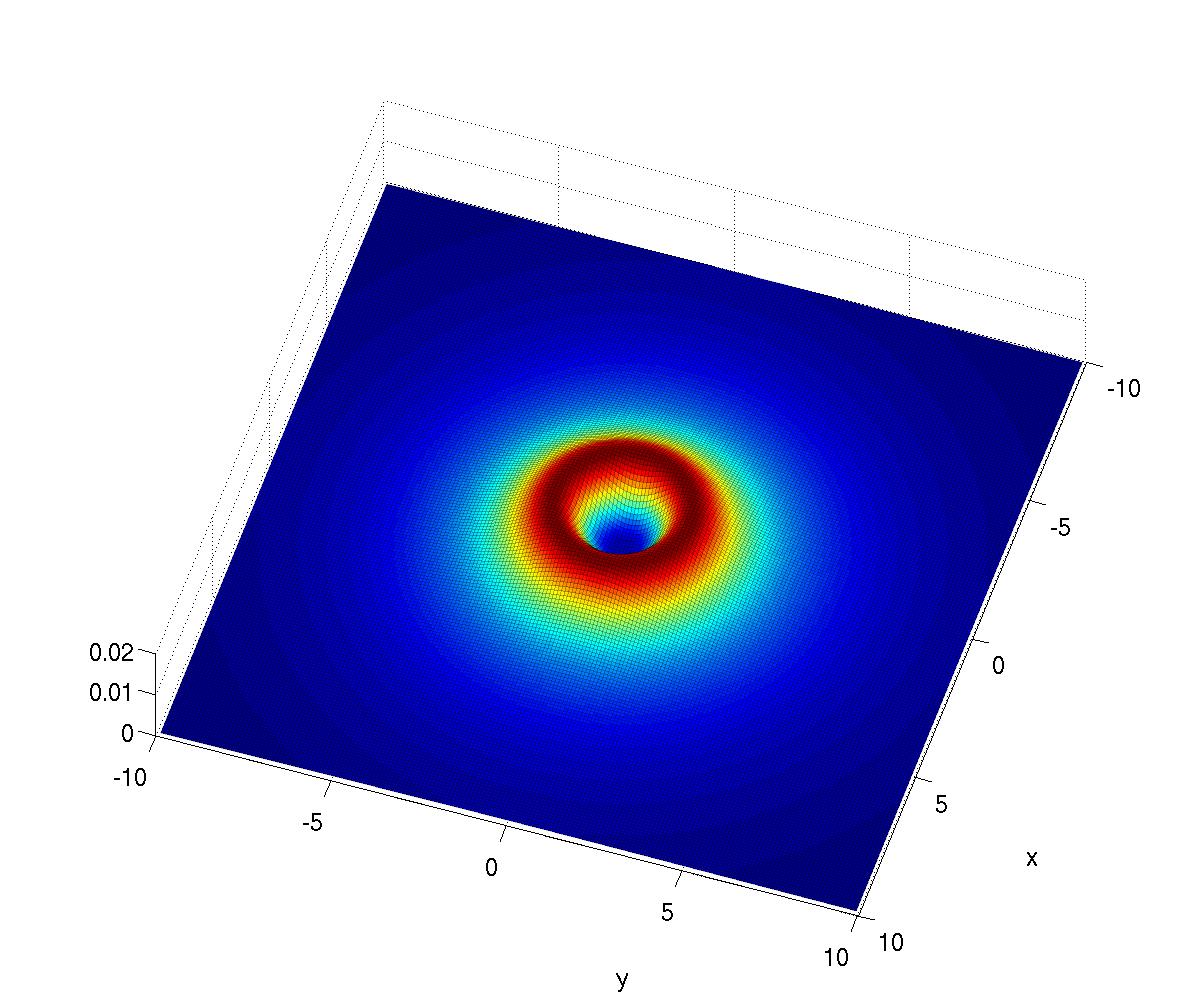}}
\subfloat[]{\includegraphics[width=0.33\linewidth]{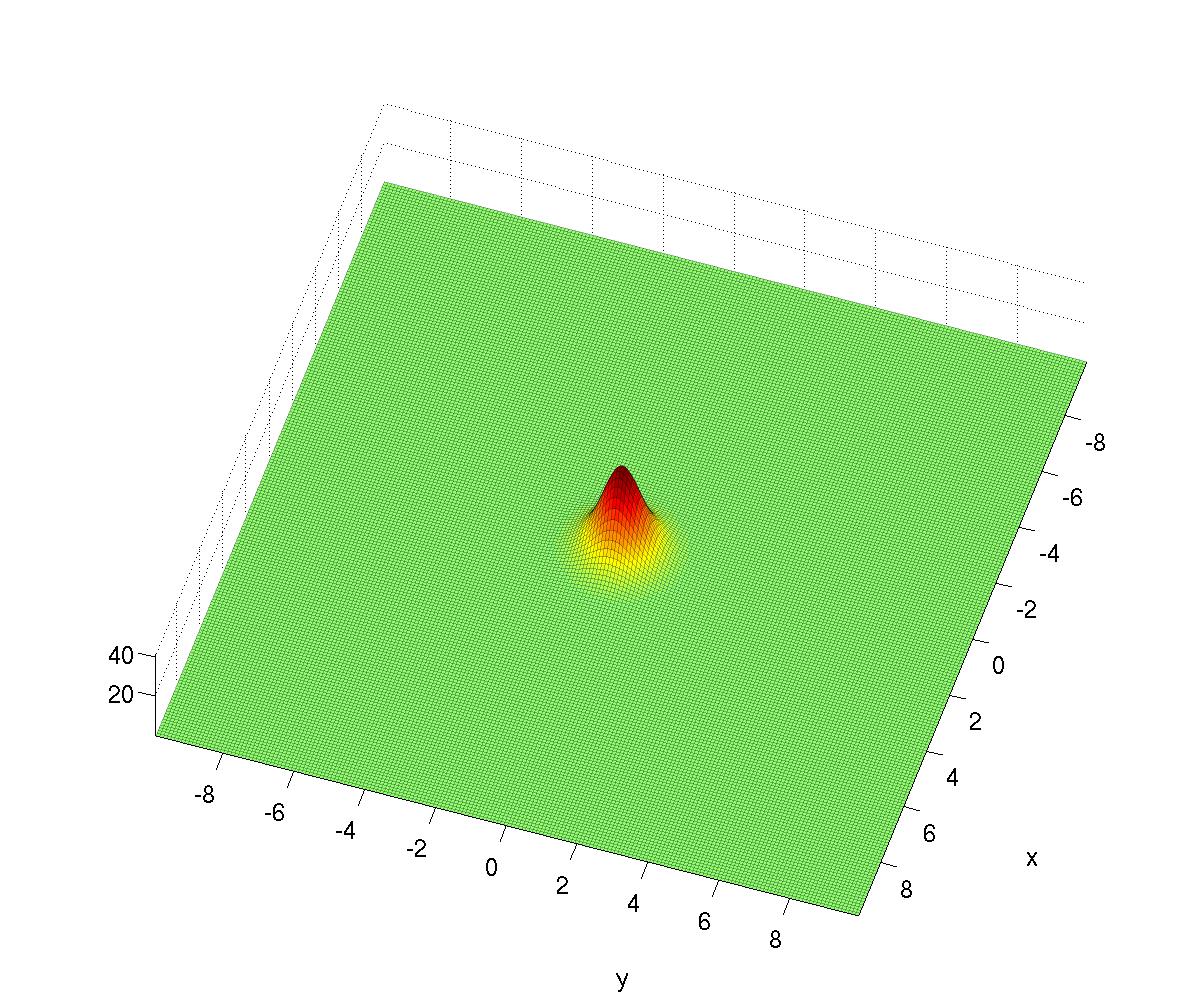}}
\subfloat[]{\includegraphics[width=0.33\linewidth]{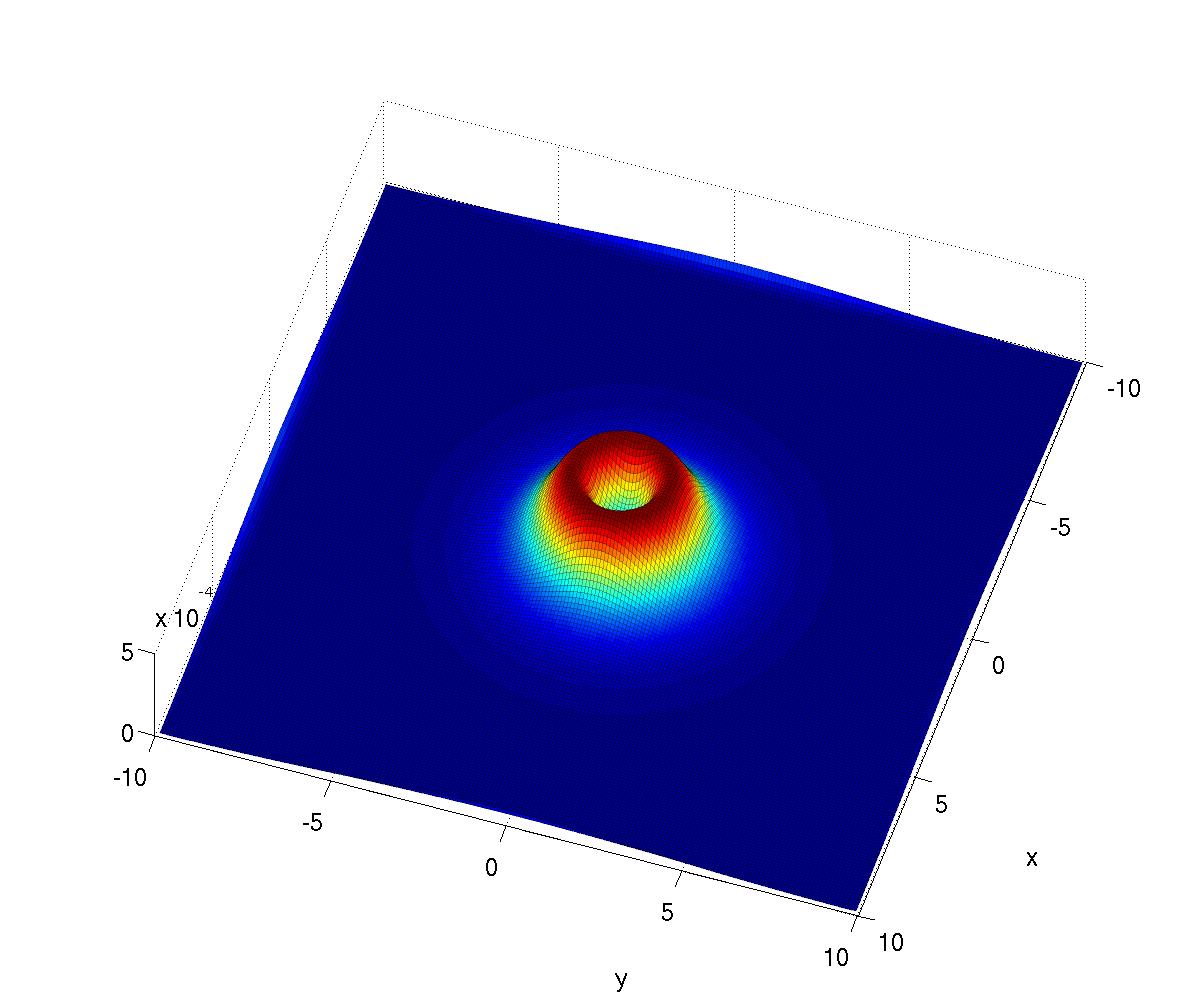}}}
\mbox{\subfloat[]{\includegraphics[width=0.33\linewidth]{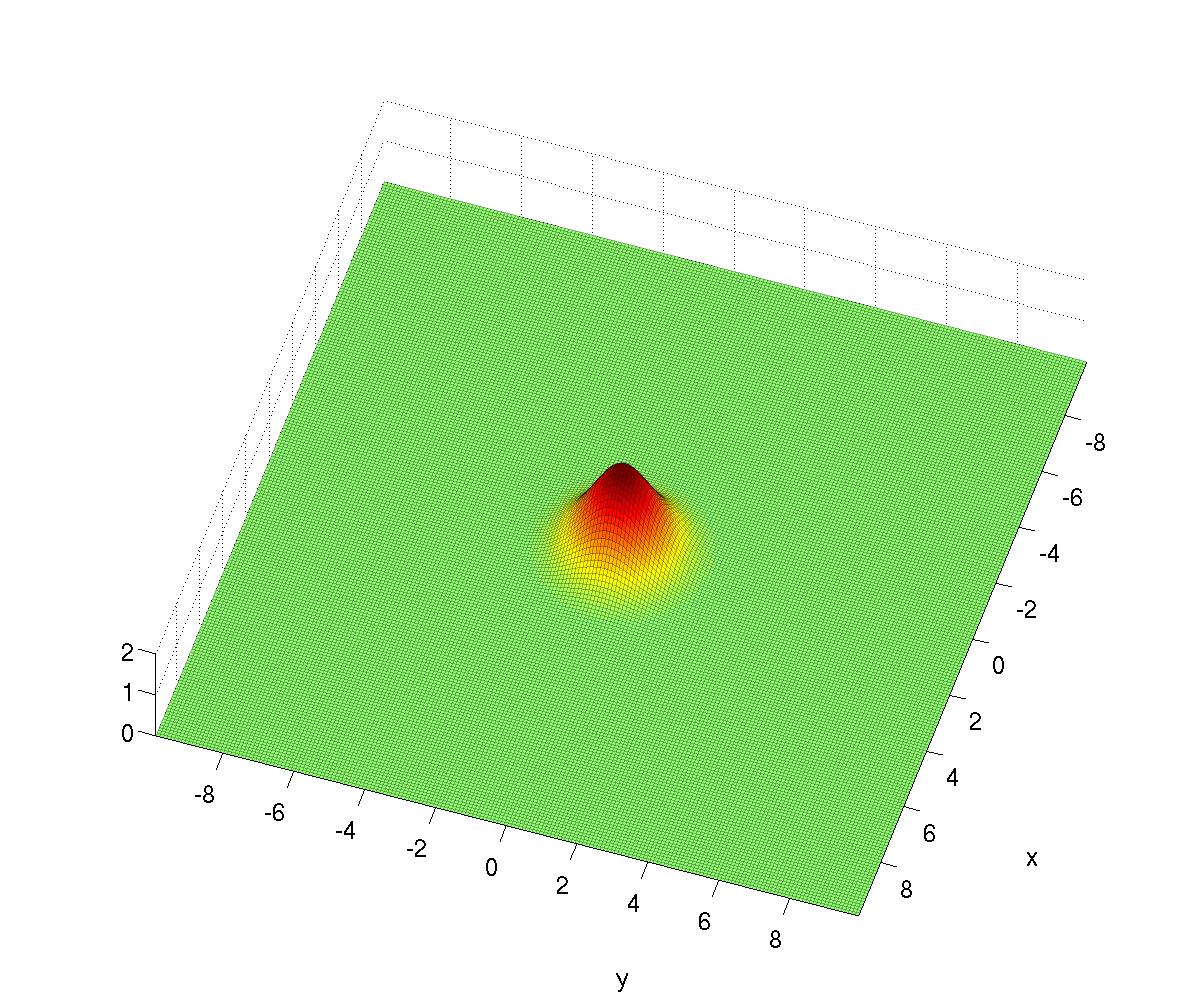}}
\subfloat[]{\includegraphics[width=0.33\linewidth]{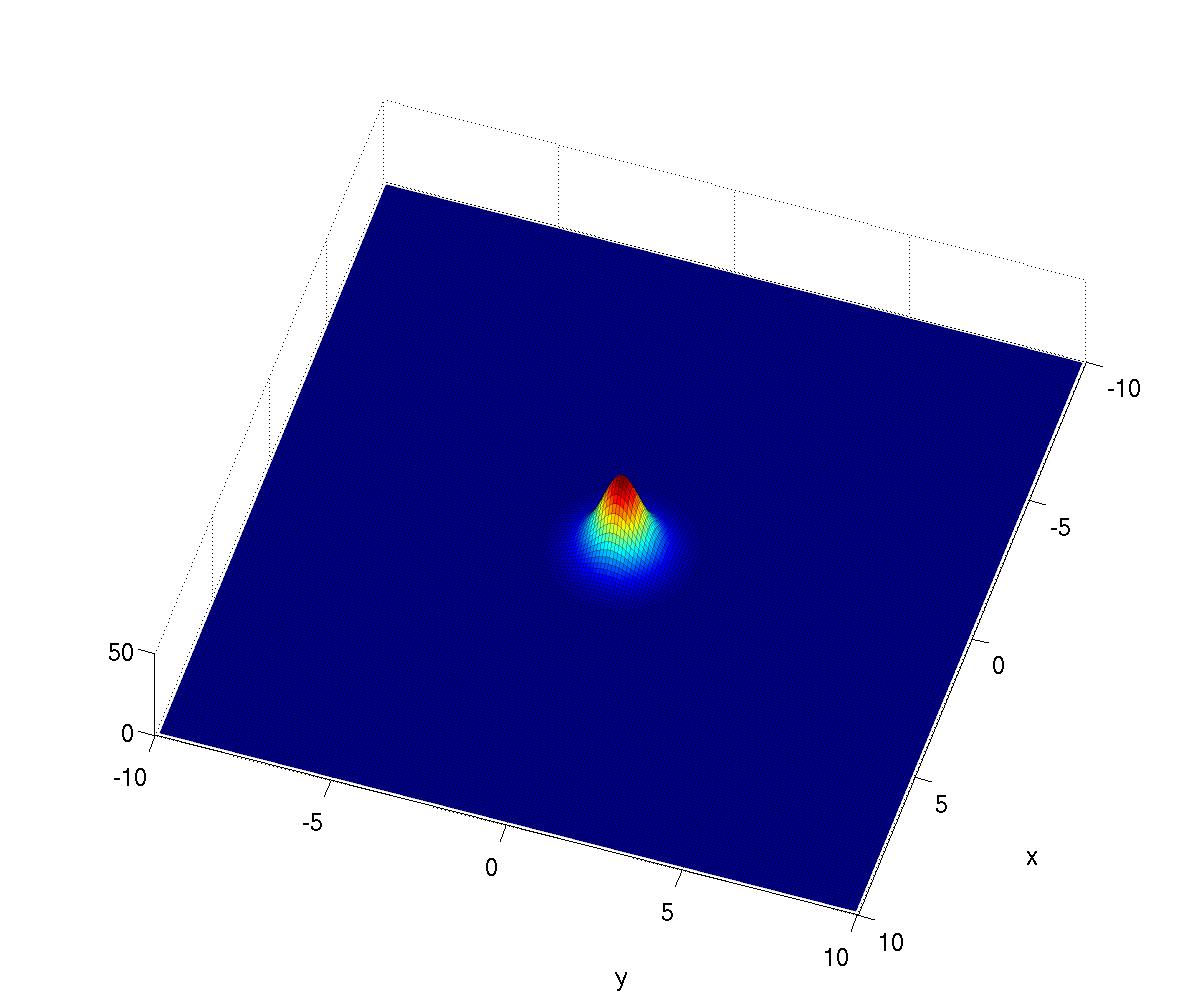}}
\subfloat[]{\includegraphics[width=0.33\linewidth]{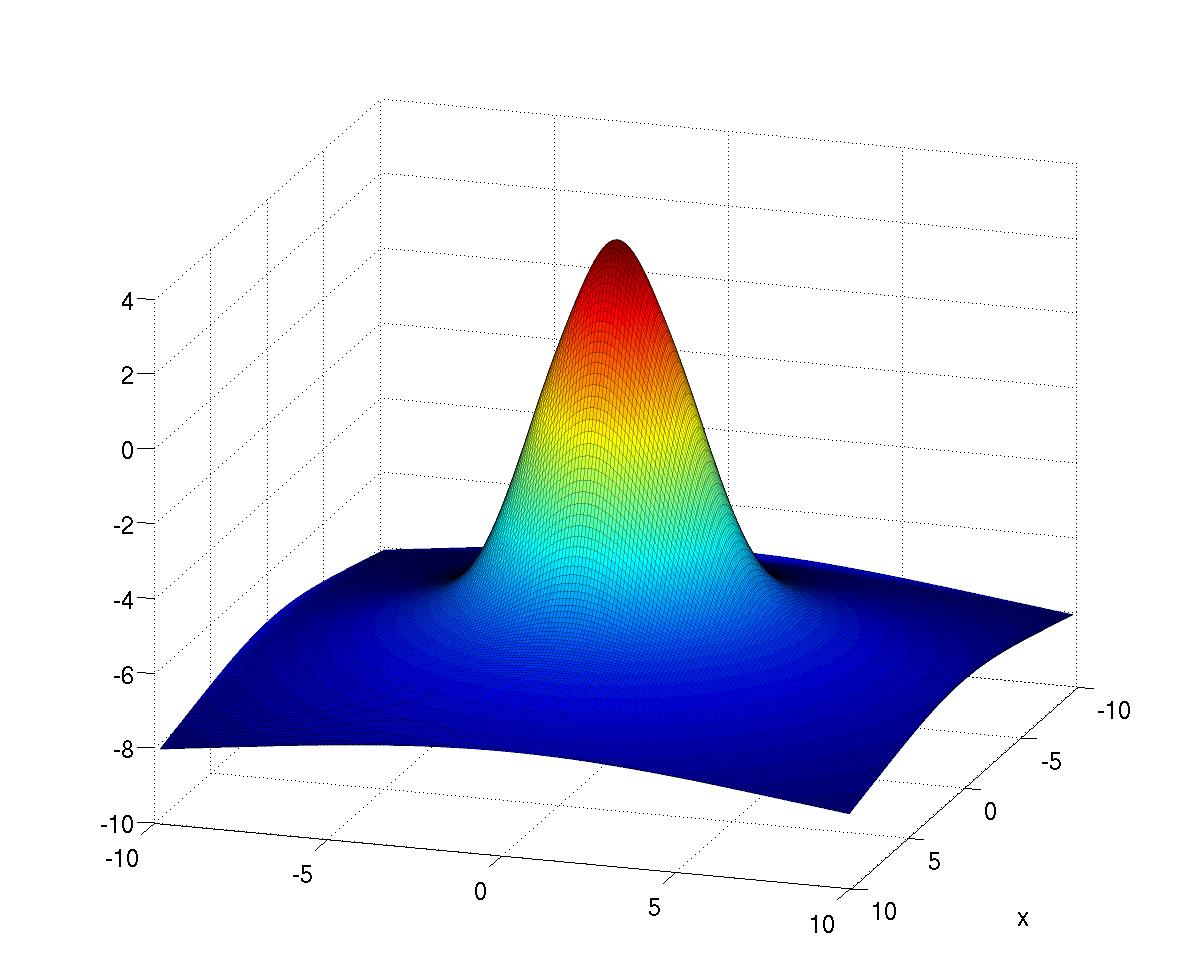}}}
\caption{Energies for the global monopole inside a Skyrmion; (a) is
  the kinetic term of the monopole (with the 
  prefactor $\mathcal{G}$), (b) is the Skyrmion kinetic energy, (c) is
  the potential for the monopole (again with the prefactor
  $\mathcal{G}$), (d) is the pion-mass term of the Skyrmion sector,
  (e) is the total energy of the configuration and (f) is the
  logarithm of the total energy. 
  The figures show slices in the $xy$ plane for $z=0$. } 
\label{fig:skm11energies}
\end{center}
\end{figure}

In Fig.~\ref{fig:skm11energies} are shown six $xy$ slices at $z=0$ of
energy densities. More precisely of the monopole kinetic energy, the
Skyrmion kinetic energy, the monopole potential, the pion mass term
(Skyrmion potential), the total energy and finally the logarithm of
the total energy. The shape of the Skyrmion is ball-like while that
of the monopole is rather like a hollow sphere. This is in
contradistinction to the monopole charge-density distribution. The
reason is that the monopole energy is multiplied by a factor
$\mathcal{G}$, which amplifies the energy outside of where the
monopole charge is located. This is no accident. This is made by
design of the function $\mathcal{G}$ in order to provide an attractive
force on the two monopole constituents. One can think of this
construction as making an energy barrier that the monopoles do not
want to cross and hence choose to stick with each other instead (even 
though they are repellent to one another). 

The reason why this calculation is so numerically challenging is that
the monopole charge is much smaller than the corresponding energy
distribution. This is because the prefactor makes the majority of the
monopole energy density sit outside of where the monopole charge is
situated.
This separation of scales requires both very large length scales (for
capturing the energy) and a very dense lattice for capturing the
charge at the center. 

\subsection{The type A, $Q=2$ Skyrmopole}\label{sec:AQ2skmm}

\begin{figure}[!tpb]
\begin{center}
\includegraphics[width=\linewidth]{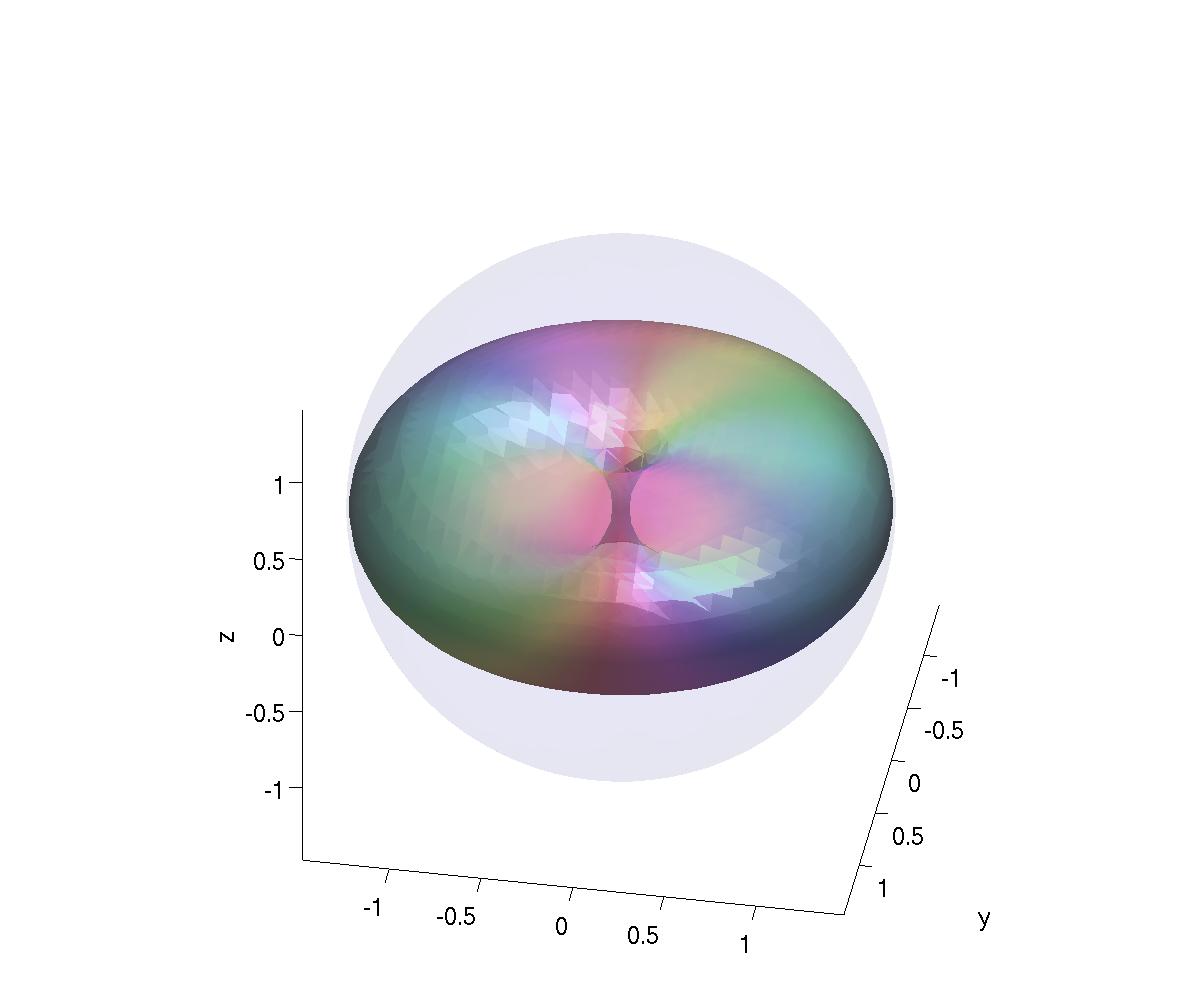}
\caption{The global two-monopole of type A inside a single
  Skyrmion. The figure shows isosurfaces of the monopole charge and
  Skyrmion charge densities, respectively, at half their respective
  maximum values. 
  The gray cloud is the Skyrmion charge isosurface and the colored
  isosurface of the shape of a torus is the monopole charge.
  The coloring is made using an HSL (hue-saturation-lightness) map
  from the monopole field such that ${\rm arg}(\Phi_1/\Phi_2)$ is
  mapped to the hue and $|\Phi_3|$ determines the lightness. 
  The parameters are chosen as: $c_2=1$, $c_4=4$, $m=1$, $v=1/4$,
  $\lambda=128$, $b=1/2$ and $\alpha=4$, and the numerically
  integrated charges are $B=0.99993$ and $Q=1.9863$. The calculation 
  is made on a $201^3$ cubic lattice. } 
\label{fig:skm21}
\end{center}
\end{figure}

We are now ready to compute the two-monopole inside a single Skyrmion;
we can call it the $Q=2$ Skyrmopole.
As the initial condition we use the Ans\"atze \eqref{eq:nhedgehog} and 
\eqref{eq:tipoa} with $d=0$. We take the profile function of the
initial condition to be
\beq
h = \tanh\left[\kappa r\right], \nonumber
\eeq
where $r$ is the radial coordinate and $\kappa$ is an appropriately 
chosen constant. 
We use the same lattice and same parameters as in the previous
subsection, i.e.~$c_2=1$, $c_4=4$, $m=1$, $v=1/4$, $\lambda=128$,
$b=1/2$ and $\alpha=4$. We again satisfy the constraint
\eqref{eq:flipconstraint}. 
In Fig.~\ref{fig:skm21} is shown the global two-monopole ($Q=2$) 
inside a single Skyrmion ($B=1$). The figure shows the 3-dimensional
isosurfaces of the Skyrmion charge and monopole charge densities,
respectively, each at their respective half-maximum values. 

\begin{figure}[!tpb]
\begin{center}
\mbox{\subfloat[]{\includegraphics[width=0.49\linewidth]{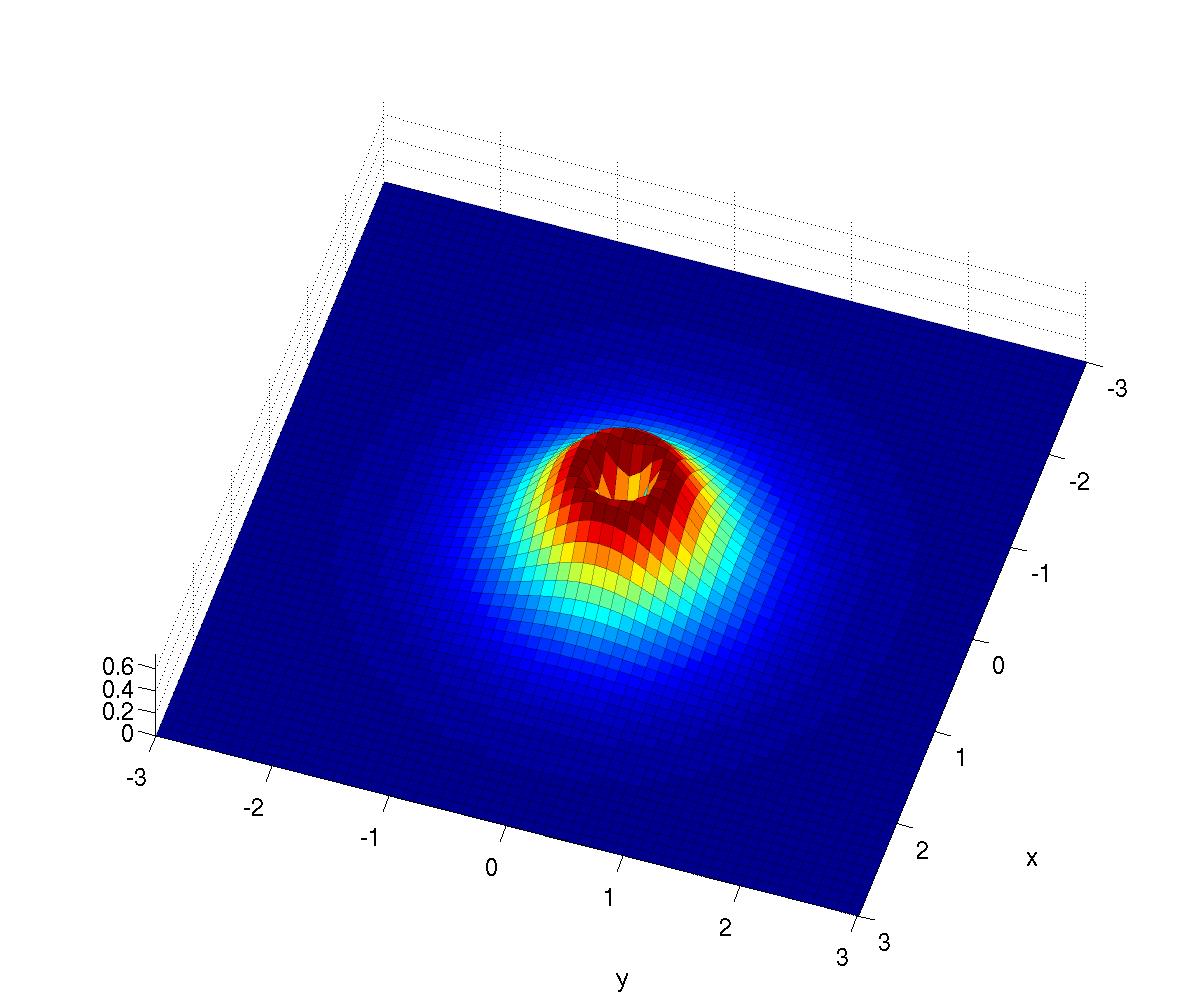}}
\subfloat[]{\includegraphics[width=0.49\linewidth]{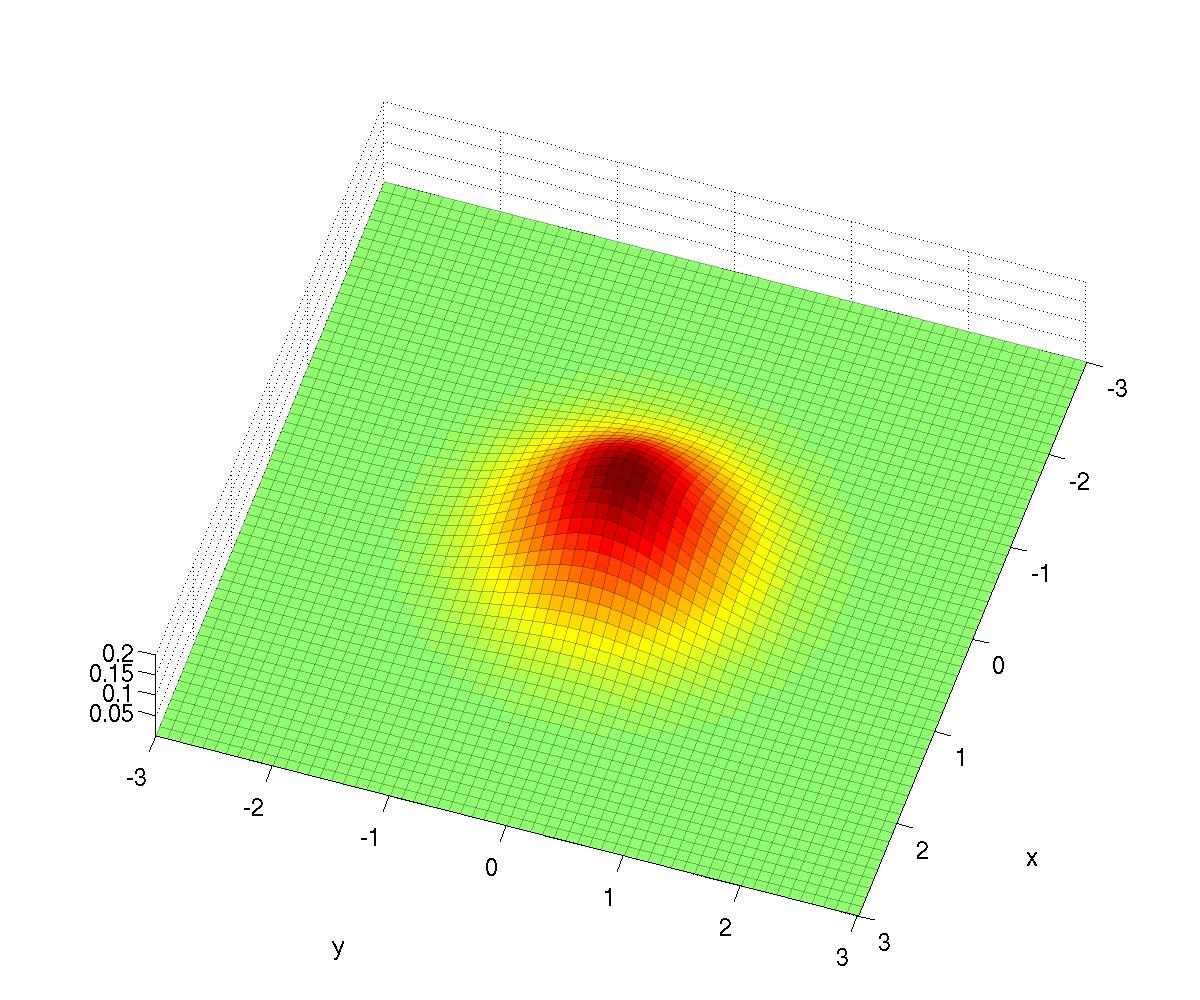}}}
\caption{Topological charges for the global two-monopole inside a
  single Skyrmion; (a) is the monopole charge and (b) is the Skyrmion
  charge. The figures show slices in the $xy$ plane for $z=0$. The 
  numerically integrated charges are $B=0.99993$ and $Q=1.9863$. } 
\label{fig:skm21charges}
\end{center}
\end{figure}

In Fig.~\ref{fig:skm21charges} are shown $xy$ slices at $z=0$ of the
monopole charge and Skyrmion charge, respectively. The Skyrmion is still
a ball while the charge distribution of the two-monopole has the shape
of a torus. Notice that the two charge distributions have almost the
same size.

\begin{figure}[!tpb]
\begin{center}
\mbox{\subfloat[]{\includegraphics[width=0.33\linewidth]{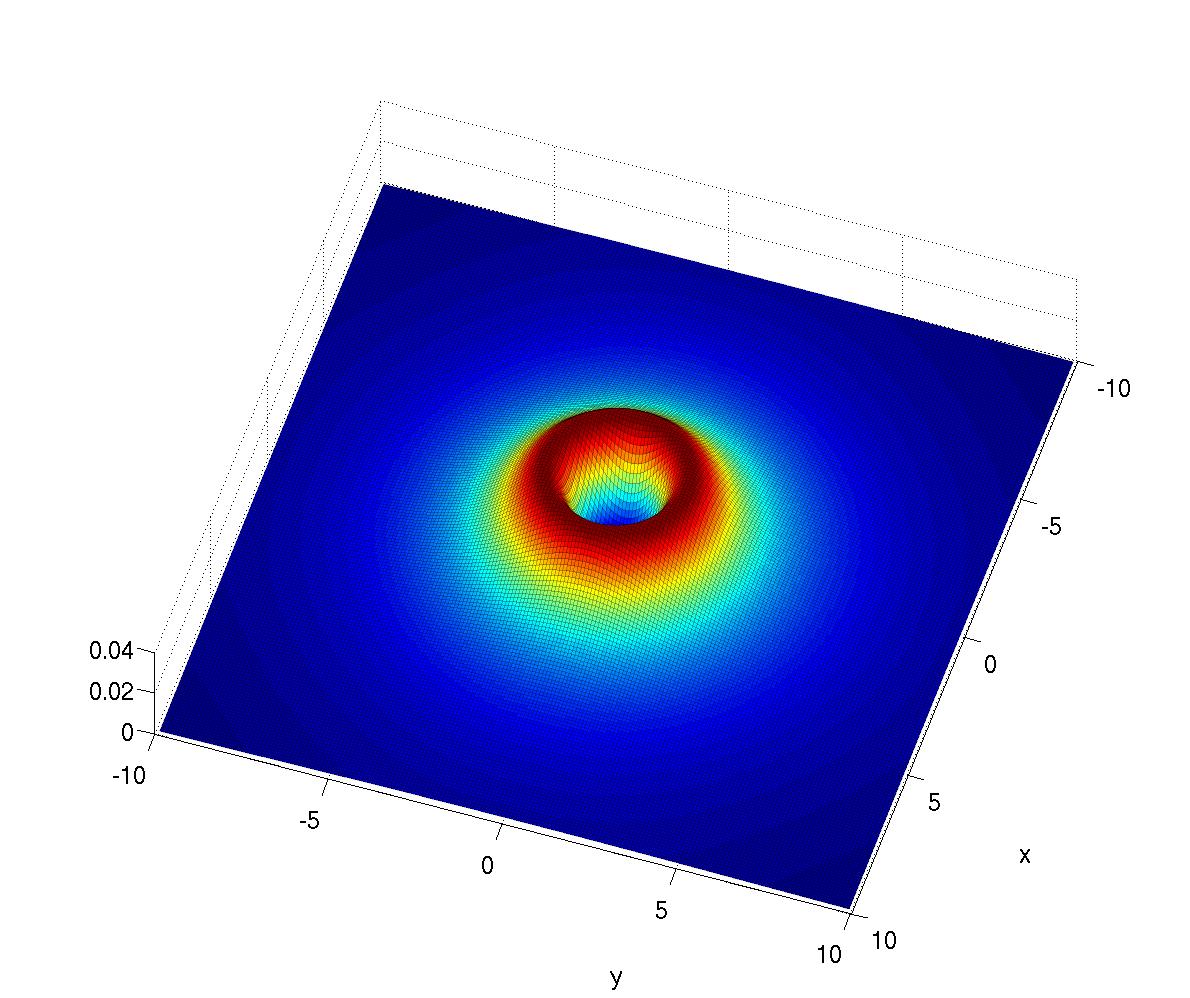}}
\subfloat[]{\includegraphics[width=0.33\linewidth]{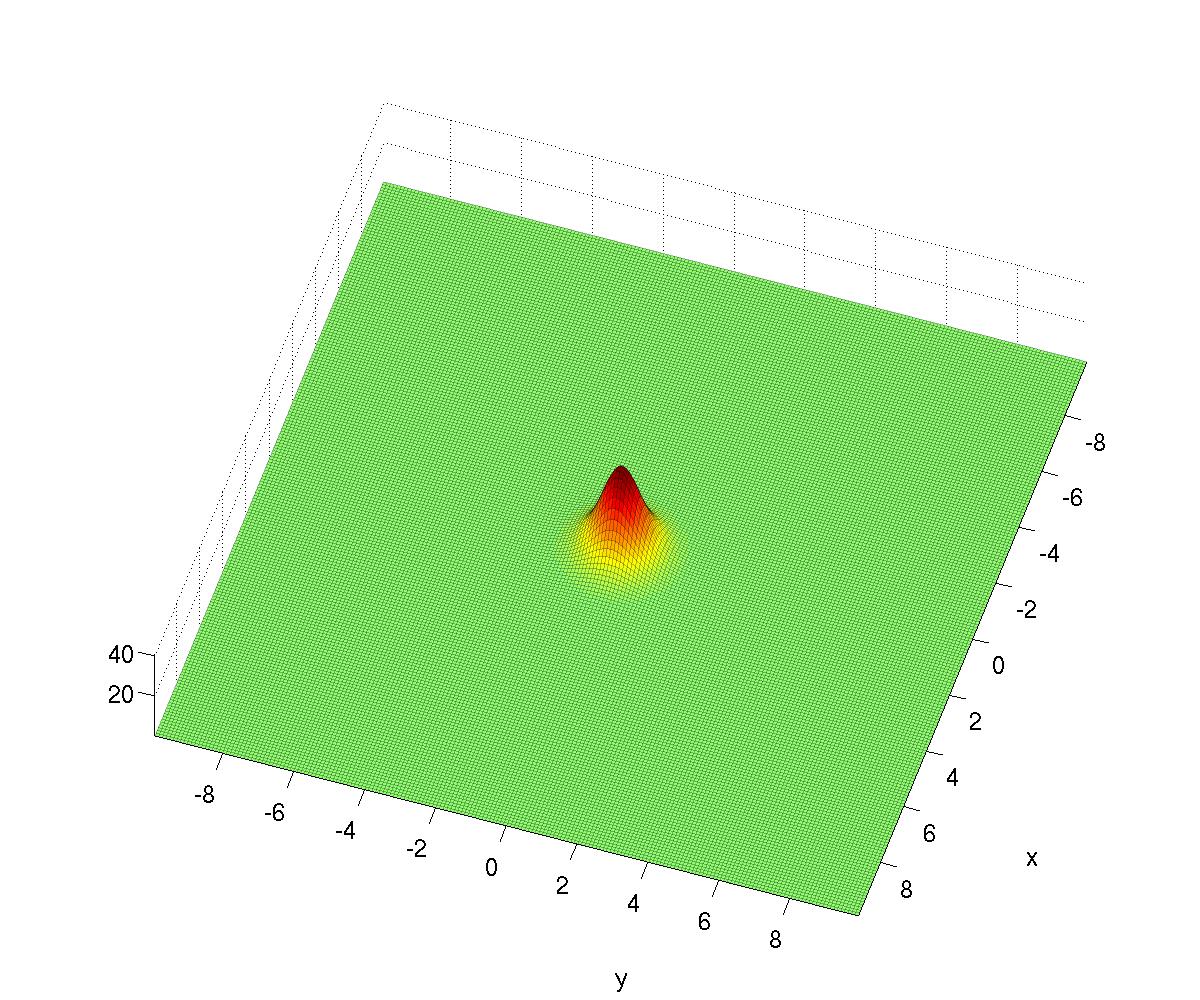}}
\subfloat[]{\includegraphics[width=0.33\linewidth]{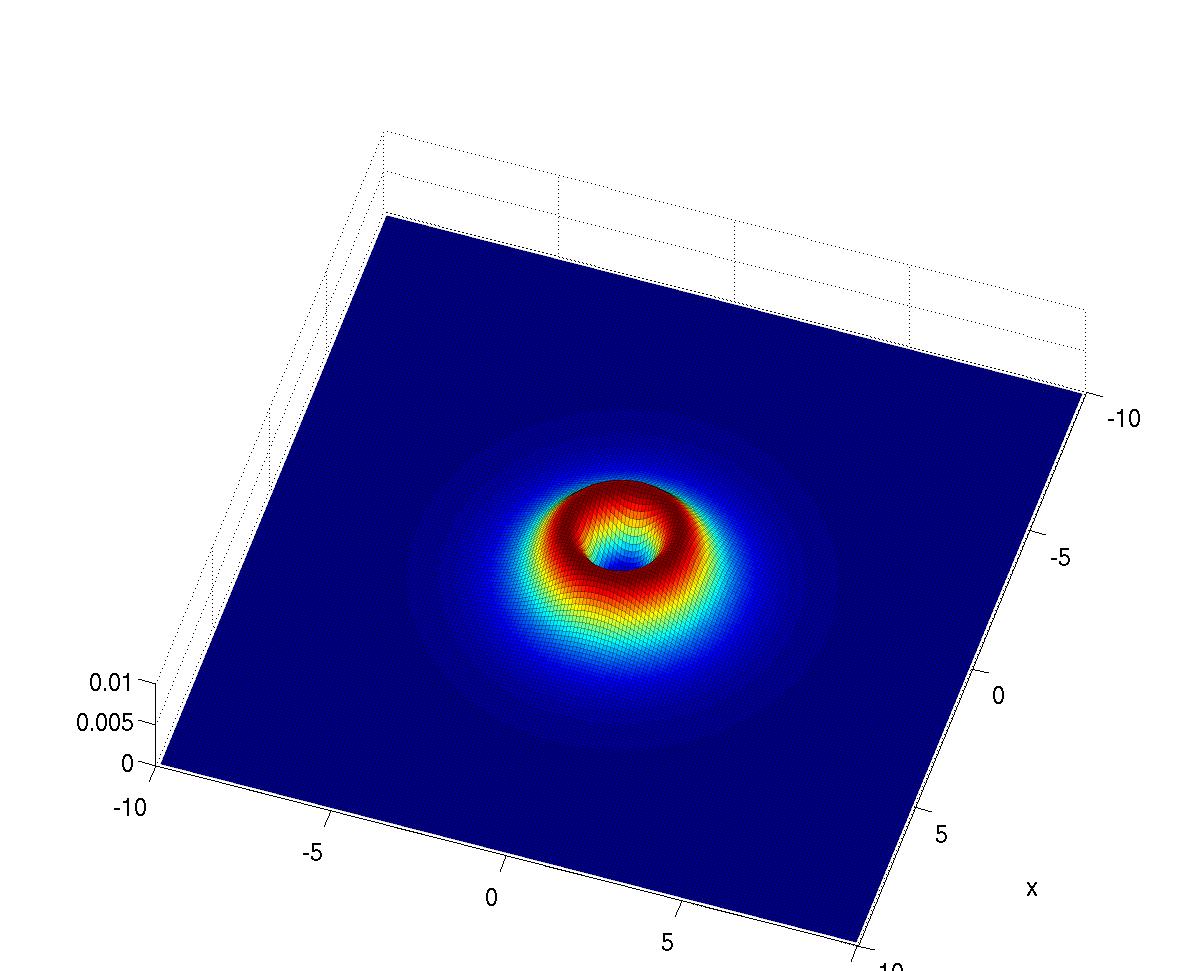}}}
\mbox{\subfloat[]{\includegraphics[width=0.33\linewidth]{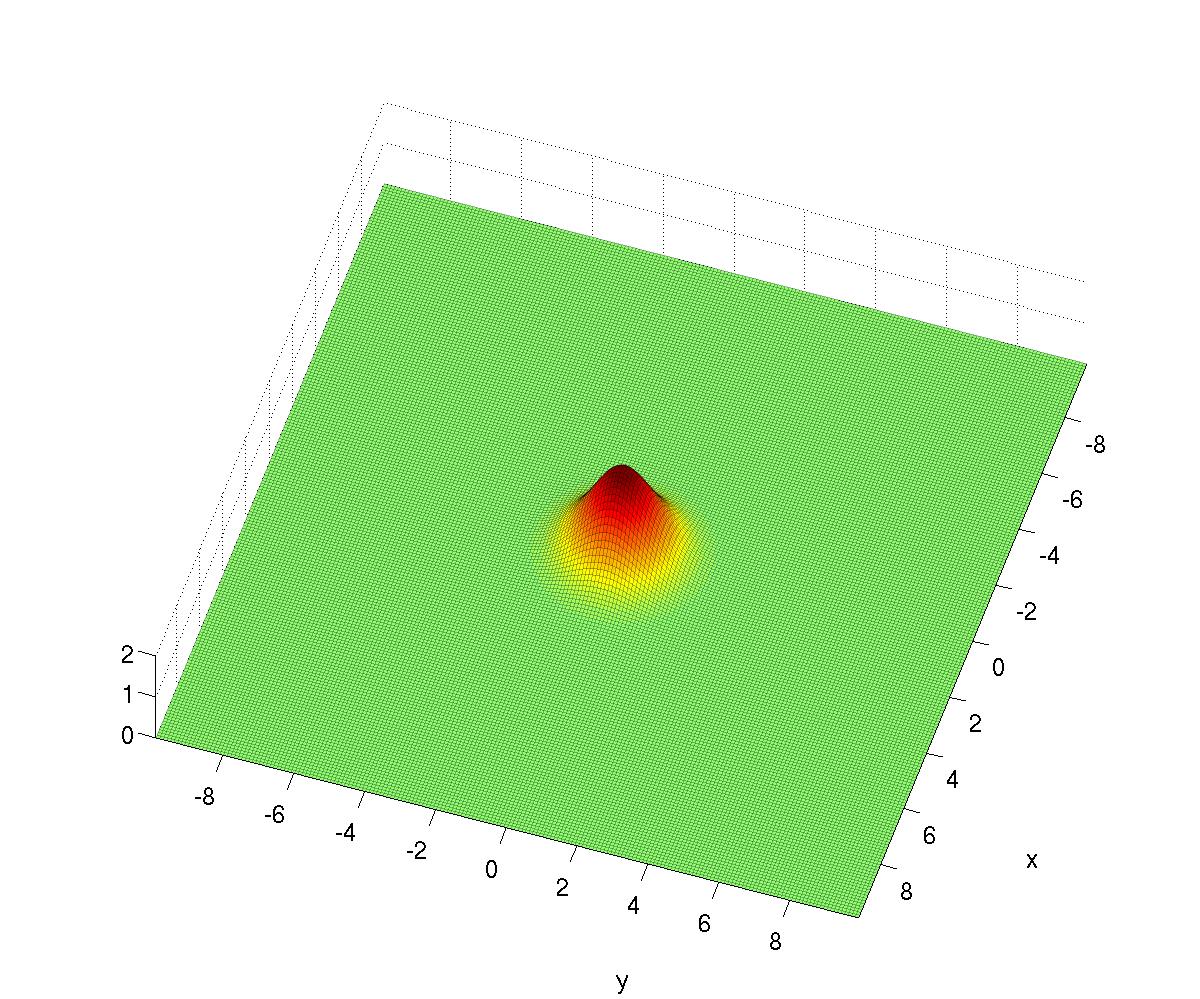}}
\subfloat[]{\includegraphics[width=0.33\linewidth]{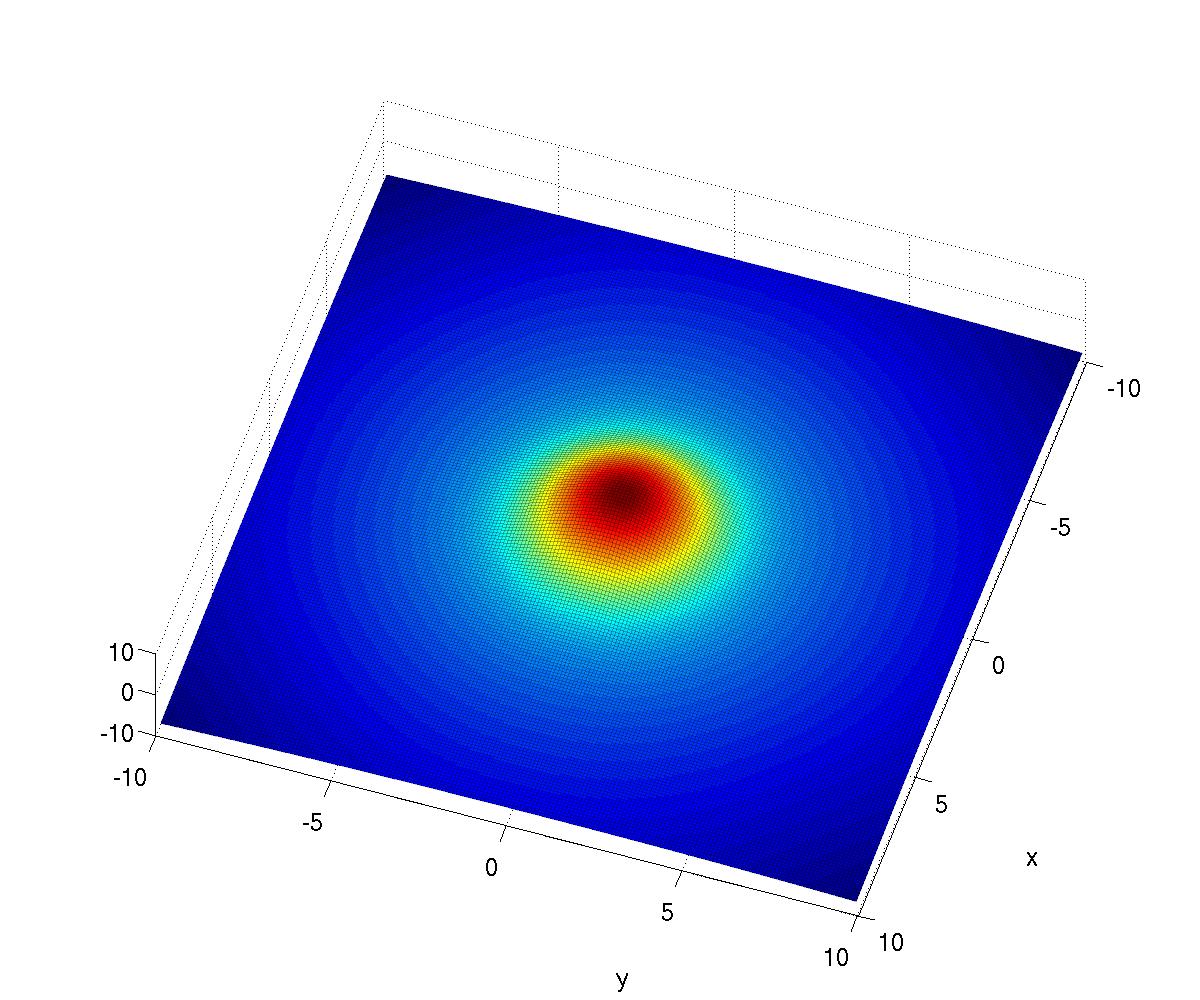}}
\subfloat[]{\includegraphics[width=0.33\linewidth]{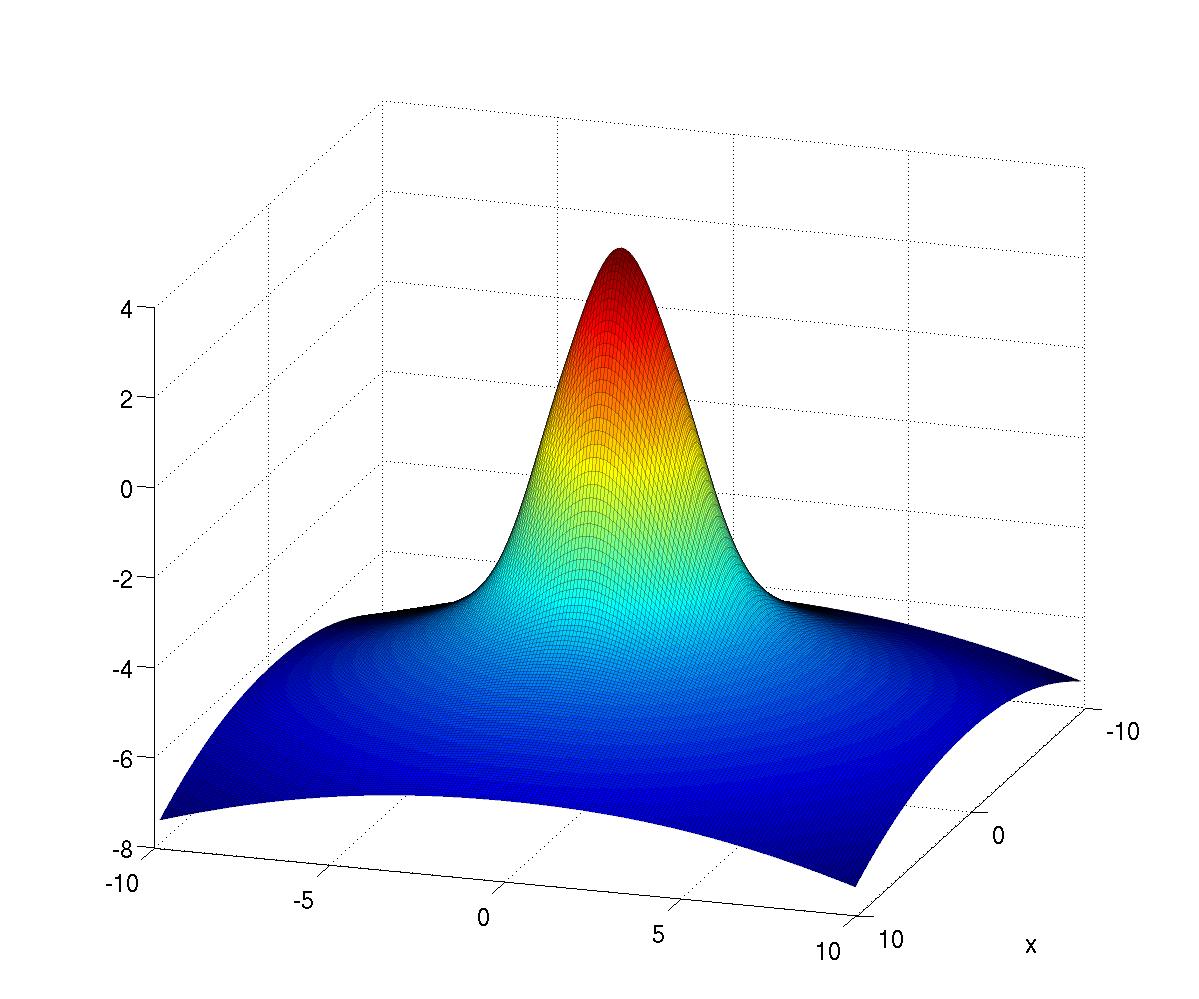}}}
\caption{Energies for the global two-monopole inside a
  single Skyrmion; (a) is the kinetic term of the monopole (with the
  prefactor $\mathcal{G}$), (b) is the Skyrmion kinetic energy, (c) is
  the potential for the monopole (again with the prefactor
  $\mathcal{G}$), (d) is the pion-mass term of the Skyrmion sector,
  (e) is the total energy of the configuration and (f) is the
  logarithm of the total energy. 
  The figures show slices in the $xy$ plane for $z=0$. } 
\label{fig:skm21energies}
\end{center}
\end{figure}

In Fig.~\ref{fig:skm21energies} are shown six $xy$ slices at $z=0$ of 
energy densities. Again the subfigures show the monopole kinetic
energy, the Skyrmion kinetic energy, the monopole potential, the pion
mass term (Skyrmion potential), the total energy and finally the
logarithm of the total energy. As was the case of the shapes in the
one-monopole case, the Skyrmion energy is ball-like while the monopole
energy has the shape of a hollow sphere. 

We numerically integrate the energy densities shown in
Fig.~\ref{fig:skm21energies} and give the results in
Tab.~\ref{tab:skm21energies}. We integrate only the solution inside
the box on which we find the solution, which has the size $20^3$.
This captures the total energy of the Skyrmion, whereas the total
energy of the two-monopole is infinite. 

\begin{table}[!ht]
  \begin{center}
    \caption{Numerically integrated energies of the type A, $Q=2$
      Skyrmopole. The integration region is the box on which the
      discritized solution is found, which has the size $20^3$.  }
    \label{tab:skm21energies}
    \begin{tabular}{l|r@{$\pm$}l}
      monopole kinetic energy & 16.18&0.11\\
      Skyrmion kinetic energy & 166.99&0.17\\
      monopole potential energy & 0.28&0.002\\
      Skyrmion potential energy & 25.49&0.03\\
      \hline
      total energy & 208.88&0.21
    \end{tabular}
  \end{center}
\end{table}

\subsection{Stability of the type A, $Q=2$ Skyrmopole}

A caveat of starting with a symmetric initial condition is that it may
be unstable and only a very long relaxation time will reveal if the
configuration is really stable or not.
In our case of the type A two-monopole, the symmetry enjoyed by the
configuration is axial symmetry. In order to show that the system of
the two-monopole residing within the Skyrmion is really stable, we use
again the Ans\"atze \eqref{eq:nhedgehog} and \eqref{eq:tipoa}, but
this time we perturb the axial symmetry by a finite $d>0$ parameter.
More specifically, we choose
\beq
d = \frac{1}{2} a h_x,
\eeq
where $a$ is a real number and $h_x$ is the lattice spacing in the $x$
direction.
We carry out this calculation on a $121^3$ lattice with
spatial-lattice spacing $h_x=h_y=h_z=0.16529$. Since the winding
phases now have two centers -- one at $x=-d$ and one at $x=d$, we
choose a profile function of the form
\beq
h = \sqrt{
  \tanh\left[\kappa\left((x-d)^2+y^2+z^2\right)^{\frac{3}{8}}\right]
  \tanh\left[\kappa\left((x+d)^2+y^2+z^2\right)^{\frac{3}{8}}\right]},
\eeq
with an appropriate value for $\kappa>0$. The fractional power of the
radius in the hyperbolic tangent is chosen empirically to better match
the $d=0$ solution.
In order to quantify the asymmetry of the configuration, we define the
following spatially-weighted monopole charge 
\beq
Q_{x^i} \equiv \int d^3x \; x^i \mathcal{Q},
\eeq
where the monopole charge density is
\beq
\mathcal{Q} = -\frac{1}{8\pi}\epsilon^{ijk}\epsilon^{abc}
\p_i\Phi^a \p_j\Phi^b \p_k\Phi^c.
\eeq

\begin{figure}[!tpb]
\begin{center}
\includegraphics[width=0.75\linewidth]{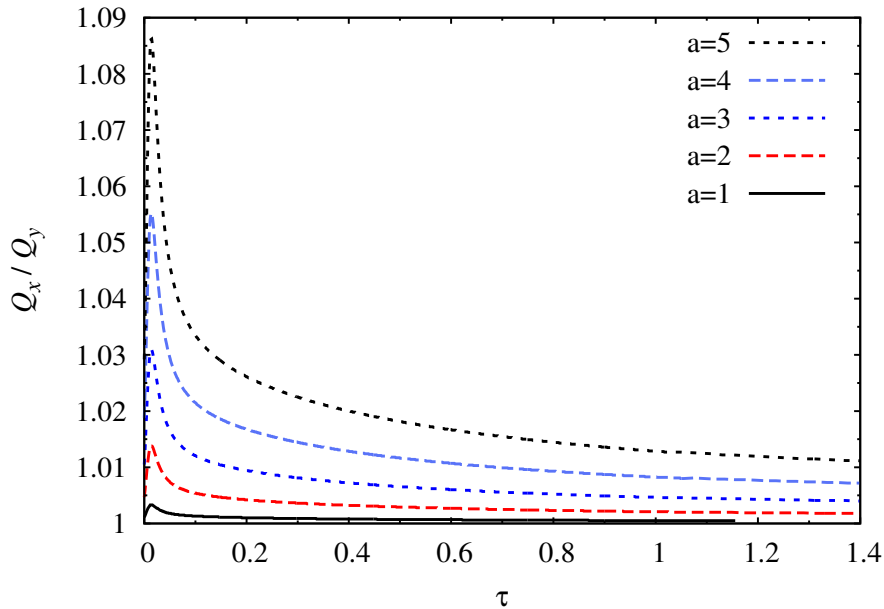}
\caption{Asymmetry in the monopole charge, $Q_x/Q_y$ as function of
  relaxation time $\tau$, for various values of the deformation
  parameter $a$. }
\label{fig:Qratio}
\end{center}
\end{figure}

In Fig.~\ref{fig:Qratio} is shown the asymmetry between the monopole
charge distribution in the $x$ and the $y$ direction for various
deformation values $a=1,\ldots,5$ as function of relaxation time
$\tau$.
In the shown relaxation-time interval, all the configurations clearly
converge towards the non-deformed configuration, i.e.~$d=0$. The more
deformed the initial configuration is, the longer it takes for the
relaxation to relax the configuration back to the non-deformed one. 

\begin{figure}[!tpb]
\begin{center}
\mbox{\subfloat[]{\includegraphics[width=0.33\linewidth]{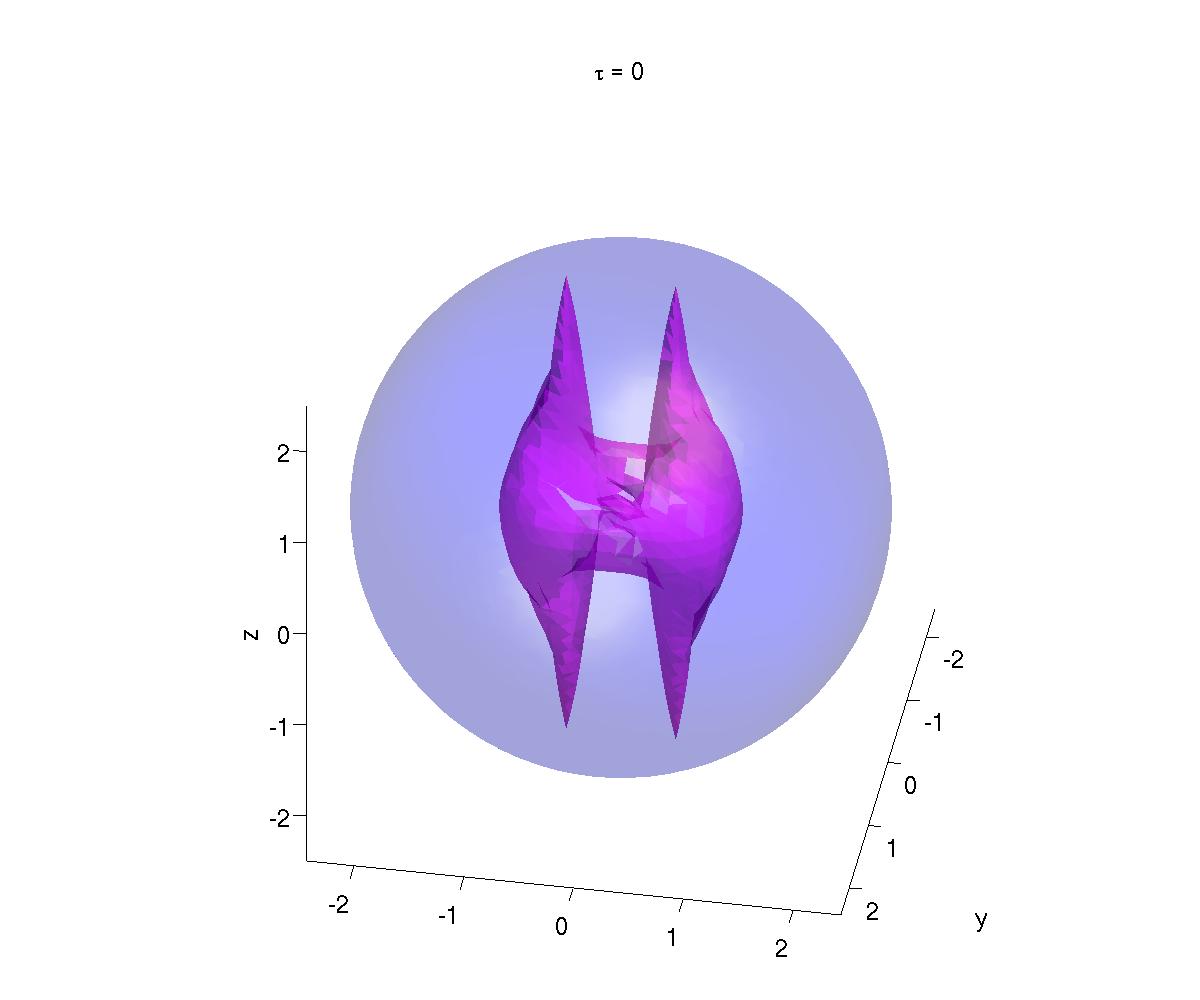}}
  \subfloat[]{\includegraphics[width=0.33\linewidth]{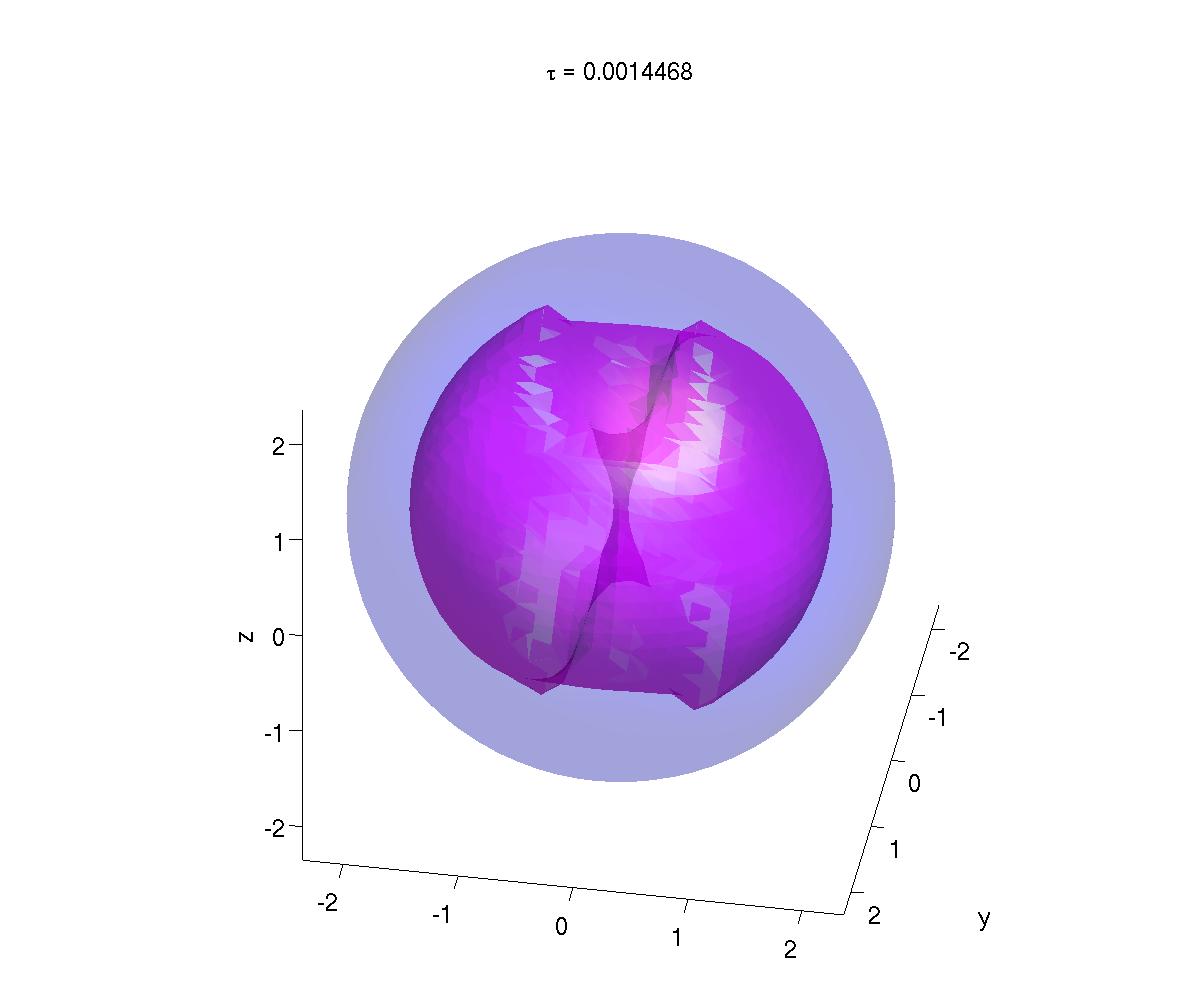}}
  \subfloat[]{\includegraphics[width=0.33\linewidth]{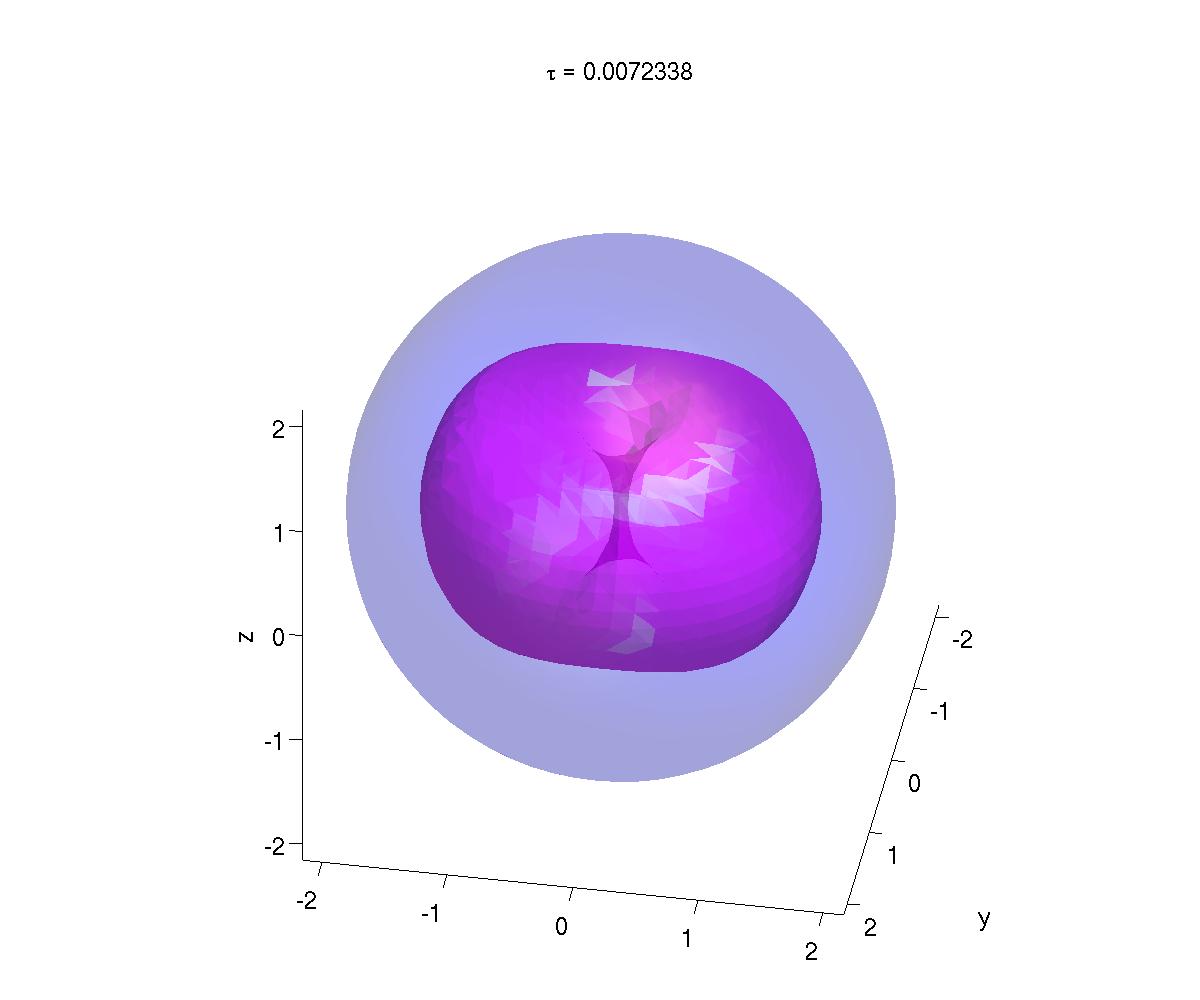}}}
\mbox{\subfloat[]{\includegraphics[width=0.33\linewidth]{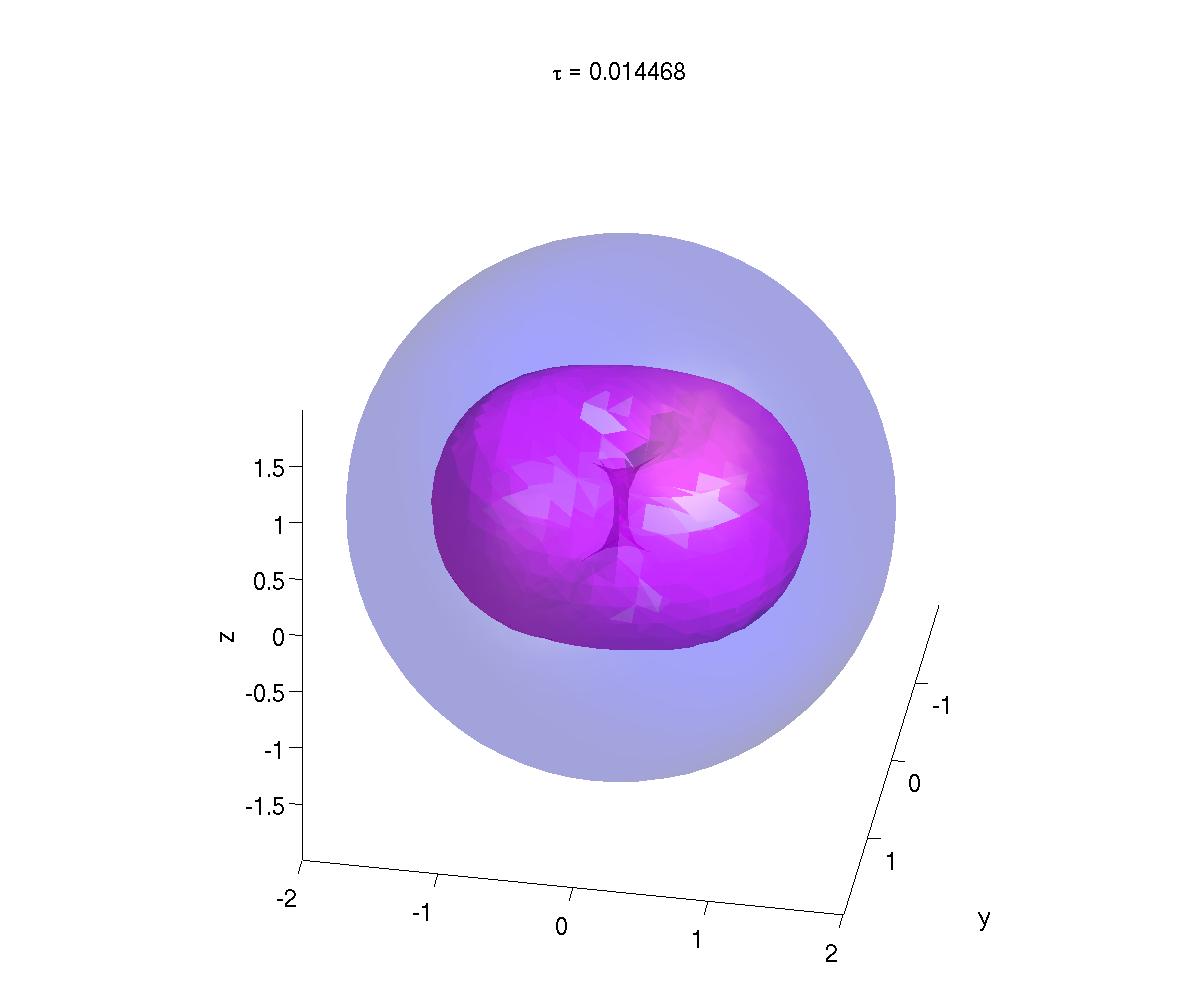}}
  \subfloat[]{\includegraphics[width=0.33\linewidth]{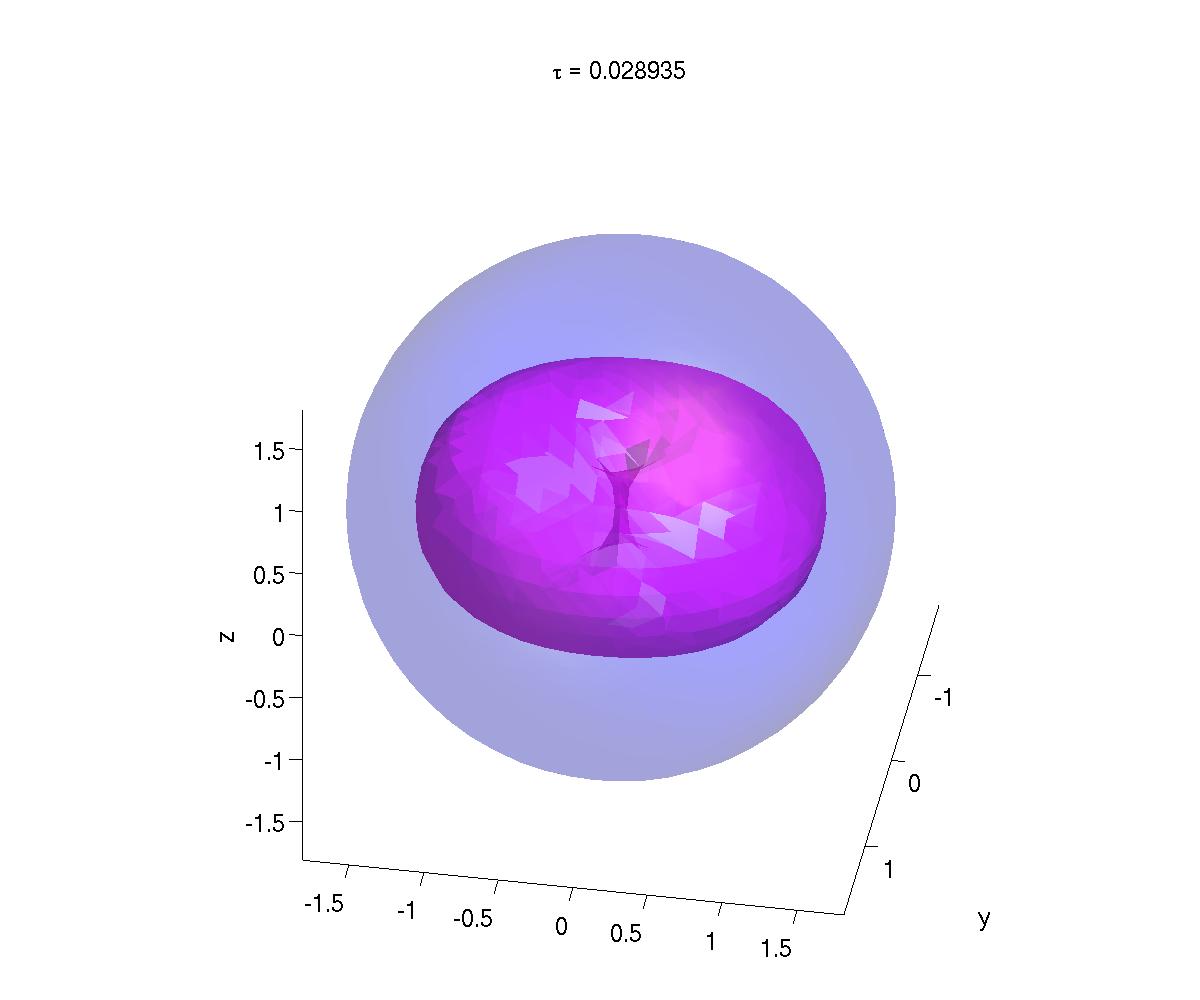}}
  \subfloat[]{\includegraphics[width=0.33\linewidth]{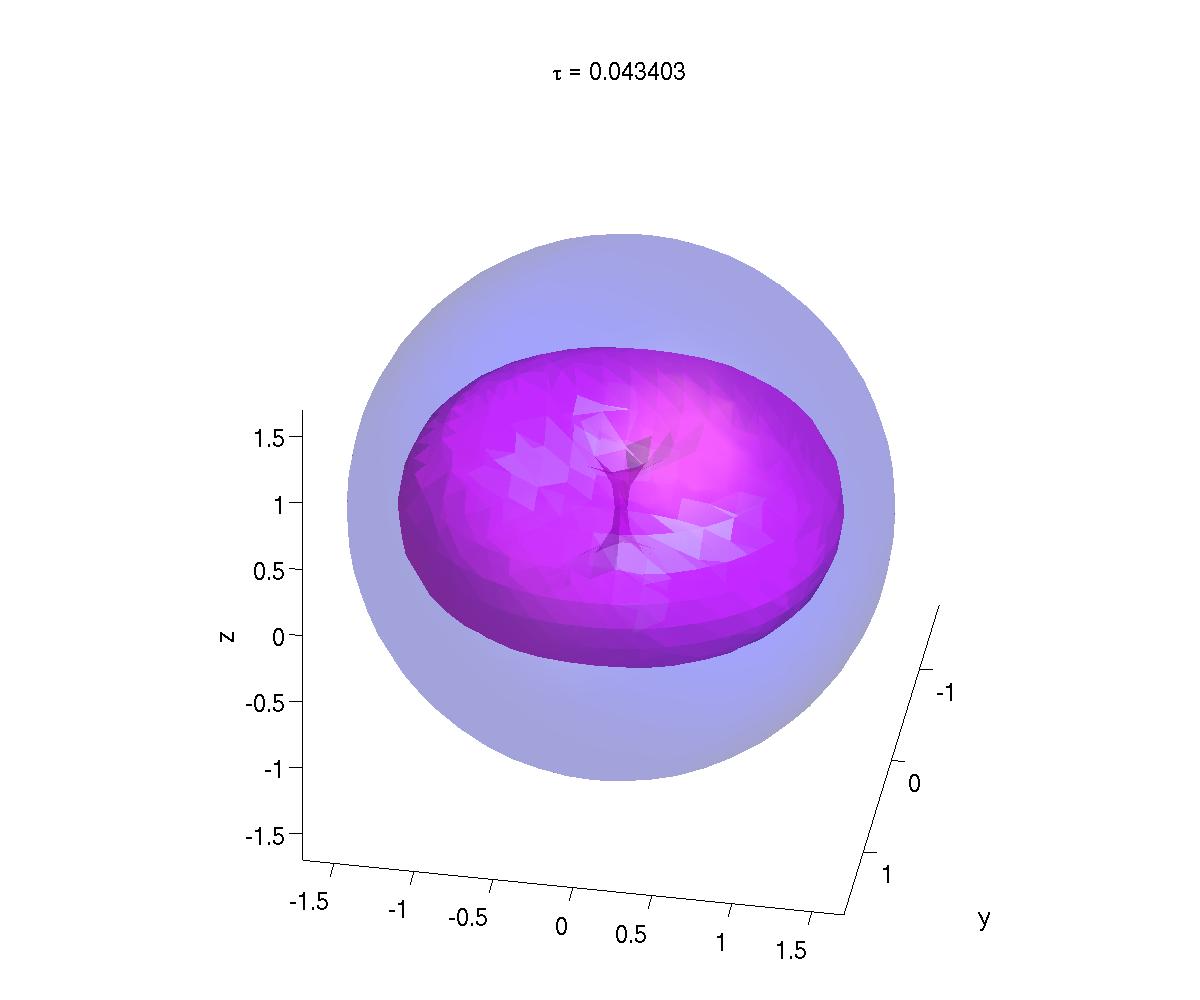}}}
\mbox{\subfloat[]{\includegraphics[width=0.33\linewidth]{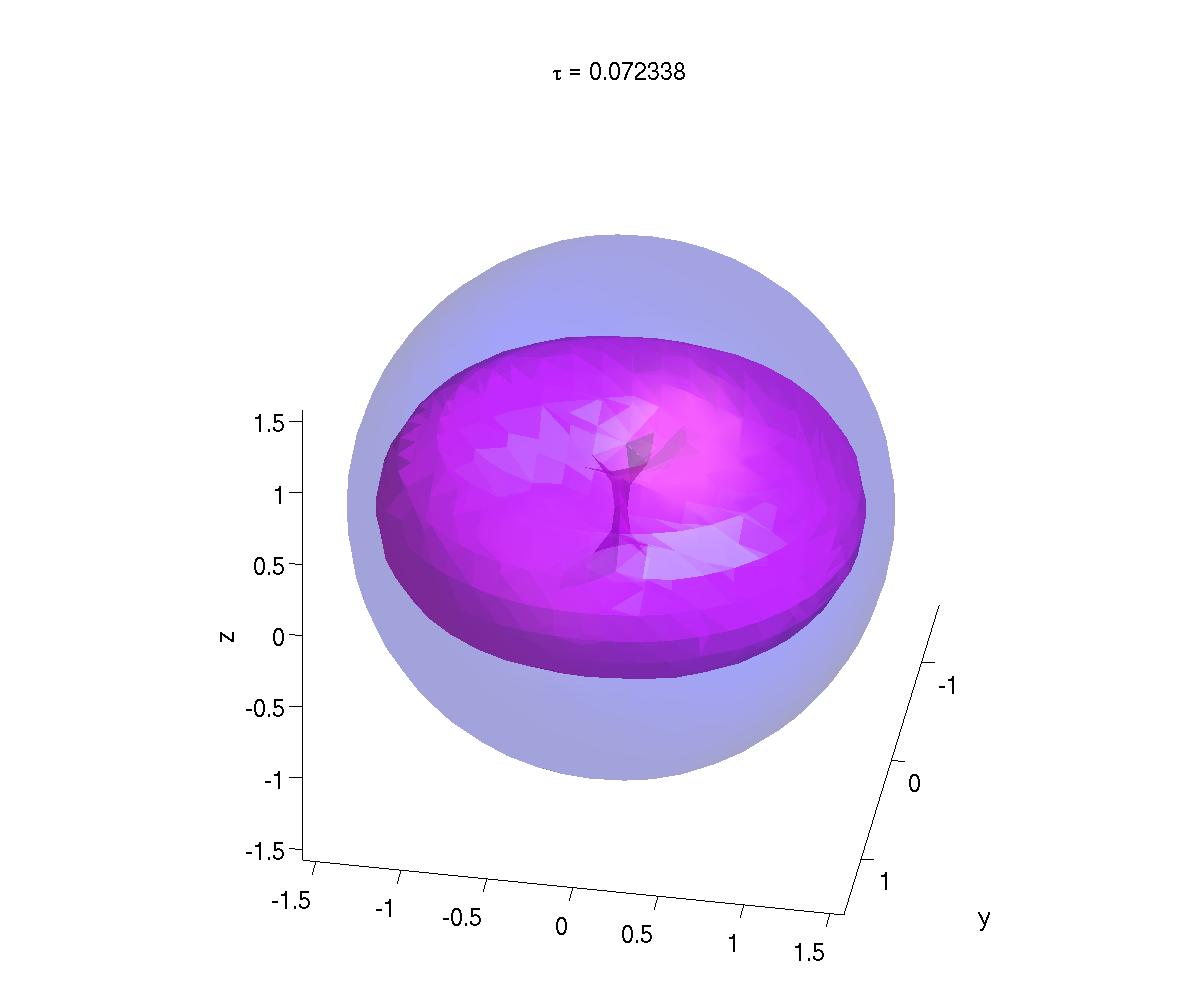}}
  \subfloat[]{\includegraphics[width=0.33\linewidth]{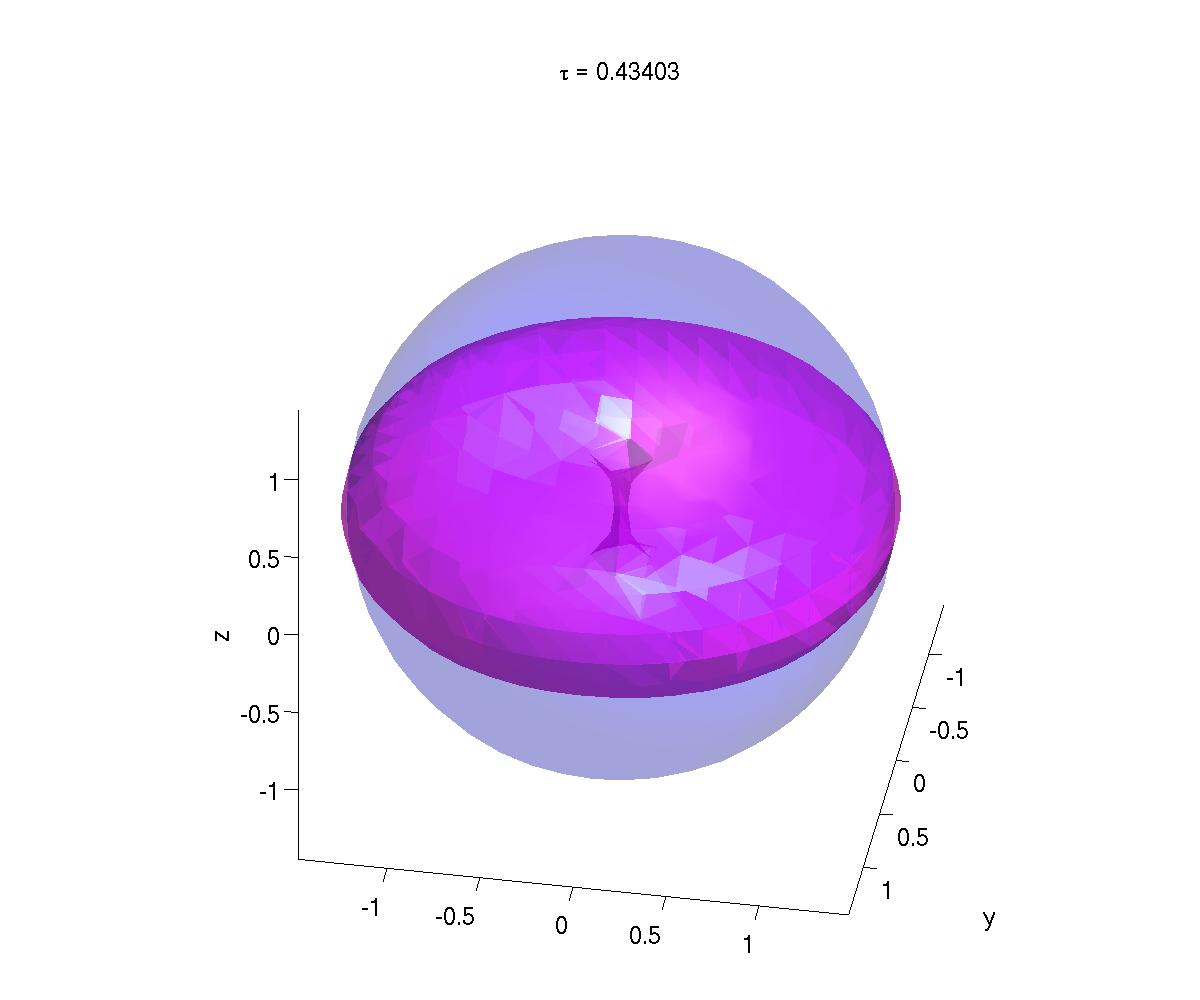}}
  \subfloat[]{\includegraphics[width=0.33\linewidth]{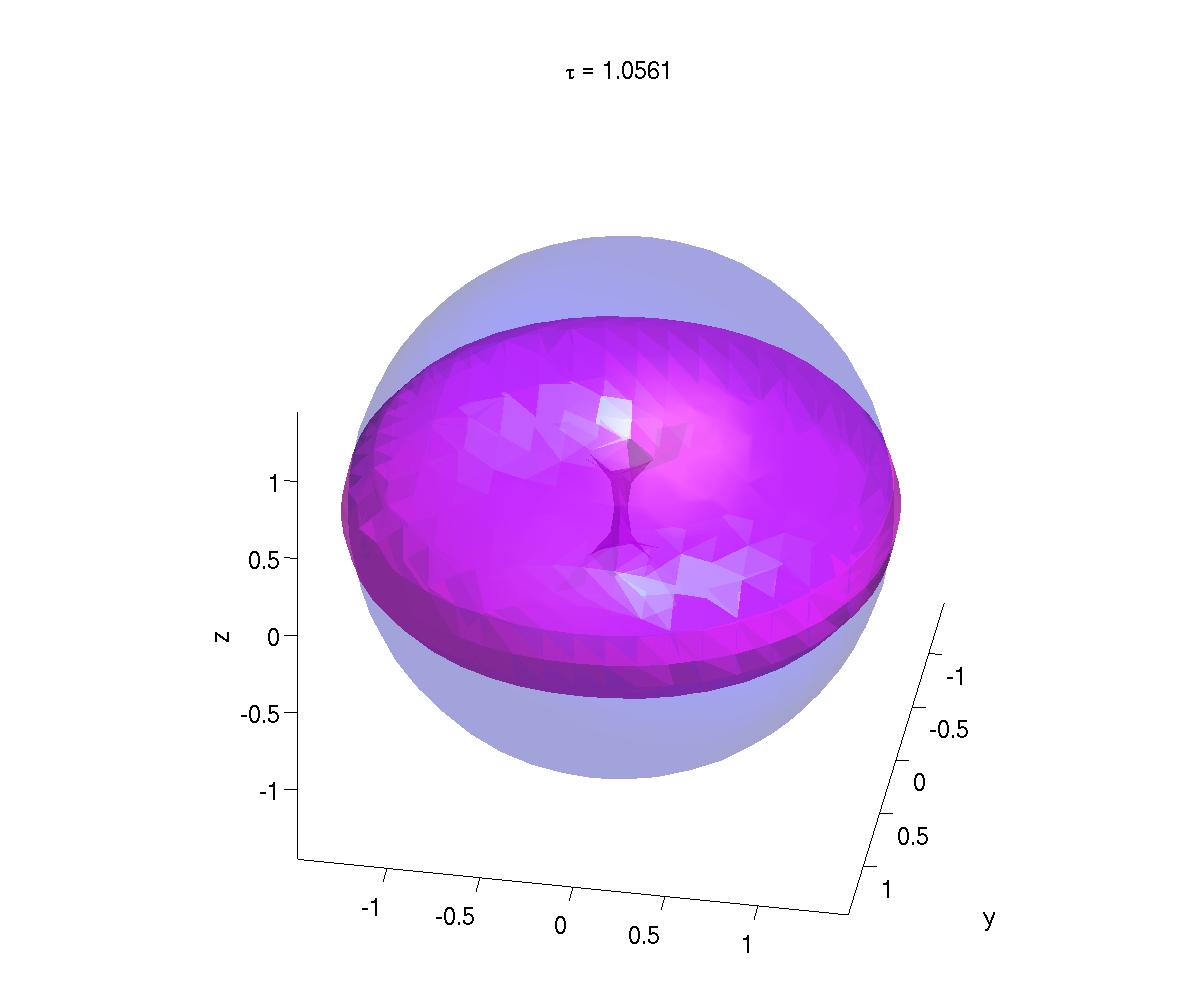}}}
\caption{The global two-monopole with a deformed initial condition
  ($a=5$) inside a single Skyrmion. The figures show isosurfaces of
  the monopole charge and Skyrmion charge densities, respectively, at
  half their respective  maximum values. 
  The blue surface is the Skyrmion charge and the magenta surface is the
  monopole charge.
  As the relaxation time progresses, the solution converges towards
  that of Fig.~\ref{fig:skm21}.
  The calculation is made on a $121^3$ cubic lattice. 
}
\label{fig:skm21a5isosurfaces}
\end{center}
\end{figure}

In Fig.~\ref{fig:skm21a5isosurfaces} is shown a series of snapshots of
the isosurfaces of Skyrmion and monopole charges as the relaxation
time progresses. This particular series is made with the deformation
parameter $a=5$. The configuration quickly converges towards a torus
shape and after a sufficiently long relaxation time, it becomes
infinitesimally close to the $a=0$ ($d=0$) solution. 

\begin{figure}[!tpb]
\begin{center}
\mbox{\subfloat[]{\includegraphics[width=0.33\linewidth]{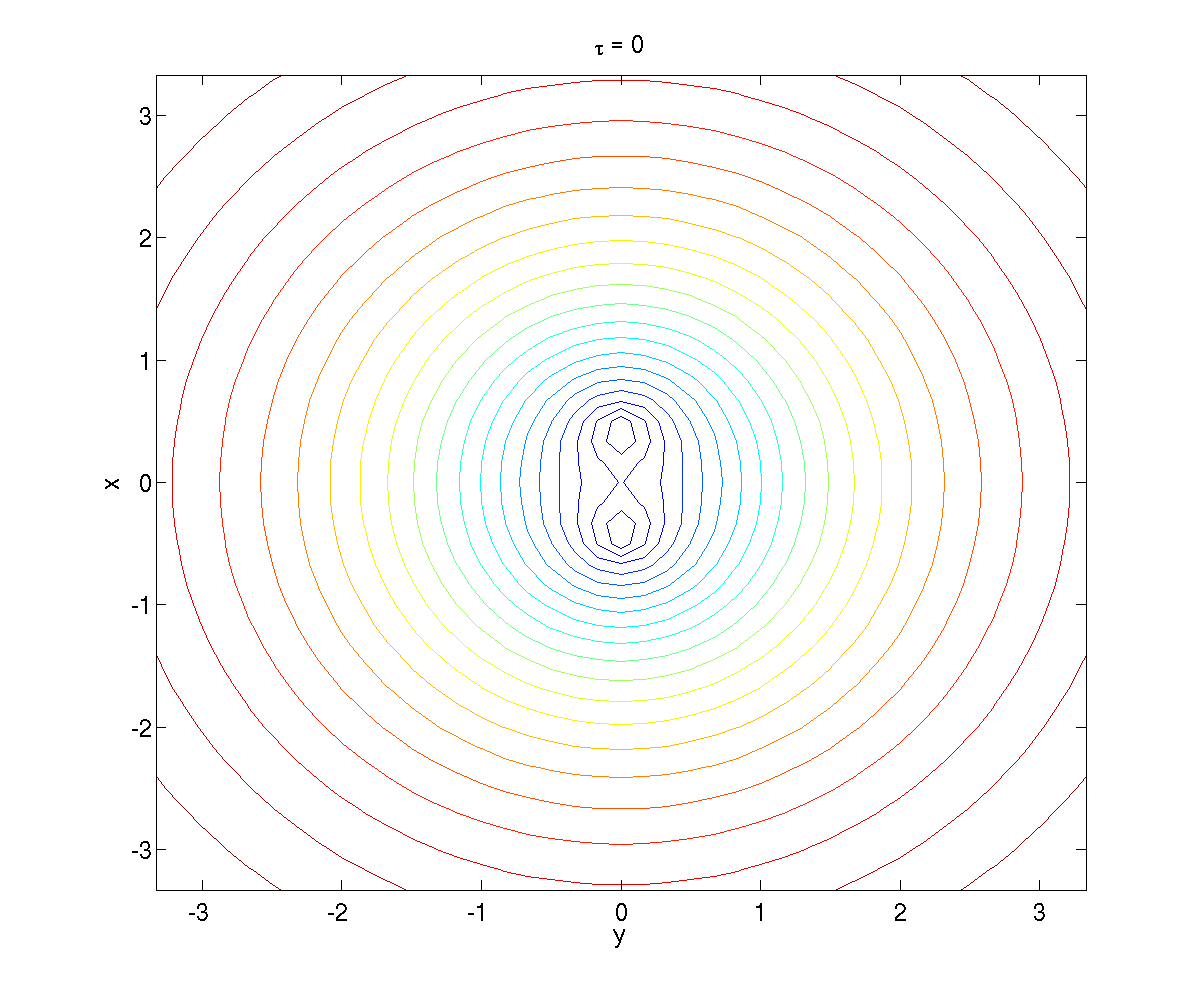}}
  \subfloat[]{\includegraphics[width=0.33\linewidth]{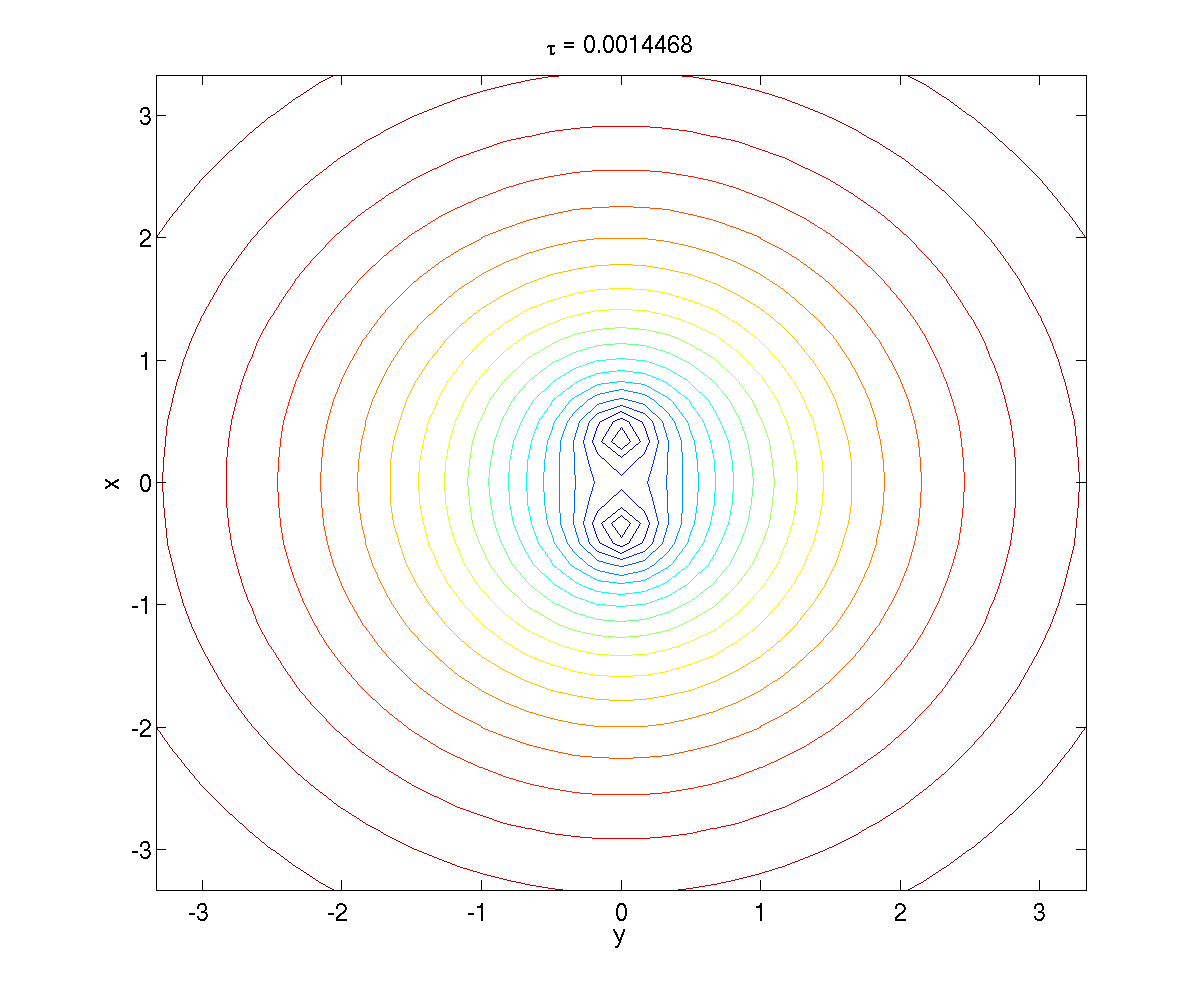}}
  \subfloat[]{\includegraphics[width=0.33\linewidth]{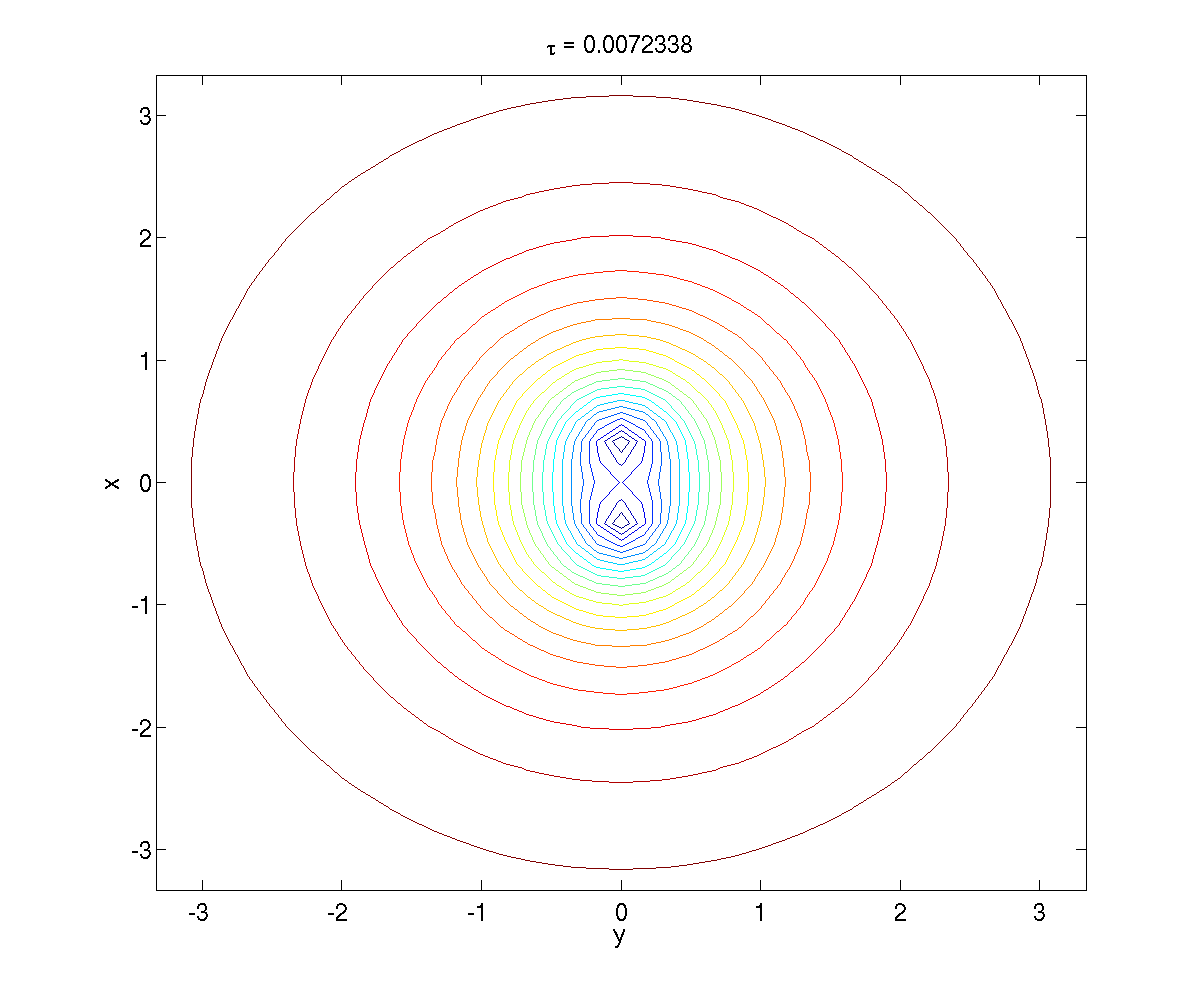}}}
\mbox{\subfloat[]{\includegraphics[width=0.33\linewidth]{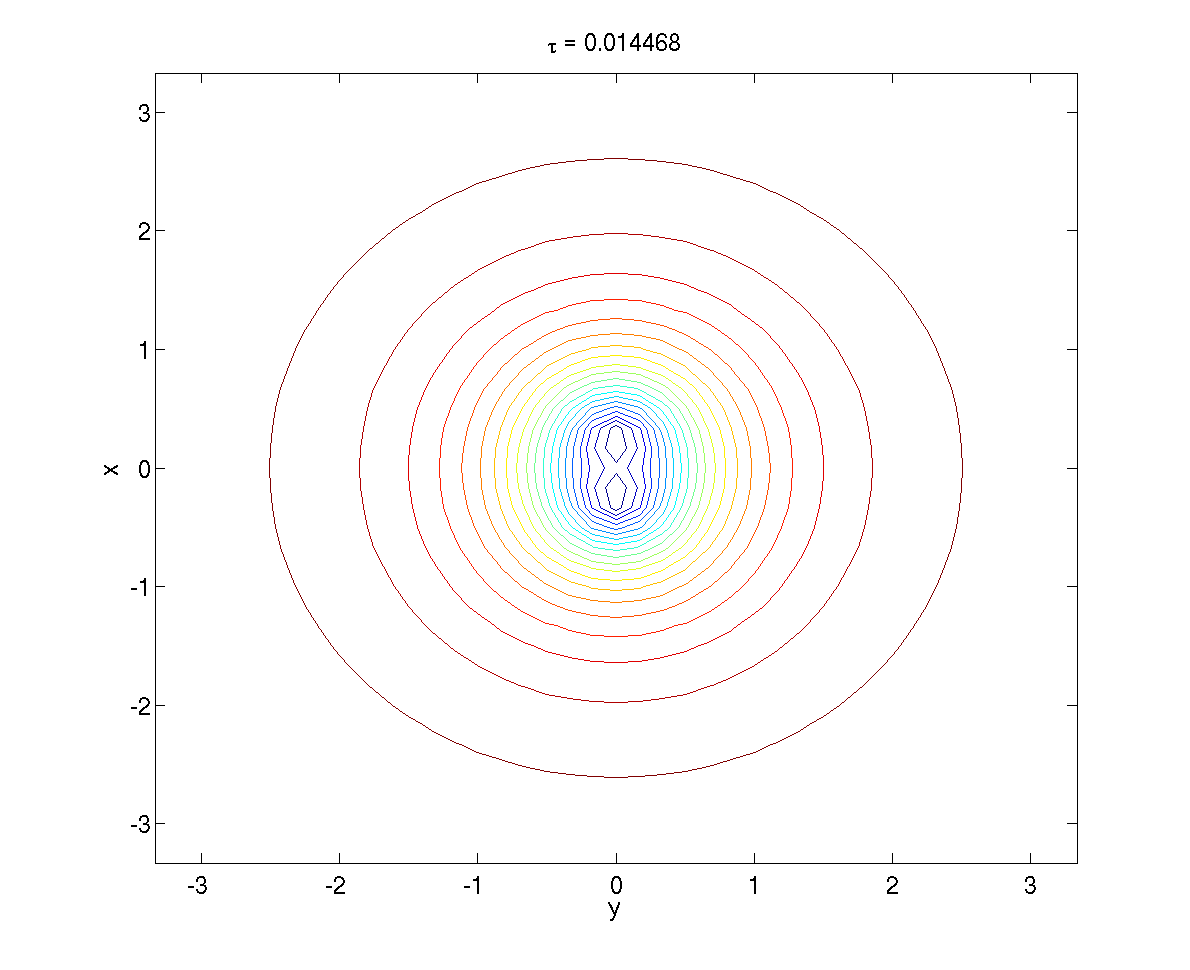}}
  \subfloat[]{\includegraphics[width=0.33\linewidth]{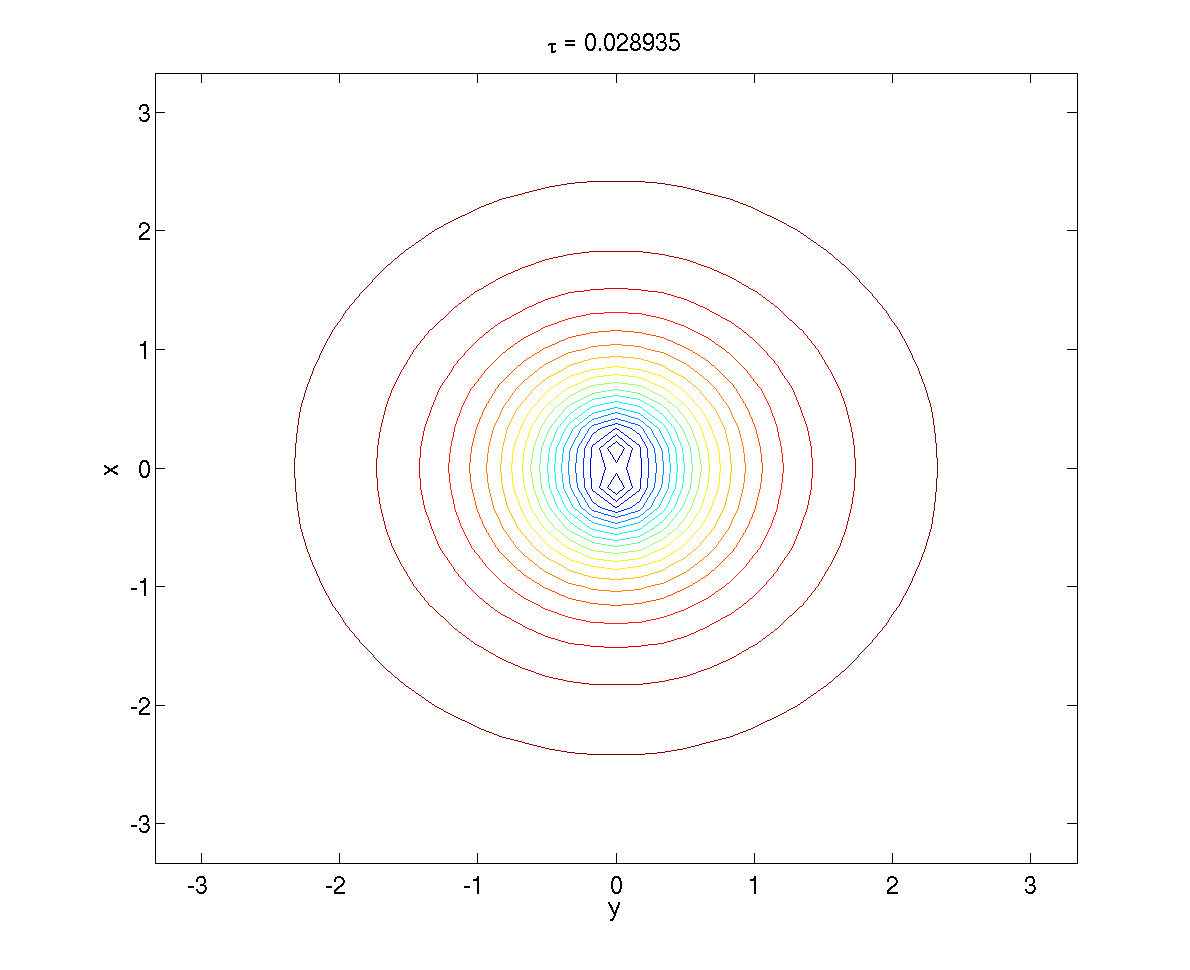}}
  \subfloat[]{\includegraphics[width=0.33\linewidth]{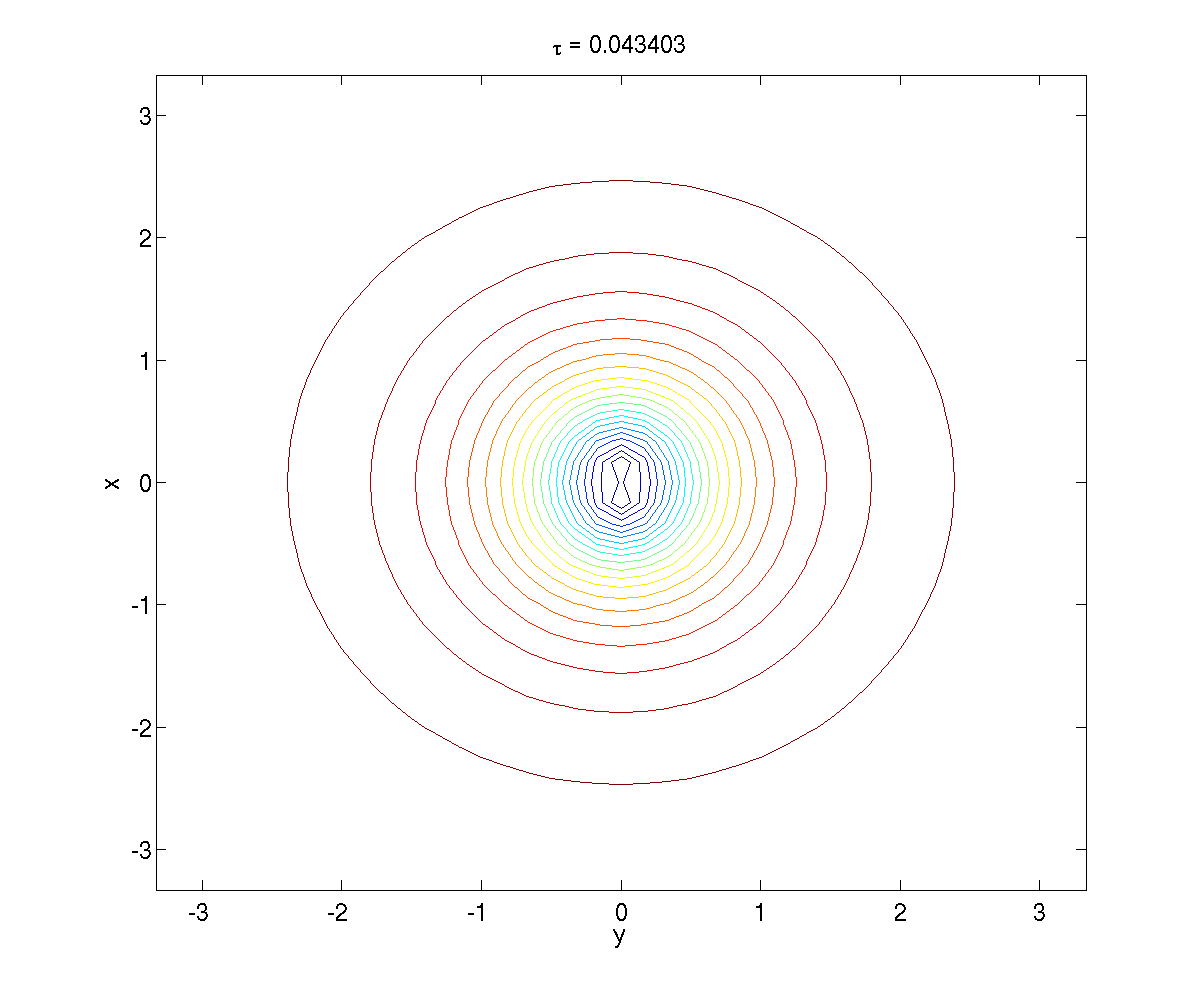}}}
\mbox{\subfloat[]{\includegraphics[width=0.33\linewidth]{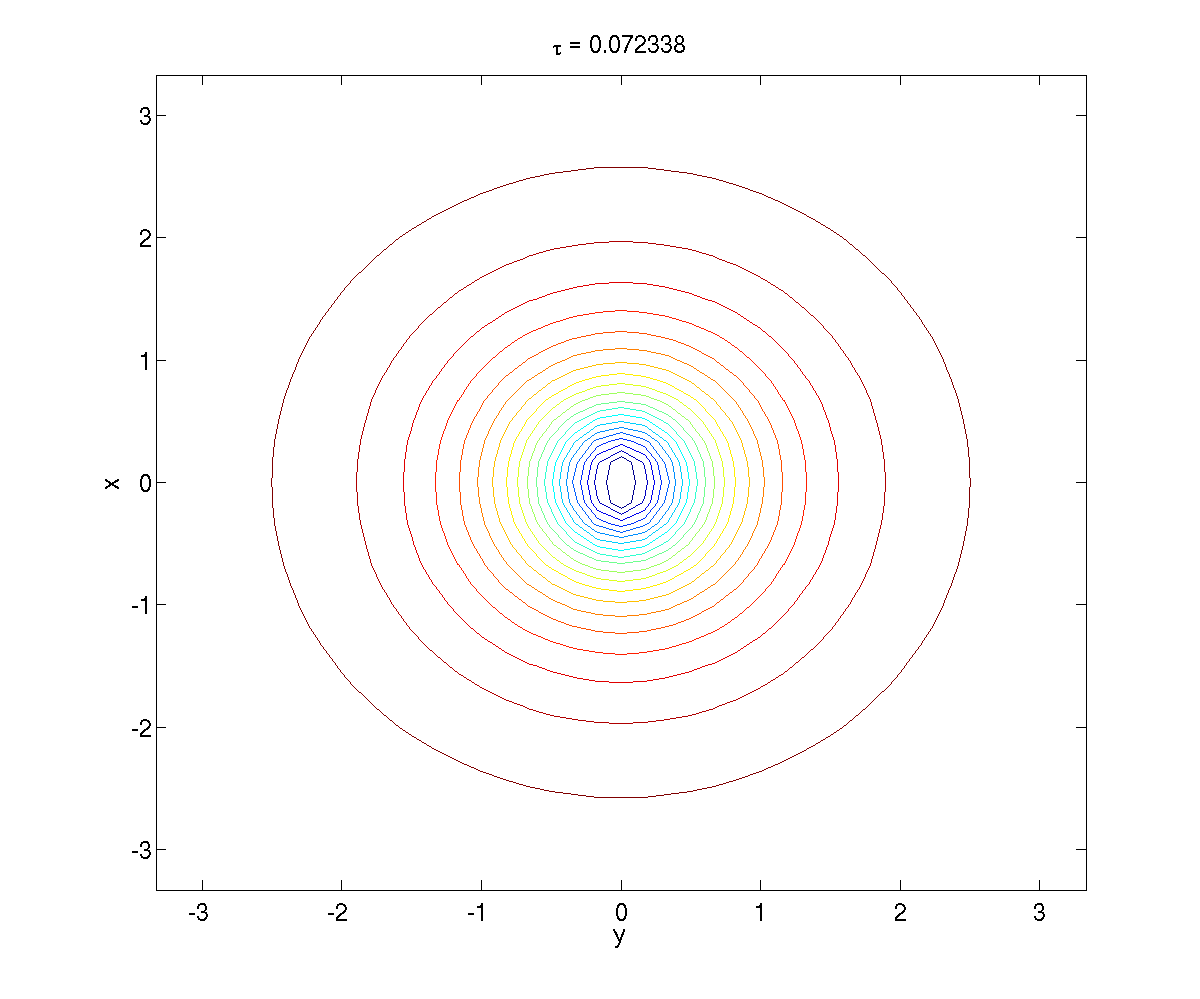}}
  \subfloat[]{\includegraphics[width=0.33\linewidth]{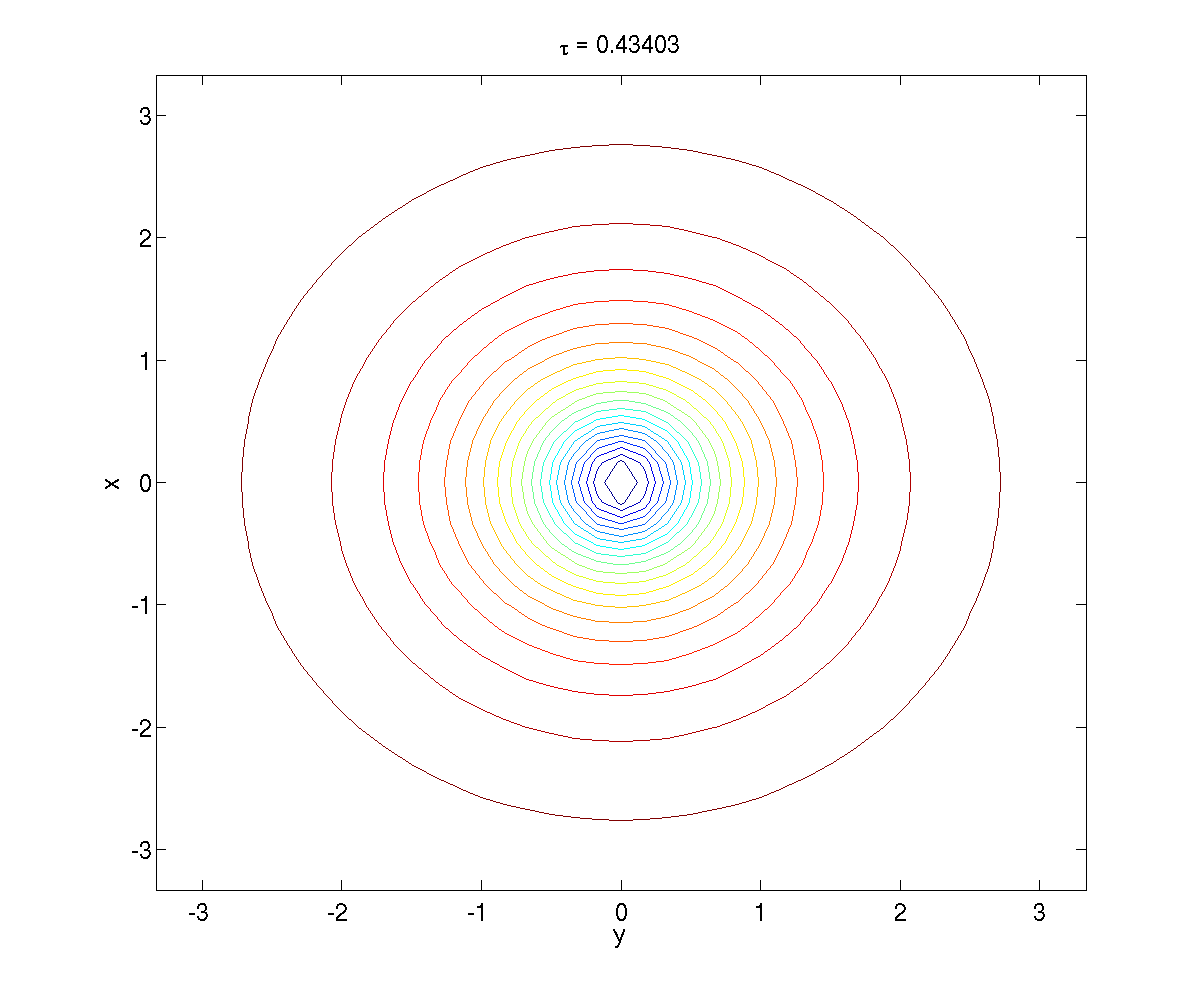}}
  \subfloat[]{\includegraphics[width=0.33\linewidth]{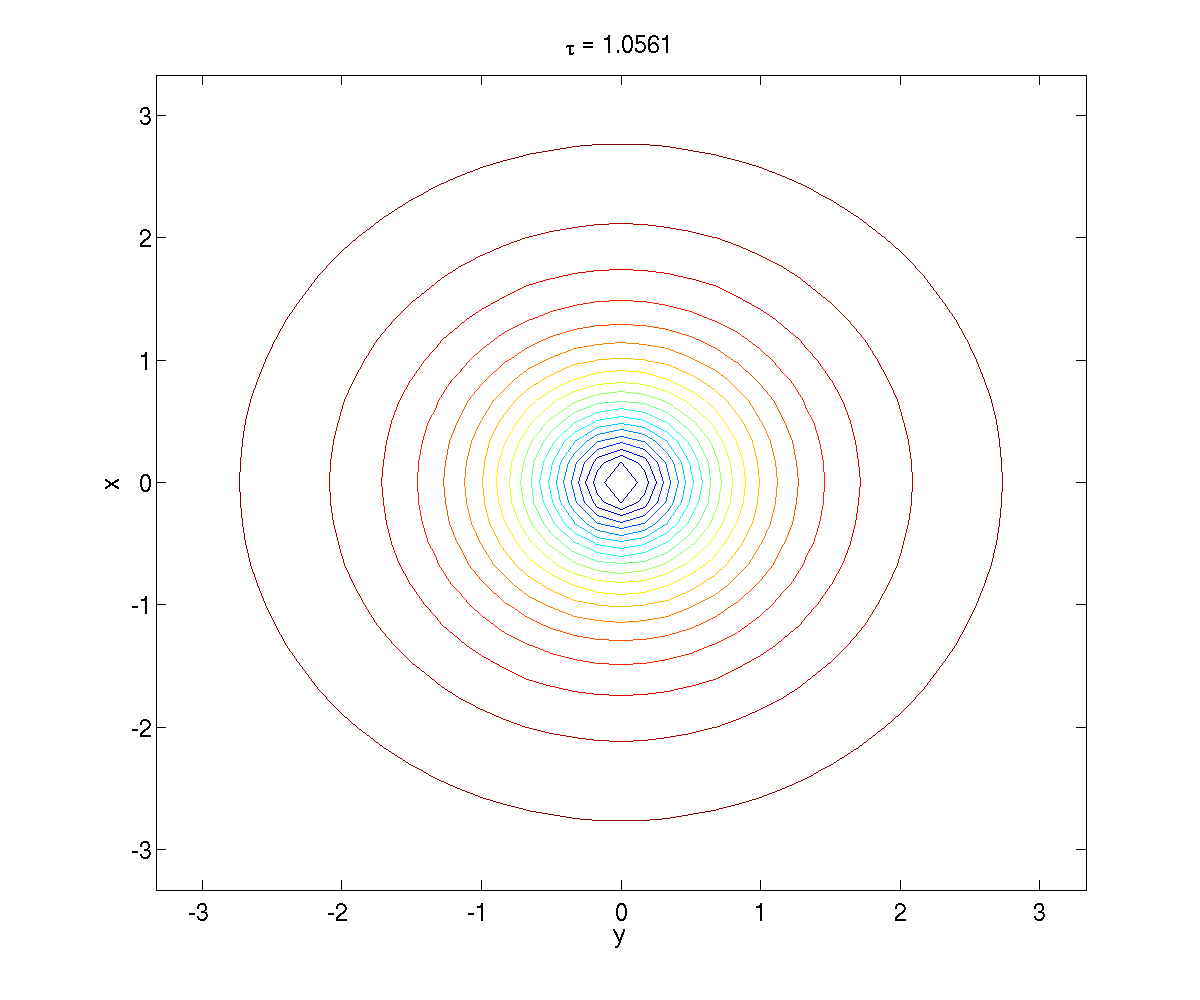}}}
\caption{The global two-monopole with a deformed initial condition
  ($a=5$) inside a single Skyrmion. The figures show contours of the
  norm of $\Phi$, i.e.~$\sqrt{\Phi_1^2+\Phi_2^2+\Phi_3^2}$ on an $xy$
  slice at $z=0$.
  The blue circles in the middle of figures (a)-(f) show two separated
  zeros in the monopole field.
  As the relaxation time progresses, the two distinct zeros clearly
  merge to a single zero and the solution converges towards that of
  Fig.~\ref{fig:skm21}. 
}
\label{fig:skm21a5normcontours}
\end{center}
\end{figure}

In Fig.~\ref{fig:skm21a5normcontours} is shown a series of snapshots
of the contours of the norm of the monopole field
\beq
\sqrt{(\Phi^1)^2 + (\Phi^2)^2 + (\Phi^3)^2}.
\eeq
as the relaxation time progresses. This particular series is again
made with the deformation parameter $a=5$.
It is observed from the innermost contour(s) that the monopole field
possesses two zeros when the relaxation algorithm begins. After the
time of Fig.~\ref{fig:skm21a5normcontours}f and before
Fig.~\ref{fig:skm21a5normcontours}g, the two zeros merge to a single
zero and hence becomes the non-deformed solution $a=0$ ($d=0$). We
take this as a proof of stability of the torus-like solution shown in
the last subsection. 

\subsection{Instability of the type B, $Q=2$ Skyrmopole}

\begin{figure}[!tpb]
\begin{center}
\mbox{\subfloat[]{\includegraphics[width=0.24\linewidth]{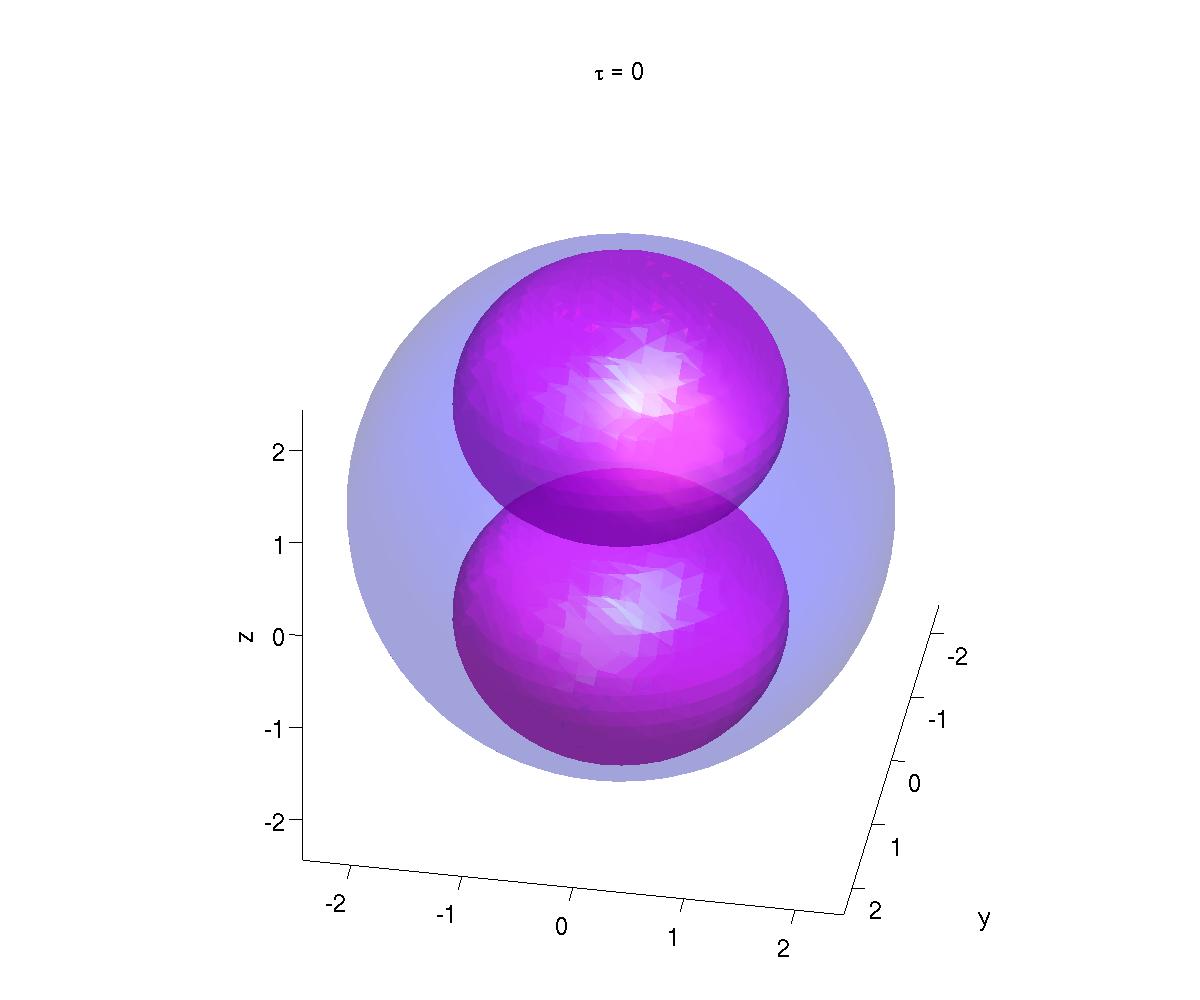}}
  \subfloat[]{\includegraphics[width=0.24\linewidth]{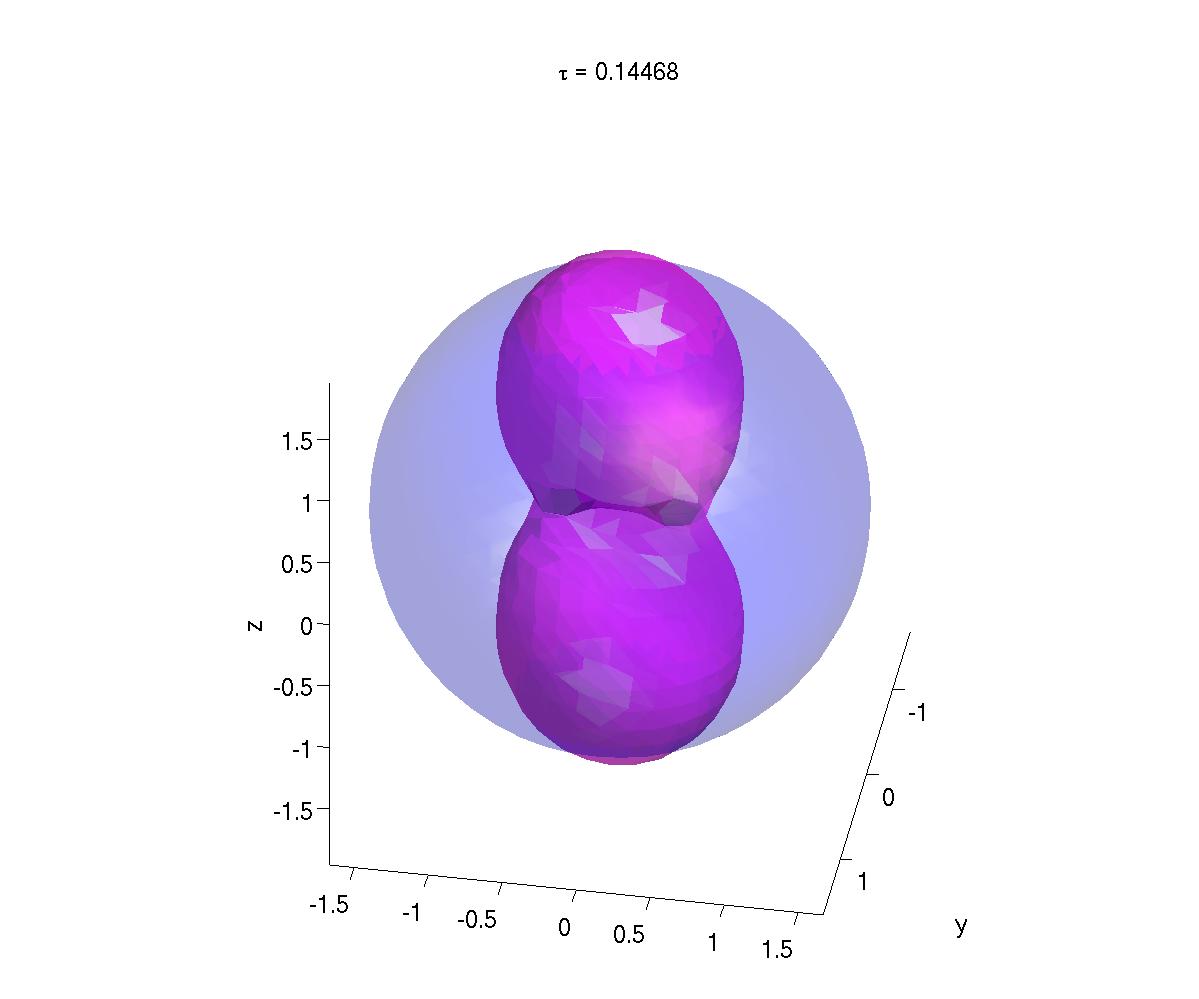}}
  \subfloat[]{\includegraphics[width=0.24\linewidth]{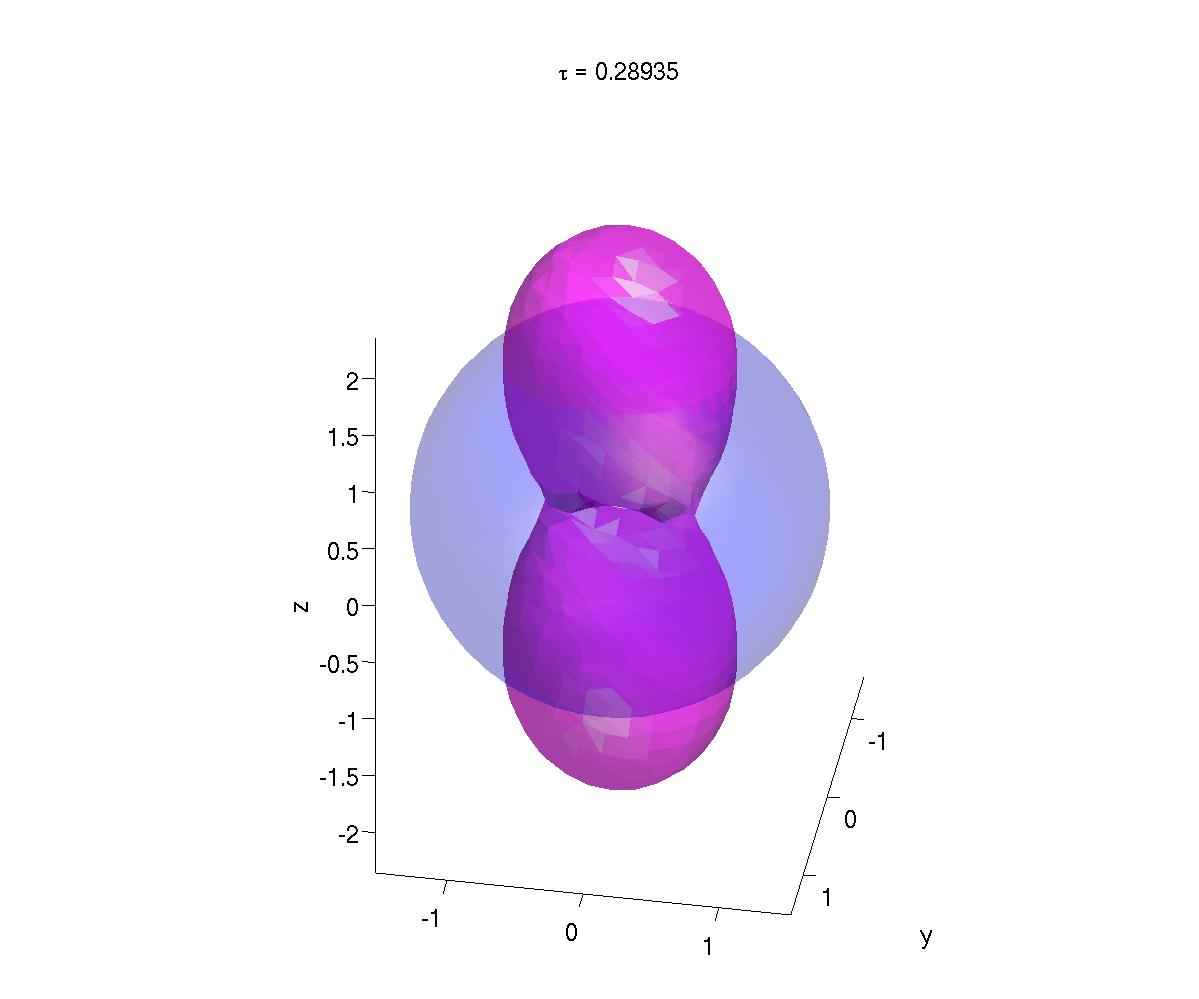}}
  \subfloat[]{\includegraphics[width=0.24\linewidth]{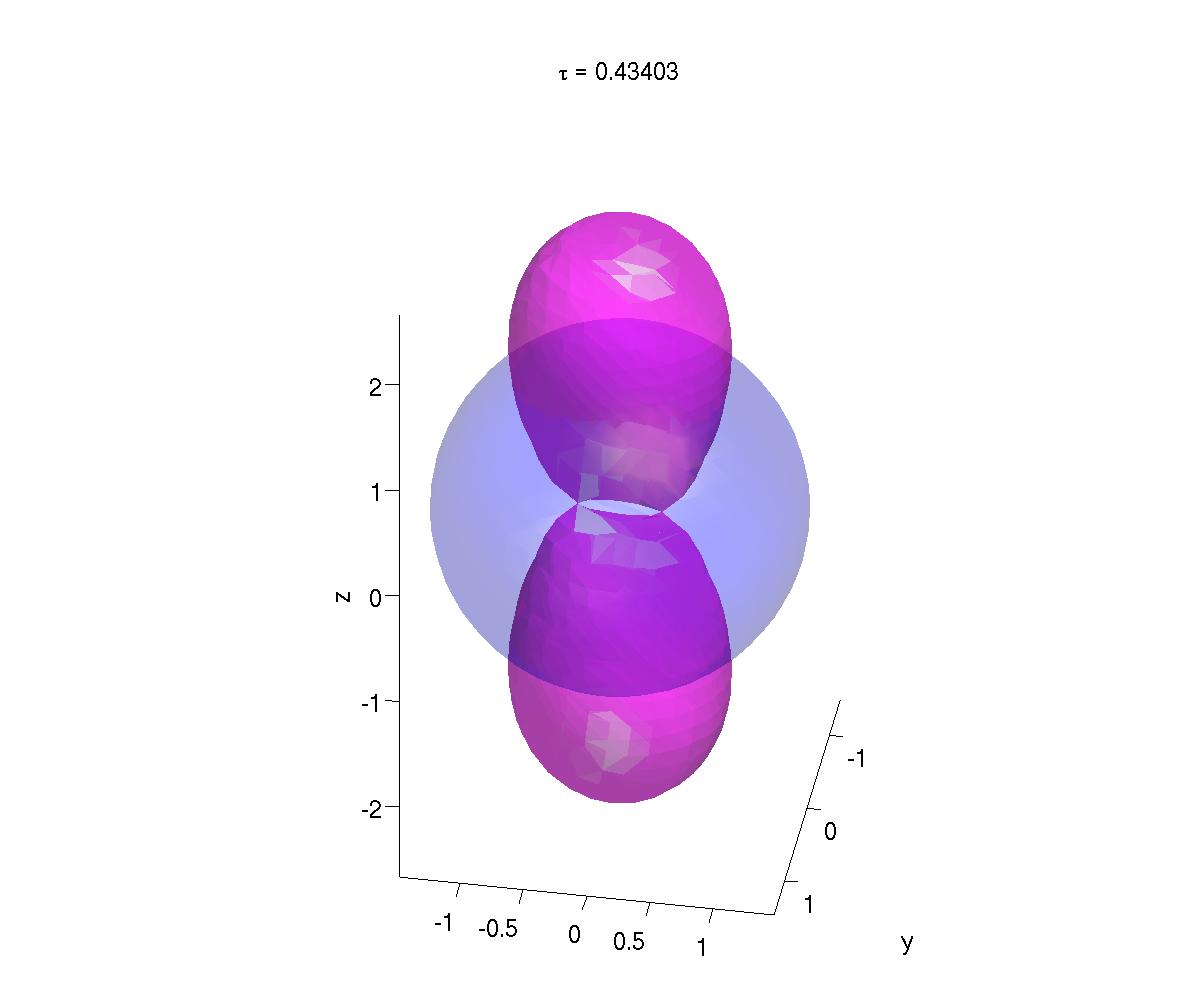}}}
\mbox{\subfloat[]{\includegraphics[width=0.24\linewidth]{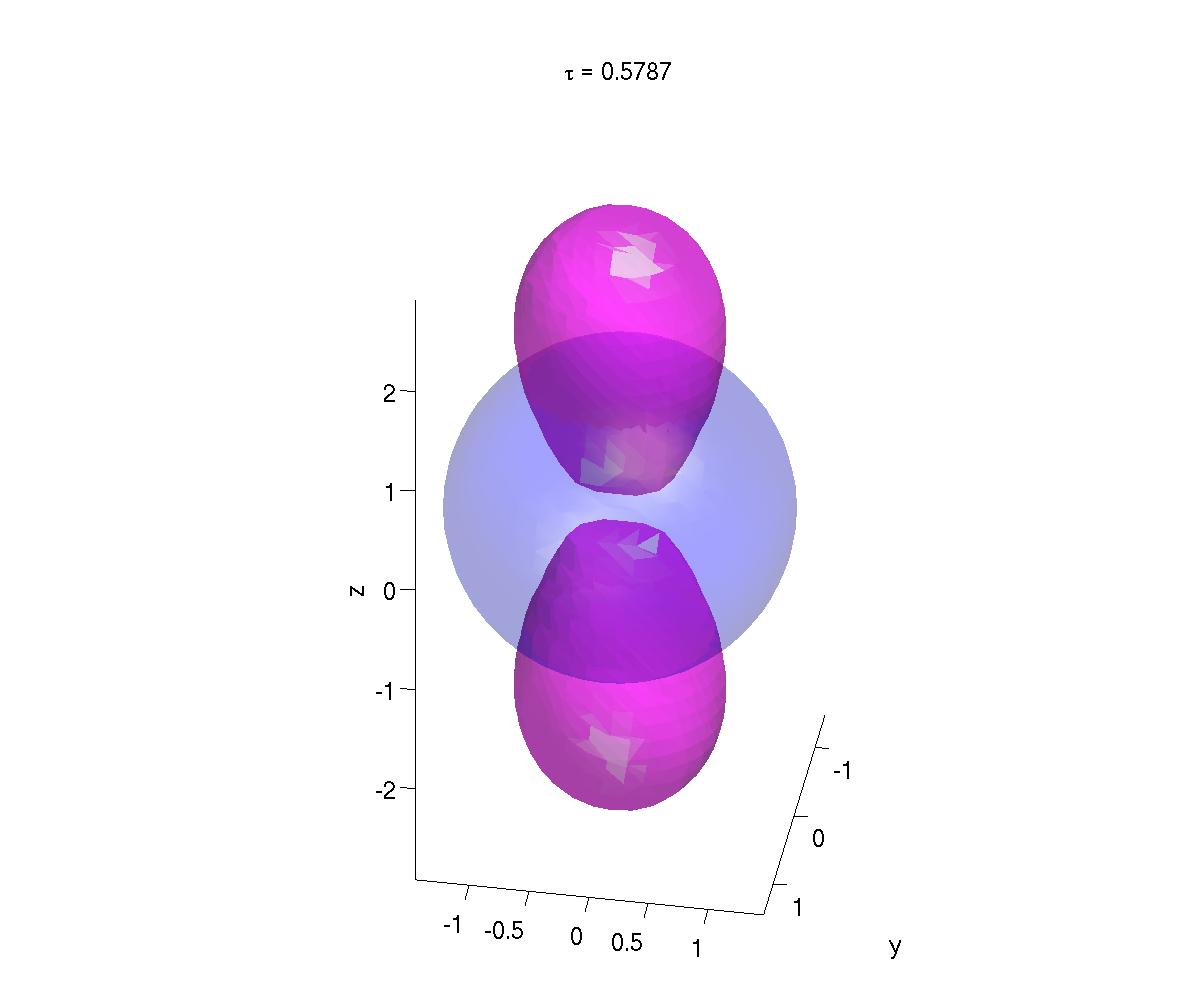}}
  \subfloat[]{\includegraphics[width=0.24\linewidth]{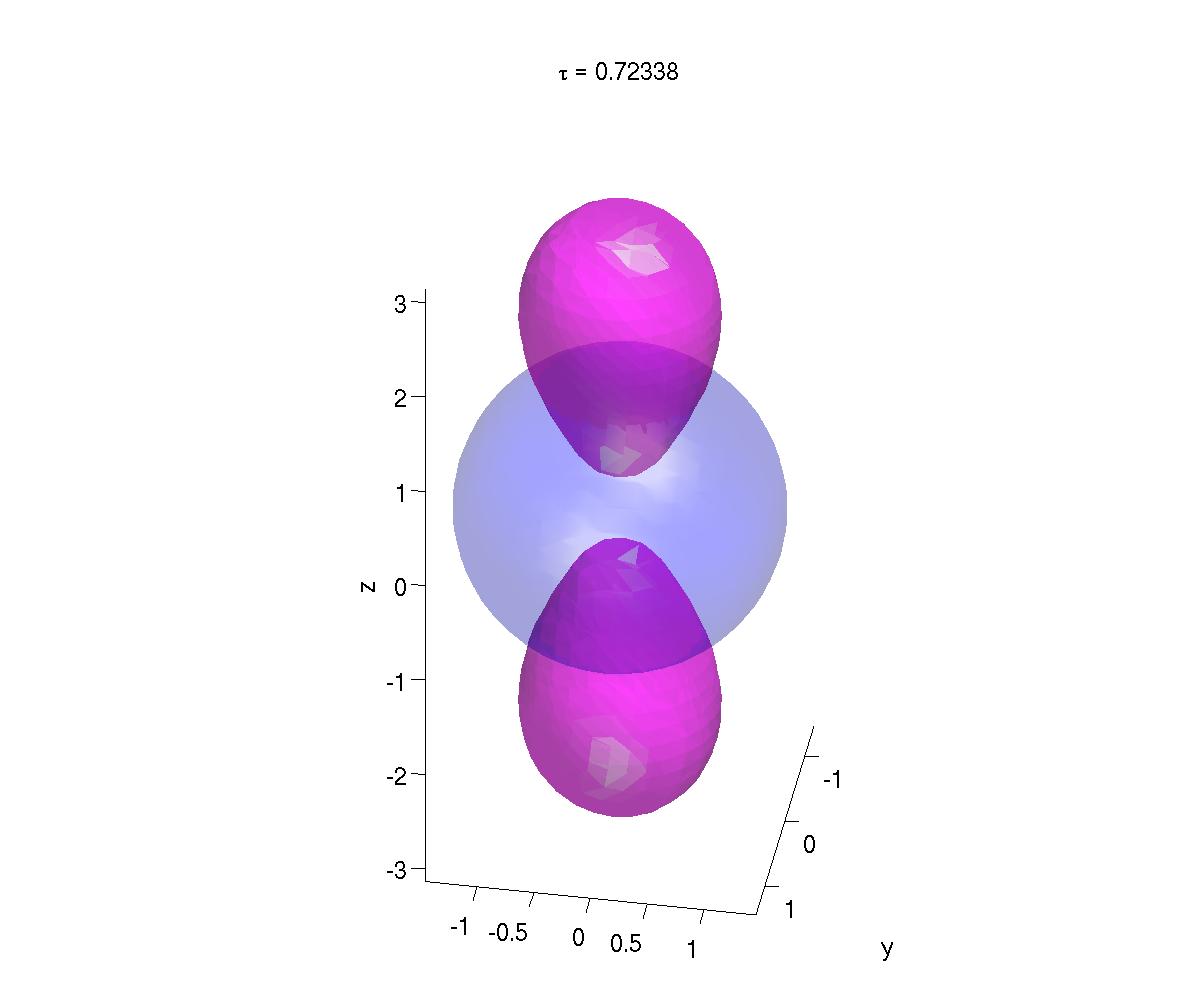}}
  \subfloat[]{\includegraphics[width=0.24\linewidth]{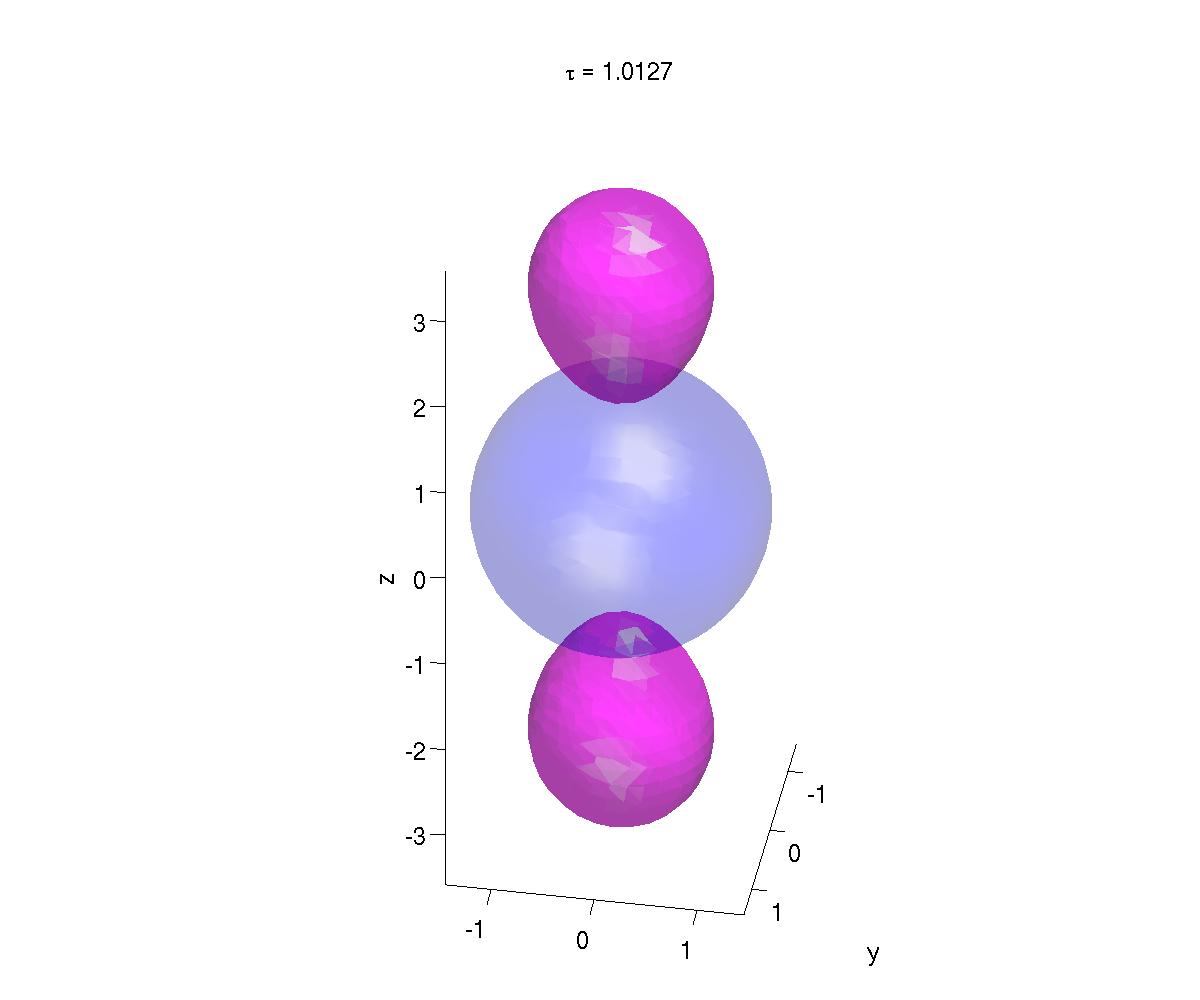}}
  \subfloat[]{\includegraphics[width=0.24\linewidth]{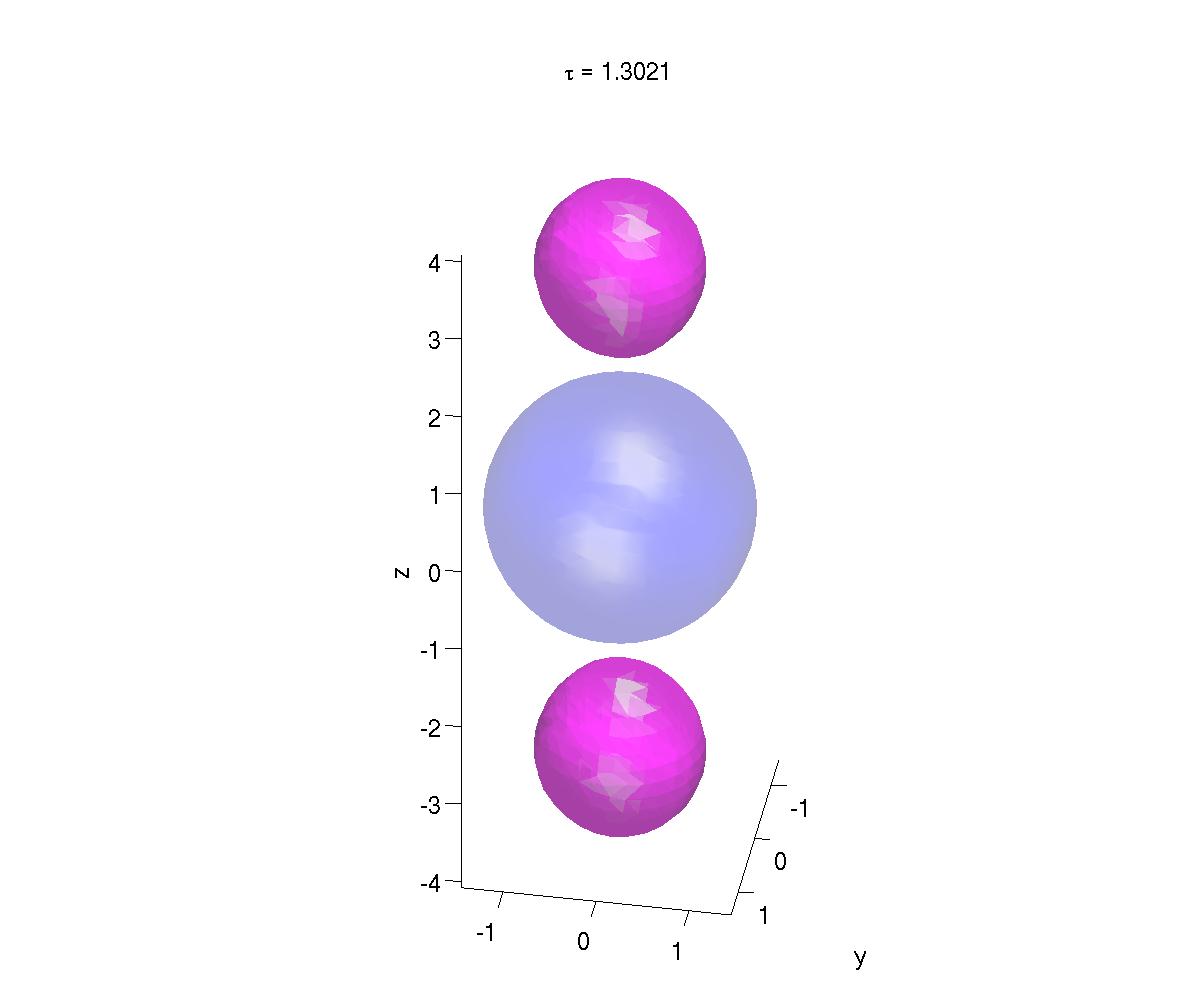}}}
\mbox{\subfloat[]{\includegraphics[width=0.24\linewidth]{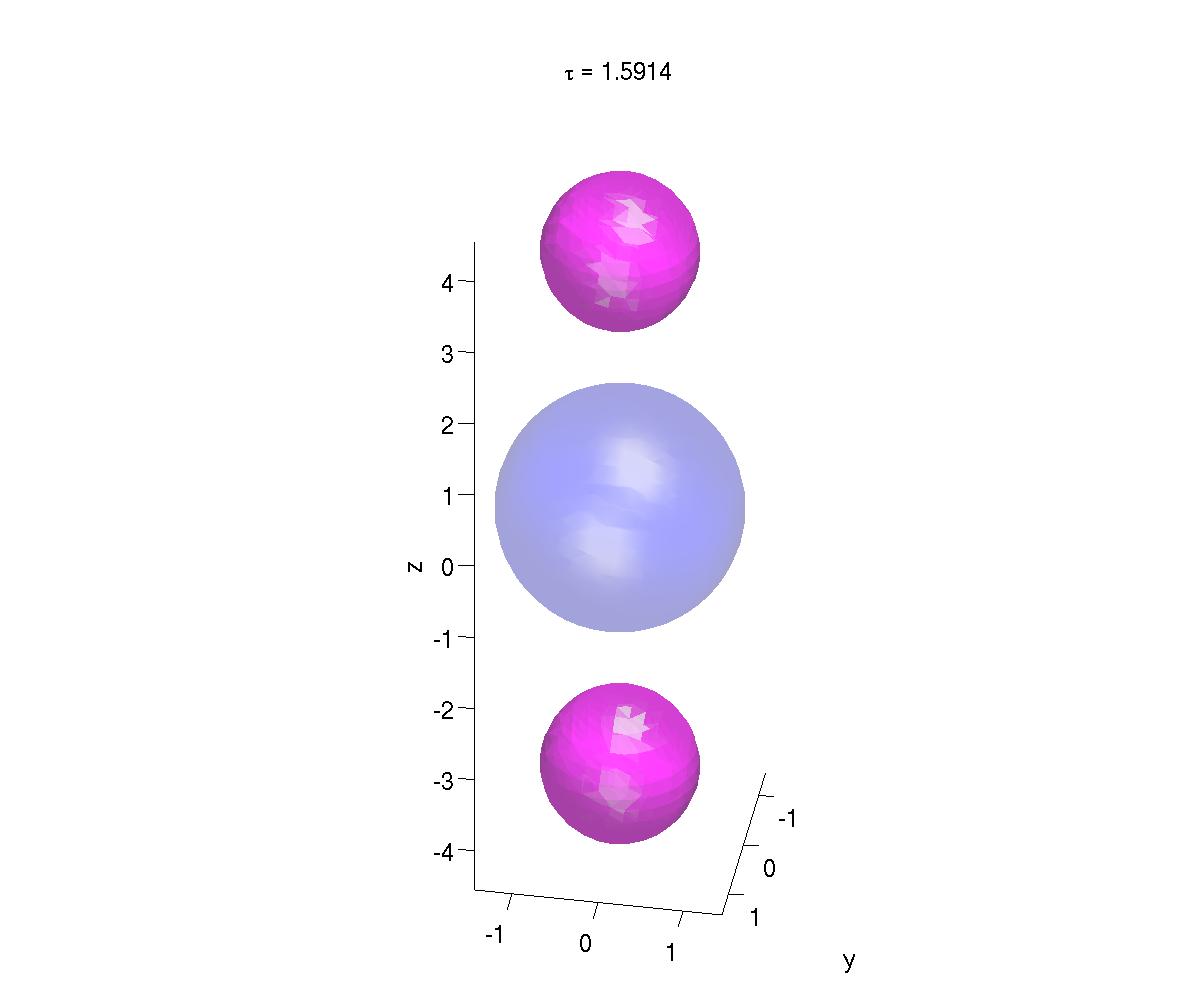}}
  \subfloat[]{\includegraphics[width=0.24\linewidth]{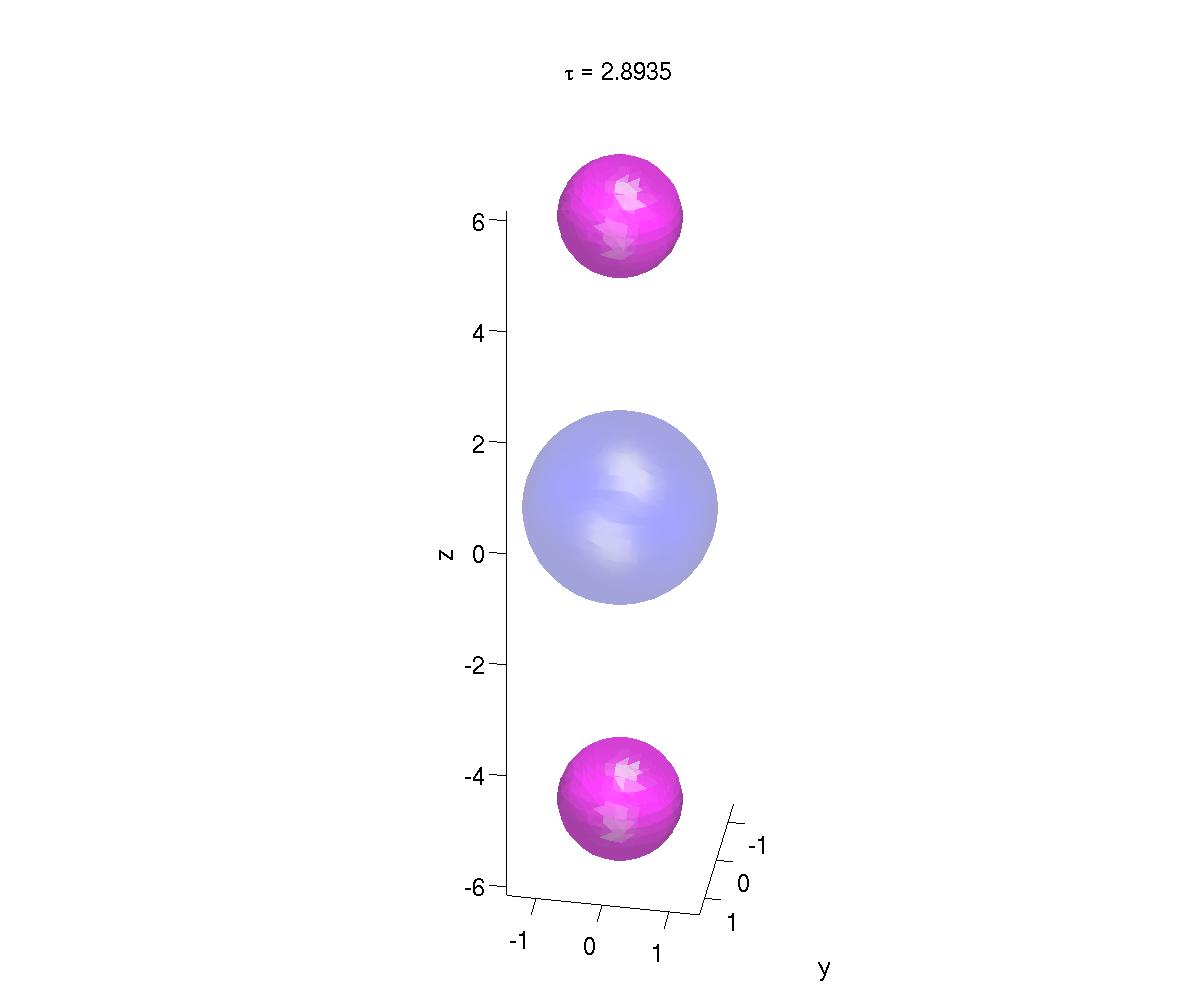}}
  \subfloat[]{\includegraphics[width=0.24\linewidth]{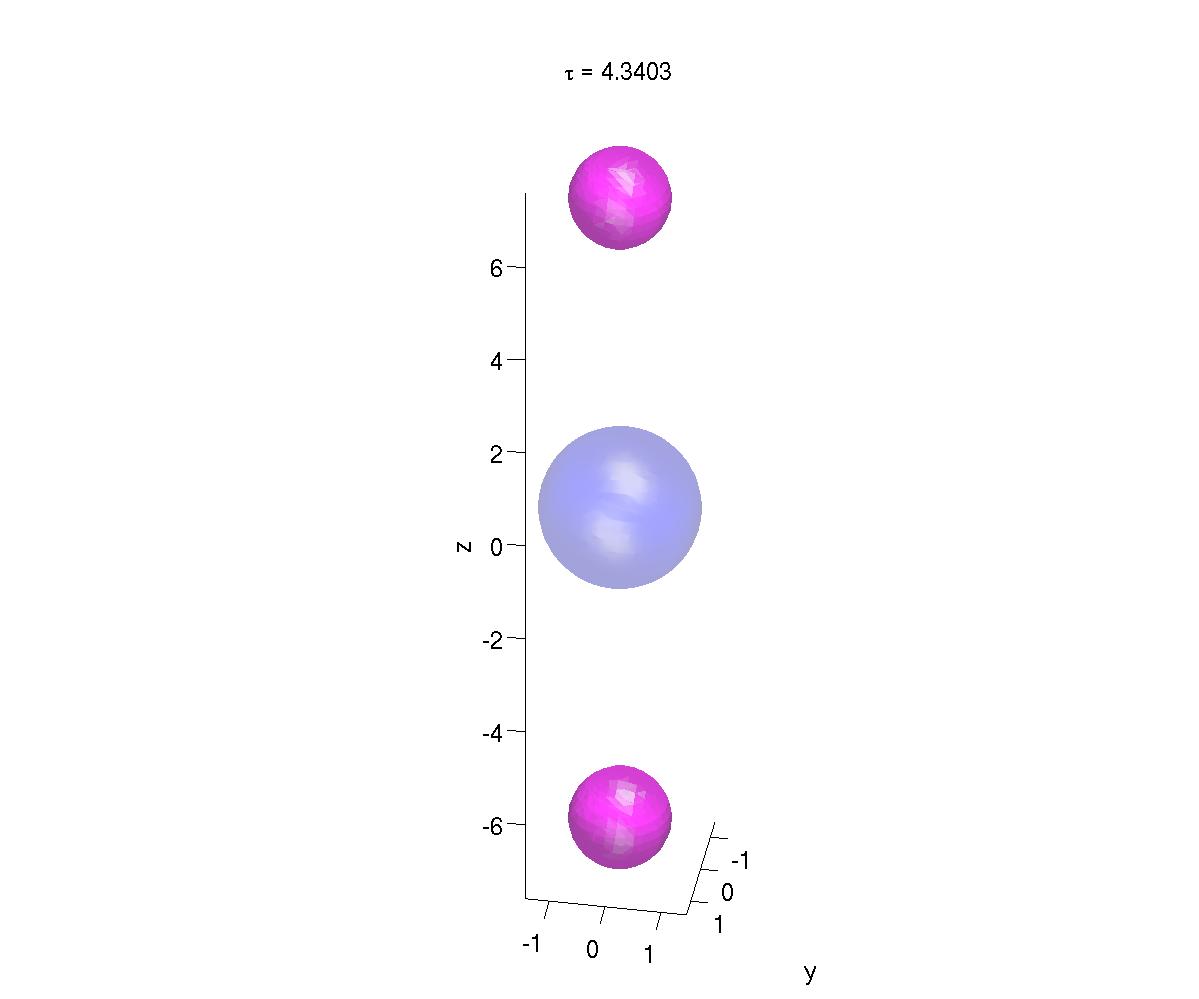}}
  \subfloat[]{\includegraphics[width=0.24\linewidth]{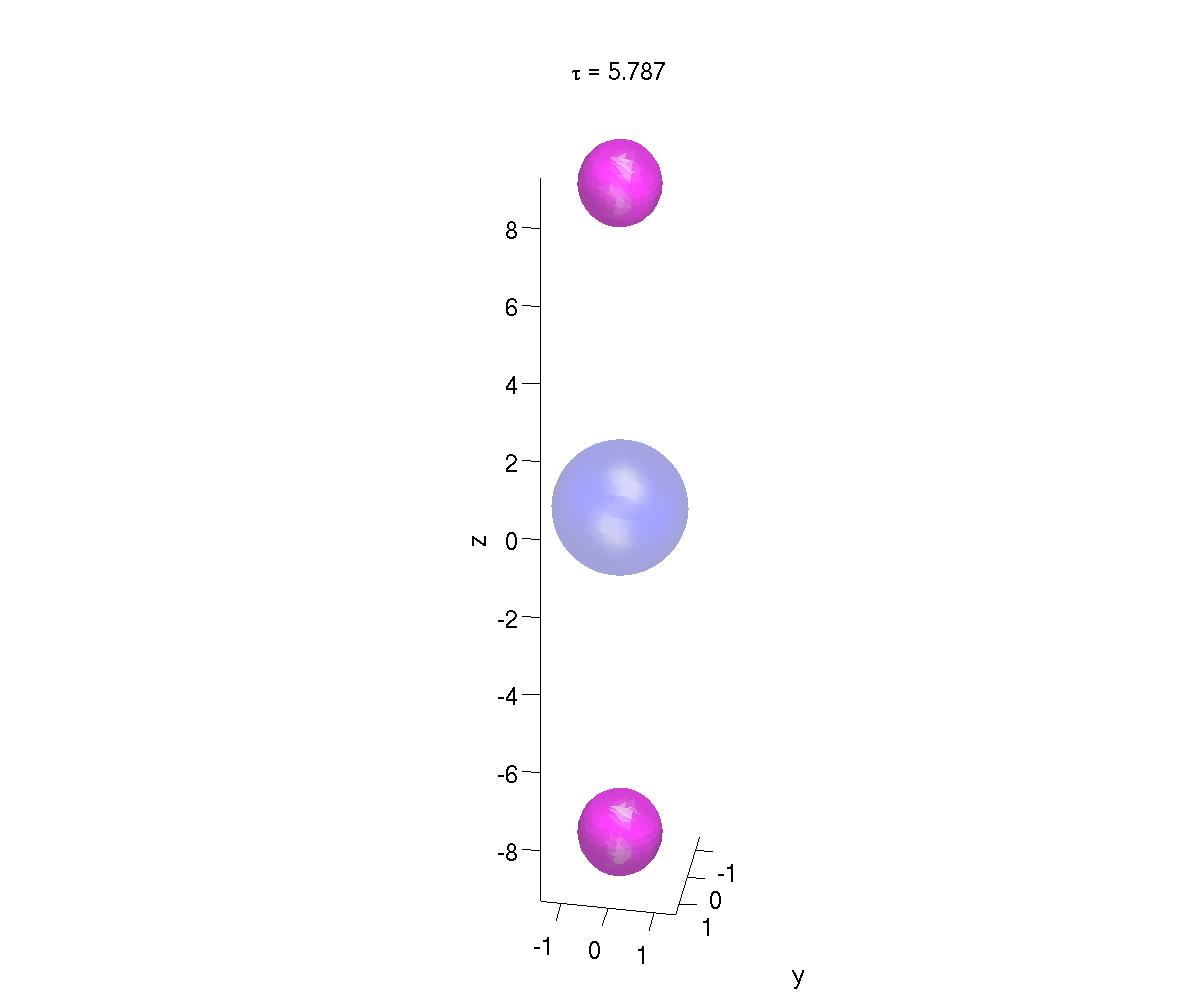}}}
\caption{The global two-monopole of type B inside a single
  Skyrmion. The figures show isosurfaces of the monopole charge and
  Skyrmion charge densities, respectively, at half their respective
  maximum values.
  The blue surface is the Skyrmion charge and the magenta surface is the
  monopole charge.
  As the relaxation time progresses, the solution diverges and the
  monopoles eventually escape off to infinity. 
  The calculation is made on a $121^3$ cubic lattice. 
  }
\label{fig:skm12a0isosurfaces}
\end{center}
\end{figure}

Although we find that the type A, $Q=2$ Skyrmopole of
Sec.~\ref{sec:AQ2skmm} is stable for the chosen values of the
parameters in the model, we also find that the corresponding type B,
$Q=2$ Skyrmopole is less stable.
We carry out the same calculation as in Sec.~\ref{sec:AQ2skmm}, but
with the initial condition given by the Ans\"atze \eqref{eq:nhedgehog}
and \eqref{eq:tipob} with $d=0$ and the parameters are again chosen as
$c_2=1$, $c_4=4$, $m=1$, $v=1/4$, $\lambda=128$, $b=1/2$ and
$\alpha=4$. We take the profile function of the initial condition to
be 
\beq
h = \tanh\left[\kappa r\right], \nonumber
\eeq
where $r$ is the radial coordinate and $\kappa$ is an appropriately
chosen constant. 
We find that for this choice of parameters, the type B configuration
is unstable.
In Fig.~\ref{fig:skm12a0isosurfaces} is shown a series of snapshots of
the isosurfaces of Skyrmion and monopole charges as the relaxation
time progresses. We observe that the two monopole constituents split
up, escape the Skyrmion and move off to infinity.
This is, however, by no means a proof that the type B configuration
cannot be stabilized. We suspect however, that by tuning the
parameters to stabilize the type B configuration, it will eventually
converge to the solution of type A. 

\section{Conclusion}\label{sec:conclusion}

In this paper we have shown that the two-monopole is unstable by
itself both in the point-charge approximation and with a smooth
profile function on a cubic lattice.
In order to stabilize the two-monopole we first considered adding an
interaction term to the monopole Lagrangian, but pointed out the
difficulties in such an approach. For this reason we followed a
different path to stabilize it.
The strategy is to create an energy barrier for the monopoles to
overcome in order for them to escape a spatial region. This energy
barrier was created as an example by a function that multiplies the
entire monopole Lagrangian and interpolates between a low value in the
center of the region where the monopoles are to be contained and a
high value outside.
We have shown with a given set of parameters, the two-monopole is
stable when the initial guess is of type A and by perturbing said
configuration, we have further shown that it returns to the found
solution after a sufficient amount of relaxation time.

This type of stable solution has one feature which, for some purposes,
may be undesirable. Namely, that the Skyrmion energy density dominates
the monopole one on the scales of the charges. Of course the
two-monopole has a diverging total energy as it is a global type of
monopole, while the Skyrmion has a finite total energy.
It may be possible to find a function $\mathcal{G}$, which multiplies
the monopole Lagrangian, such that the monopole energy locally
dominates over the Skyrmion energy. The problem of the, in this paper,
chosen type of function, is that the constraint
\eqref{eq:flipconstraint} needs to be satisfied.
A different function that would not have such a constraint can be
constructed as
\beq
\mathcal{G}(n_4) =
\frac{1}{2}\left(\frac{1+b-n_4^2}{2+b}\right)^\alpha,
\eeq
although this may be less stable than the function chosen in this
paper because the energy barrier drops to the same value outside the
Skyrmion as inside it.
This type of function can be investigated in future works.

One potential application of these stable multimonopole systems would
be as potential models of dark matter halos
\cite{Evslin:2012fe,Evslin:2013mga}, as their density profiles
asymptotically scale as $1/r^2$, automatically yielding flat rotation
curves.
The long distance divergences also need to be eliminated in this
context, but various mechanisms for this have been proposed in the
literature.  

Finally, of course global multi-monopoles with charge $Q>2$ may equally
well be considered in our construction. We leave this for future
investigations.

\section*{Acknowledgments}

S.~B.~G.~thanks Roberto Auzzi for discussions. 
J.~E.~is supported by NSFC grant 11375201. 
S.~B.~G.~thanks the Recruitment Program of High-end Foreign Experts for 
support.

\appendix
\section{The unstable two-monopole}\label{app:mm}

\begin{figure}[!tpb]
\begin{center}
\mbox{\subfloat[]{\includegraphics[width=0.24\linewidth]{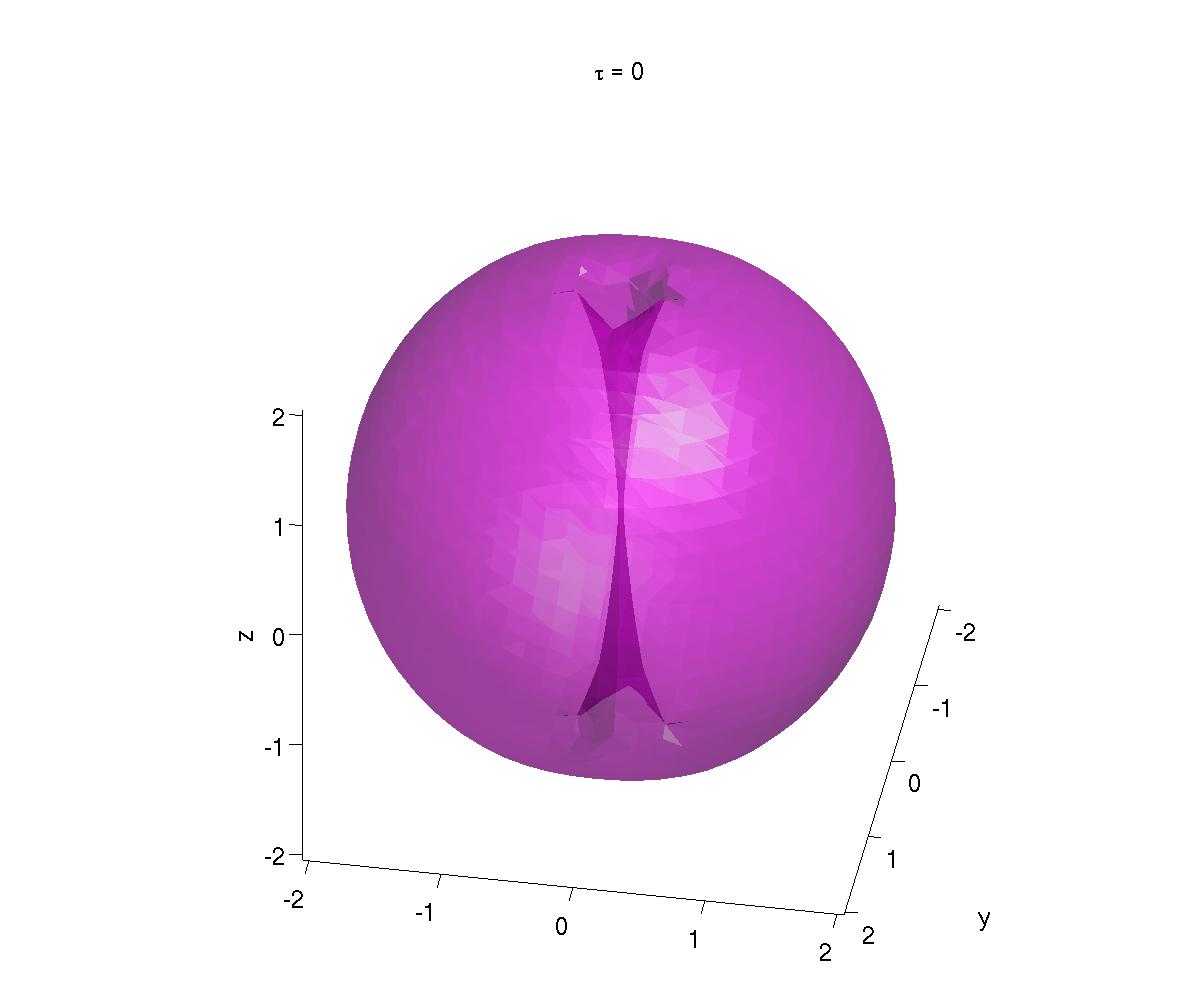}}
  \subfloat[]{\includegraphics[width=0.24\linewidth]{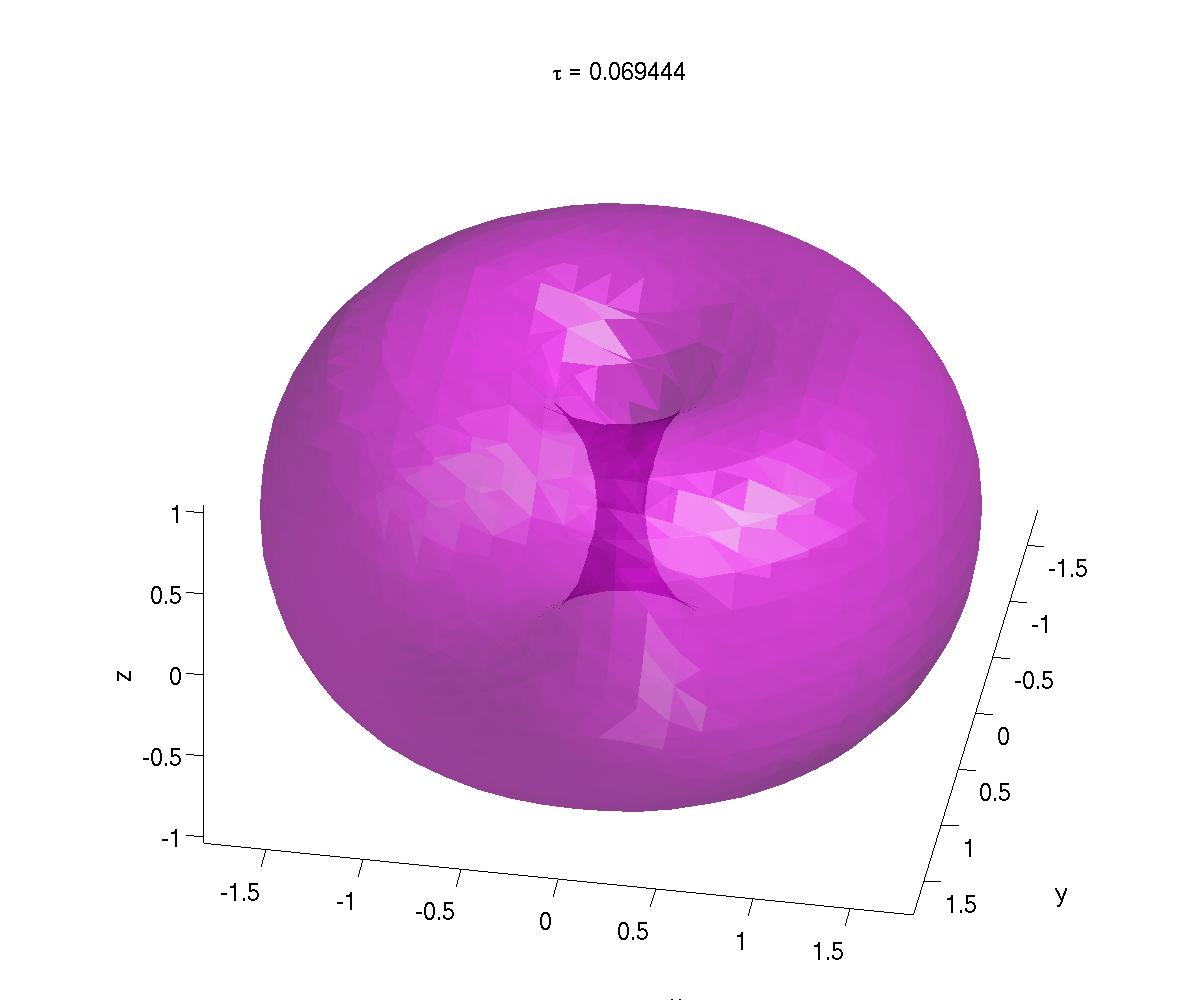}}
  \subfloat[]{\includegraphics[width=0.24\linewidth]{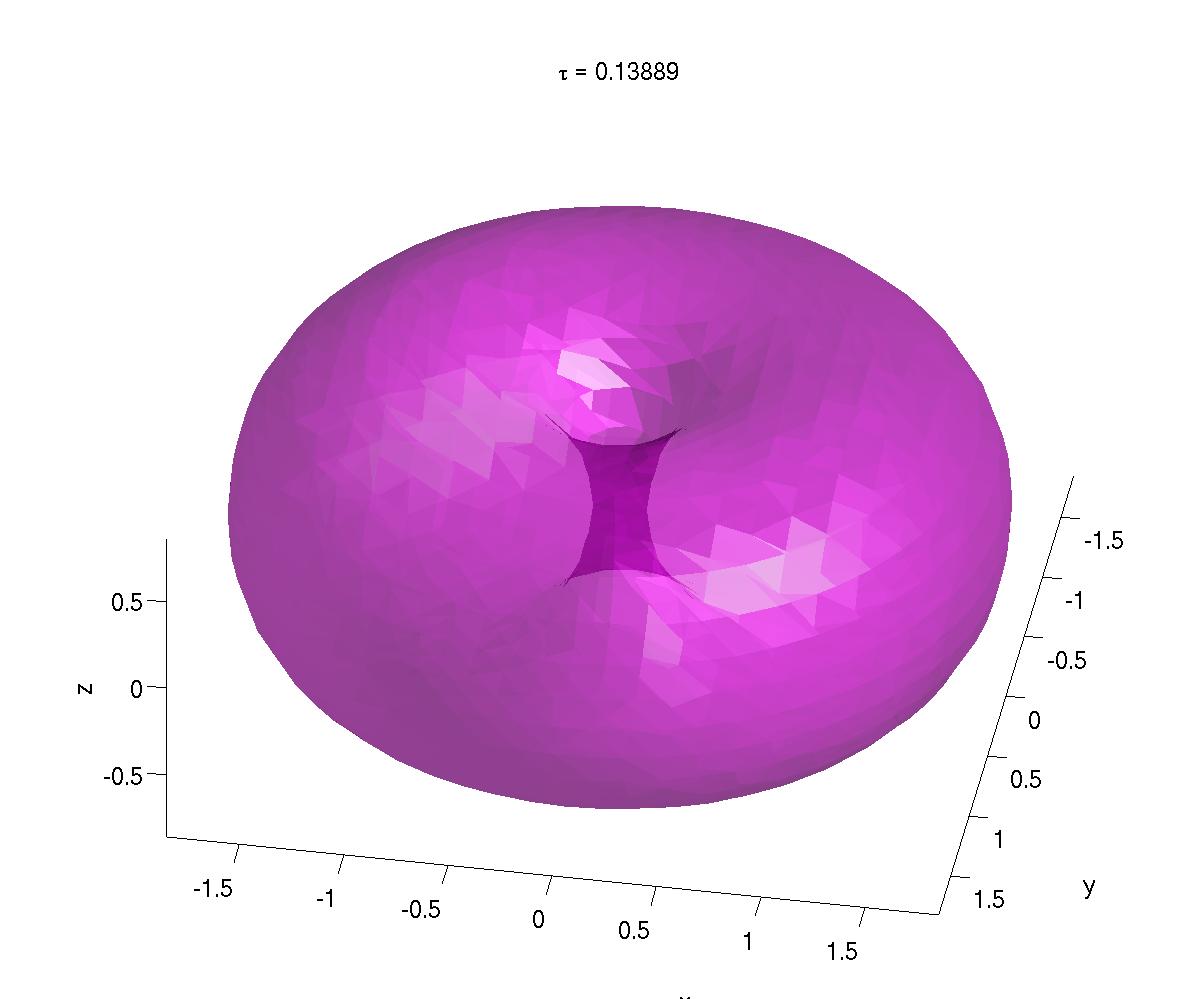}}
  \subfloat[]{\includegraphics[width=0.24\linewidth]{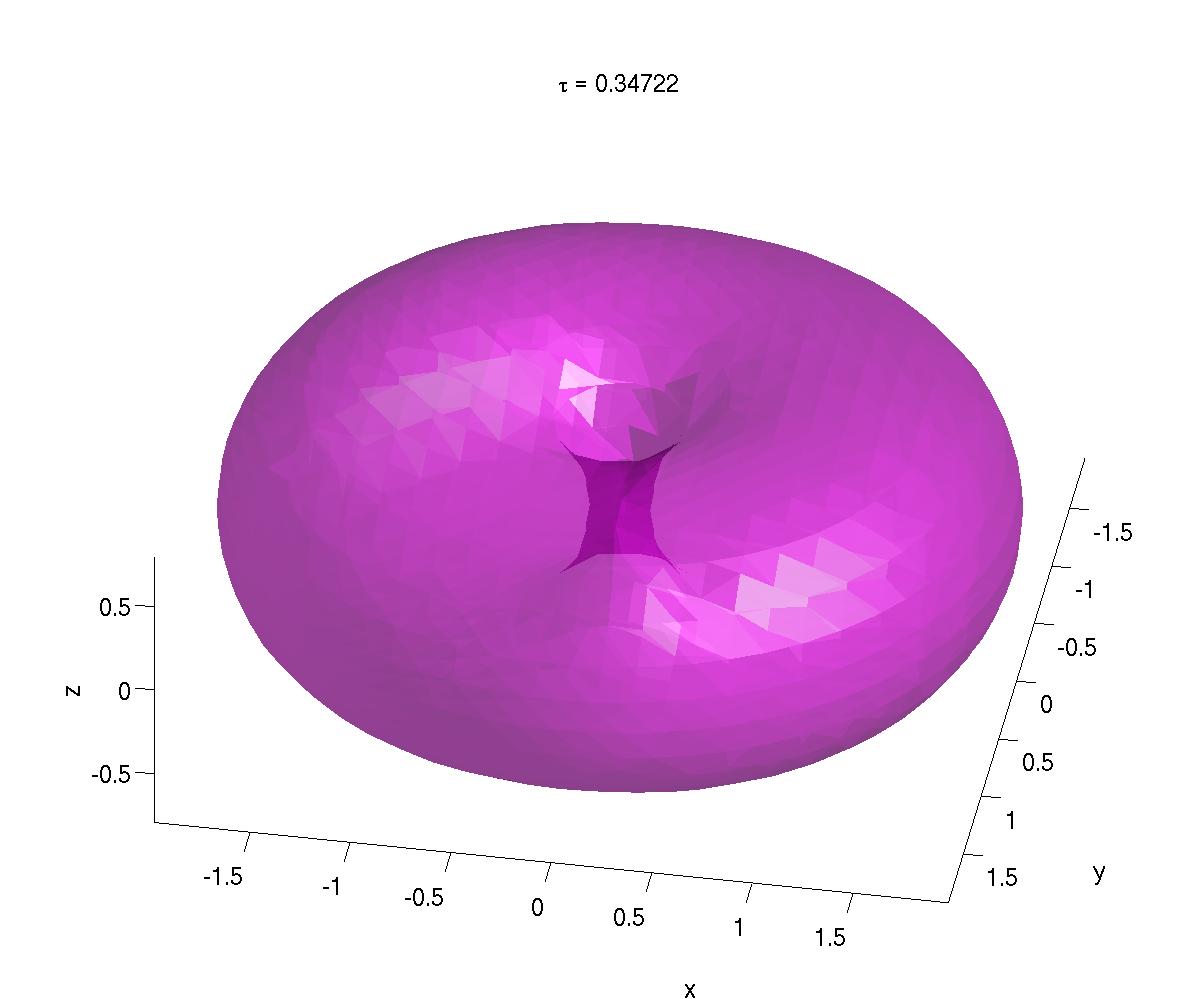}}}
\mbox{\subfloat[]{\includegraphics[width=0.24\linewidth]{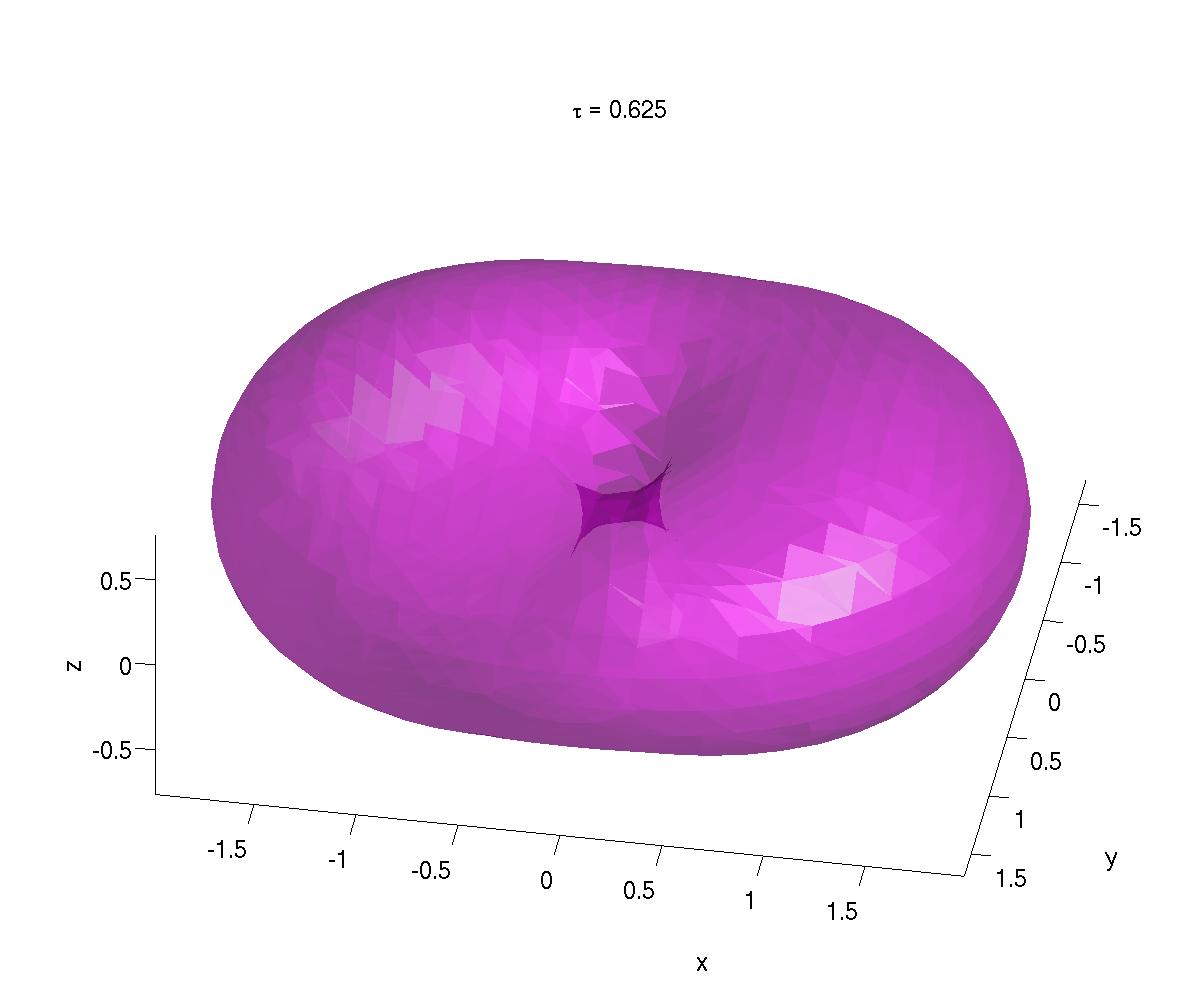}}
  \subfloat[]{\includegraphics[width=0.24\linewidth]{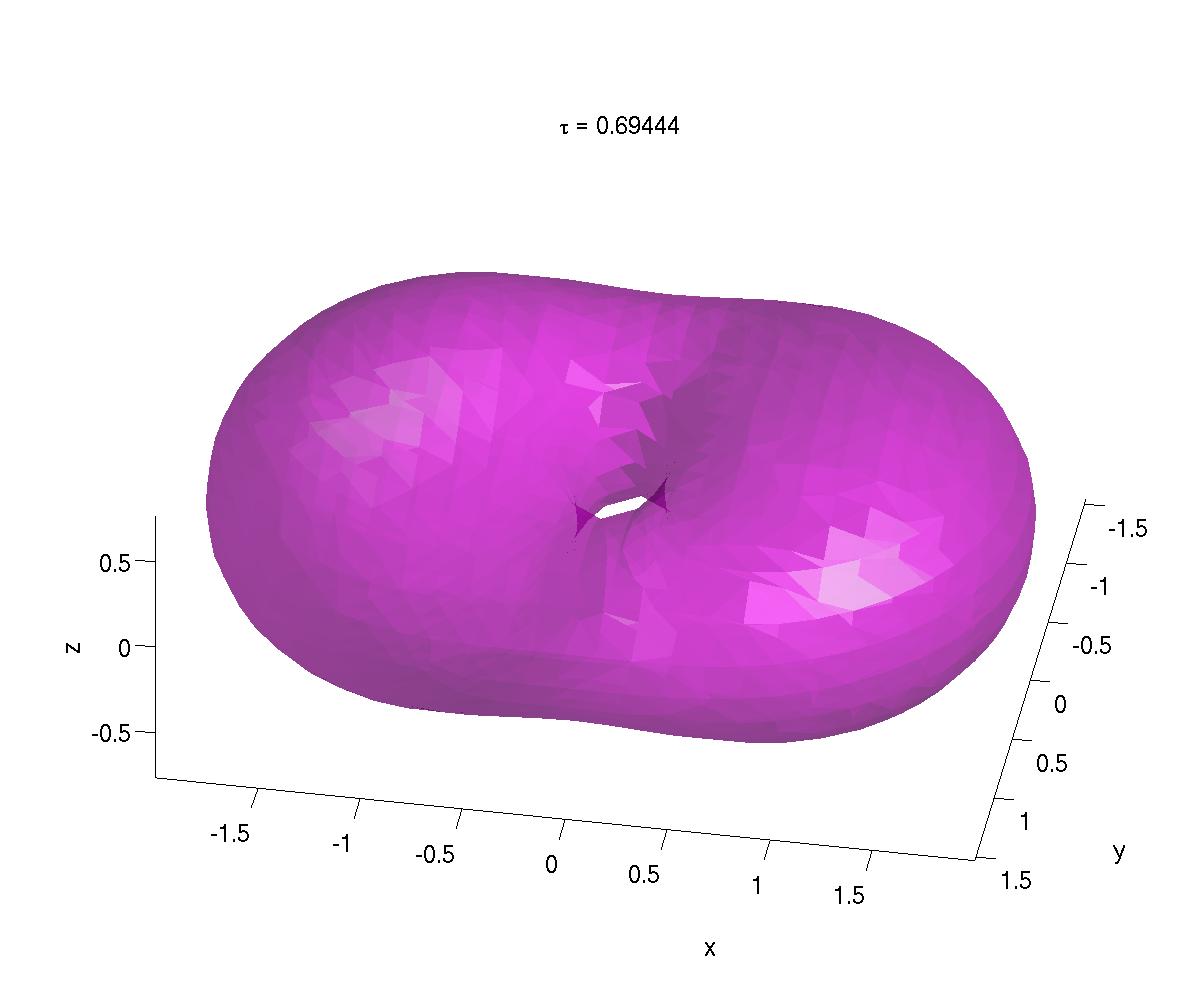}}
  \subfloat[]{\includegraphics[width=0.24\linewidth]{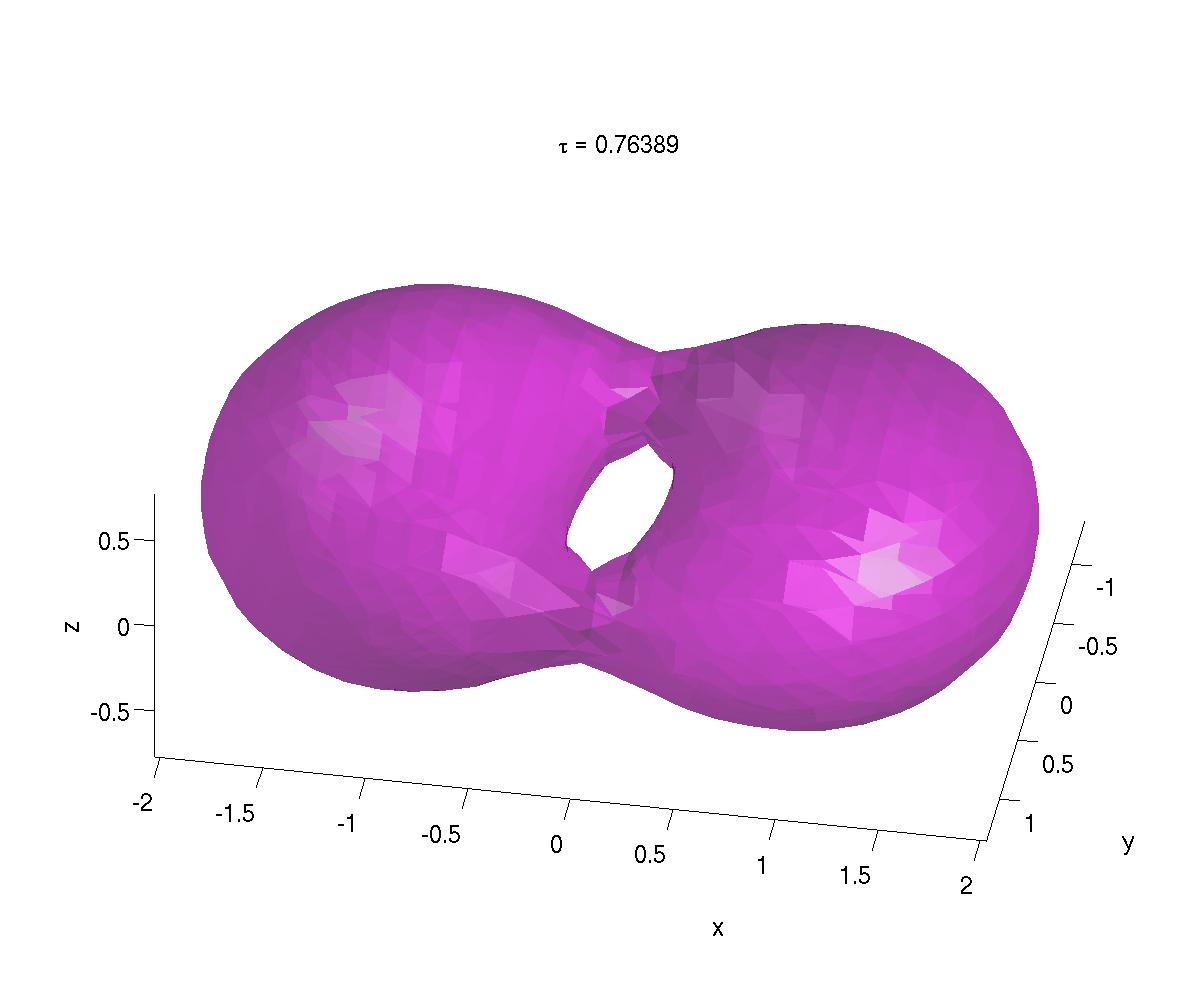}}
  \subfloat[]{\includegraphics[width=0.24\linewidth]{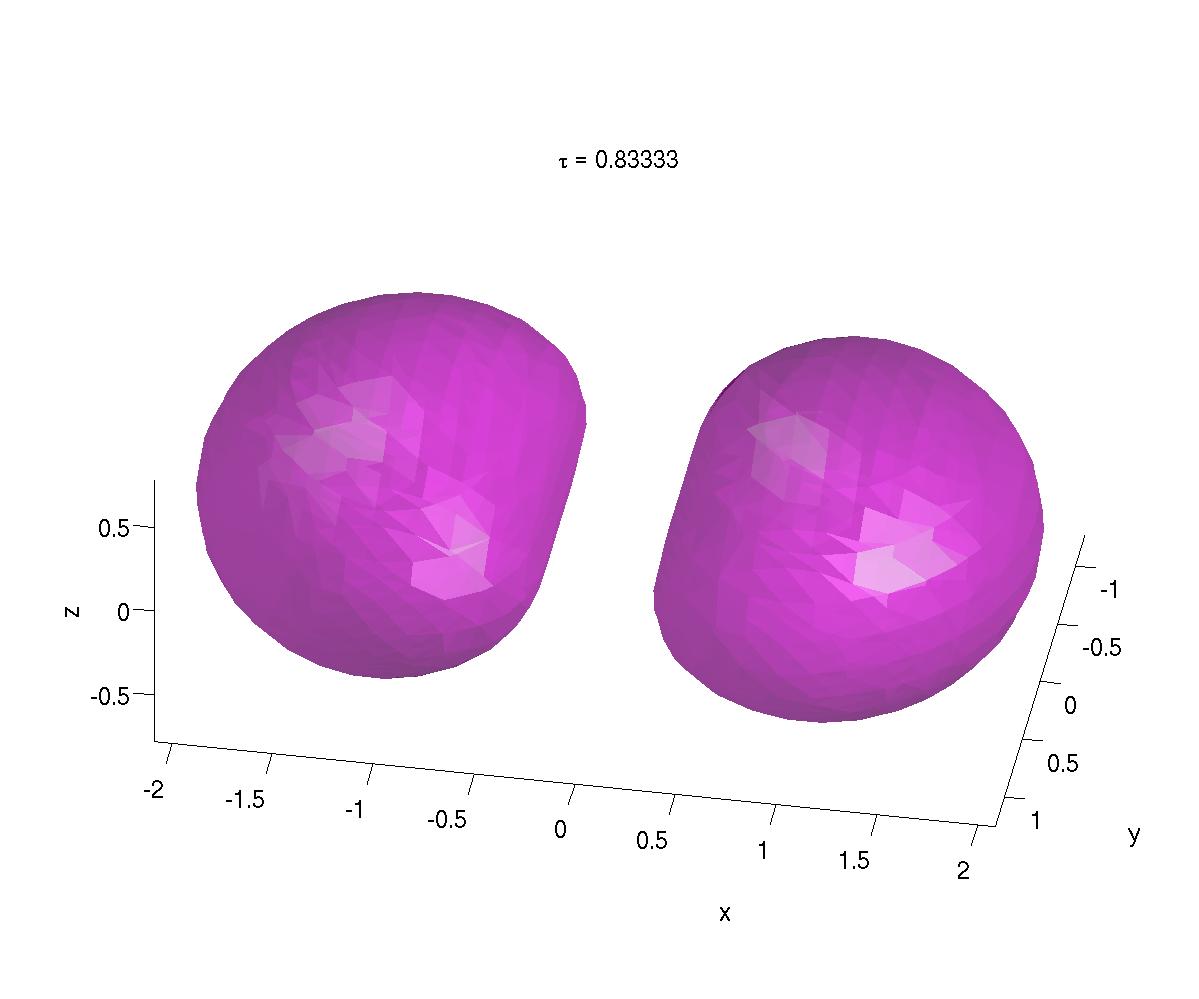}}}
\mbox{\subfloat[]{\includegraphics[width=0.24\linewidth]{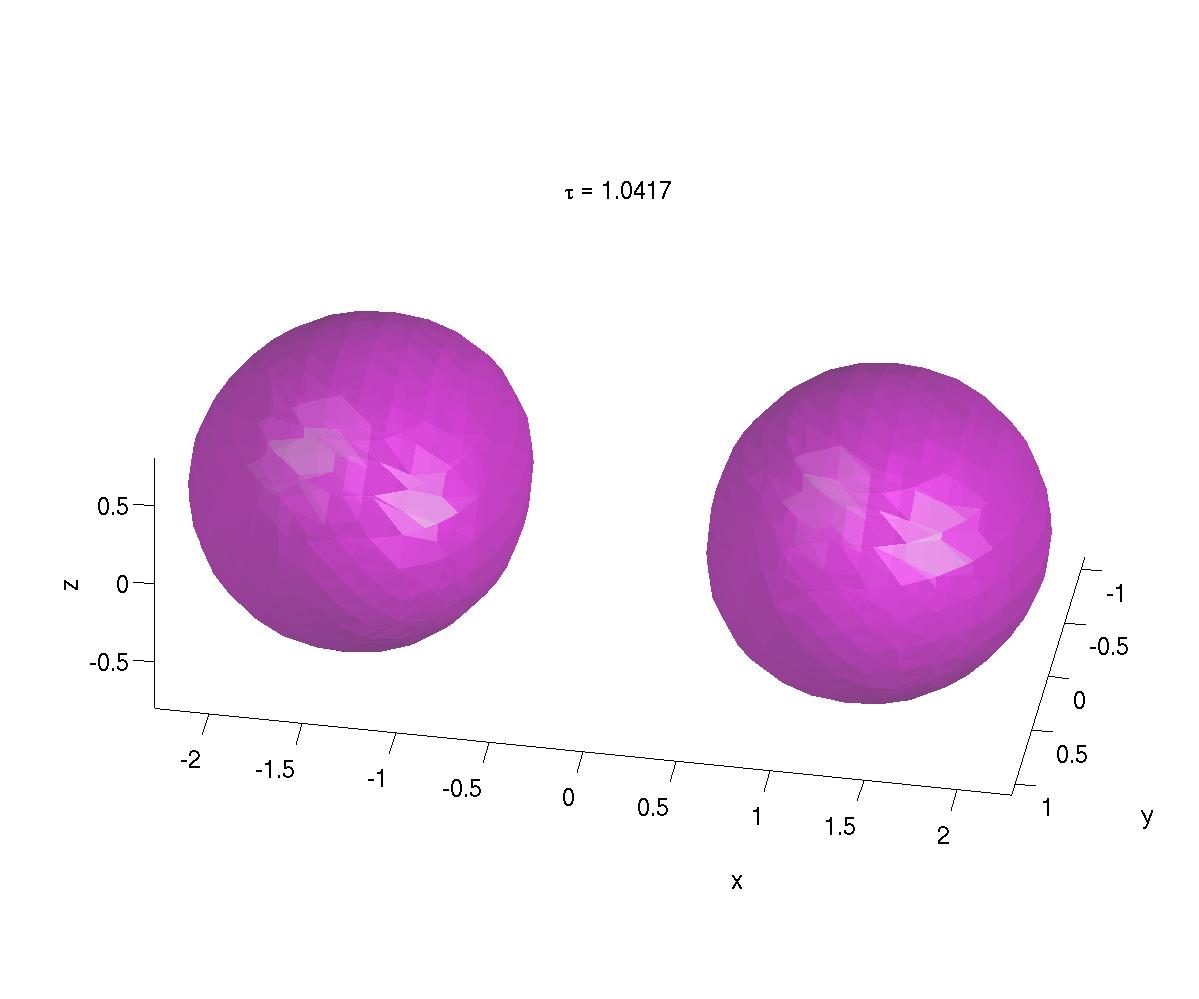}}
  \subfloat[]{\includegraphics[width=0.24\linewidth]{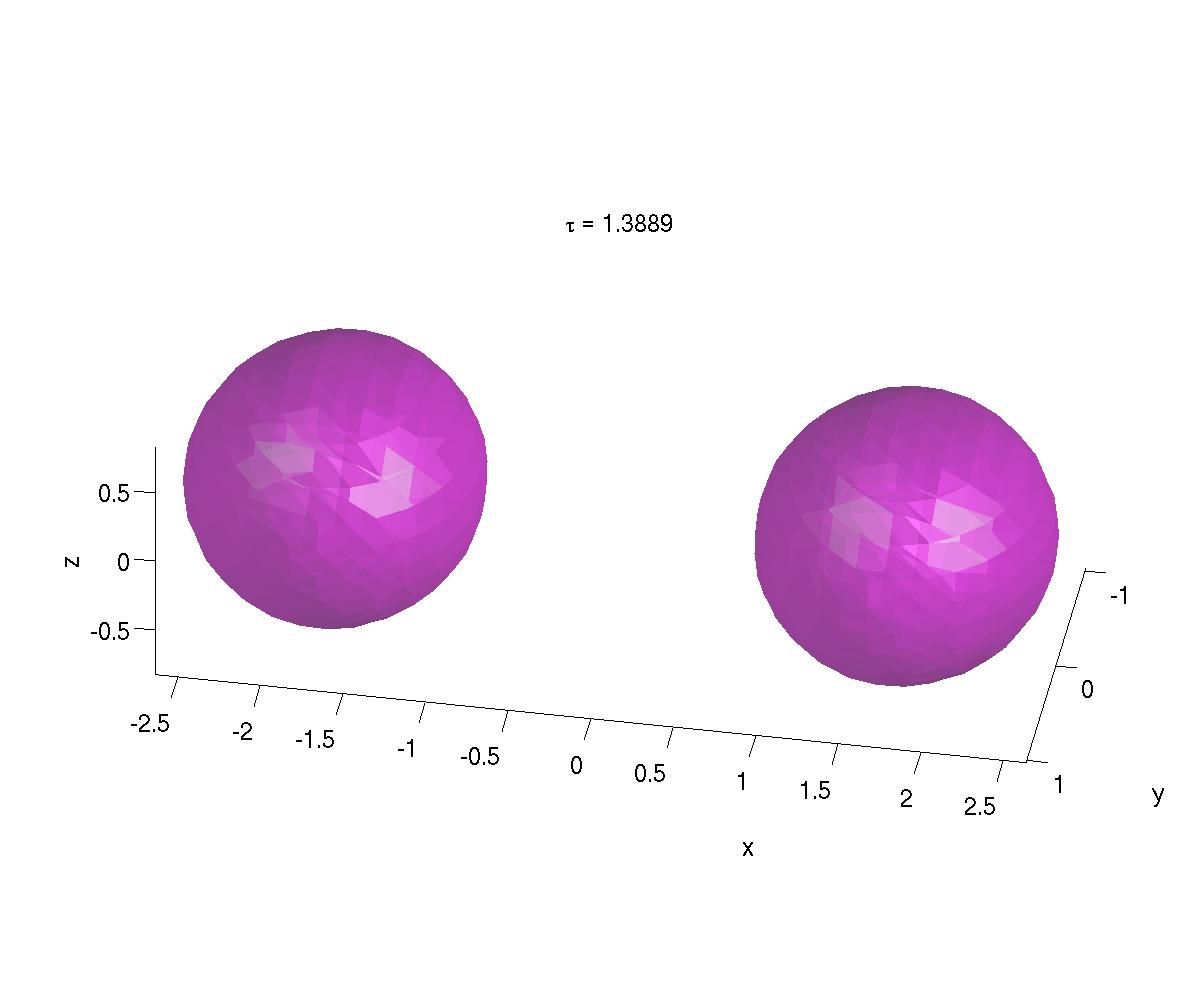}}
  \subfloat[]{\includegraphics[width=0.24\linewidth]{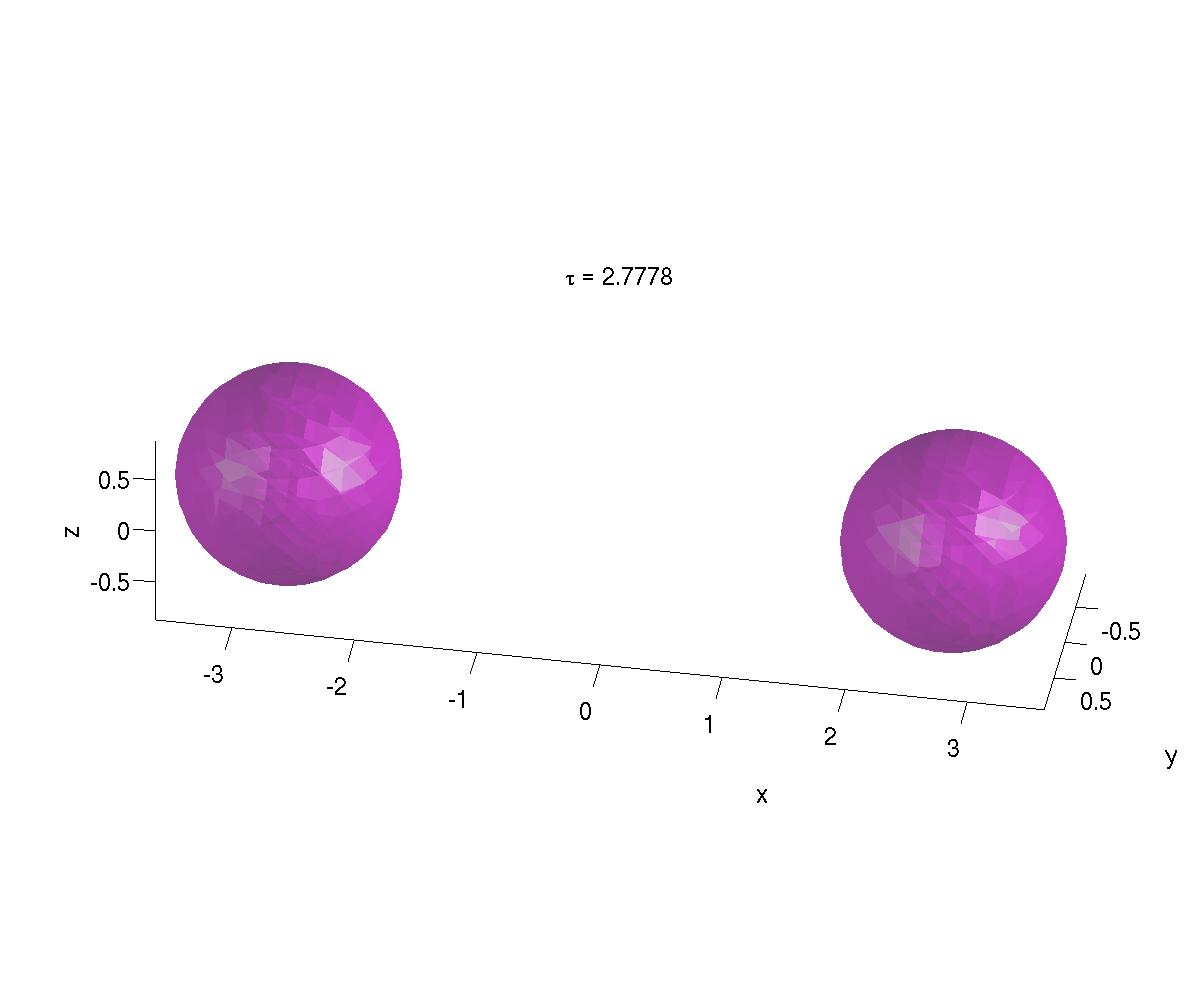}}
  \subfloat[]{\includegraphics[width=0.24\linewidth]{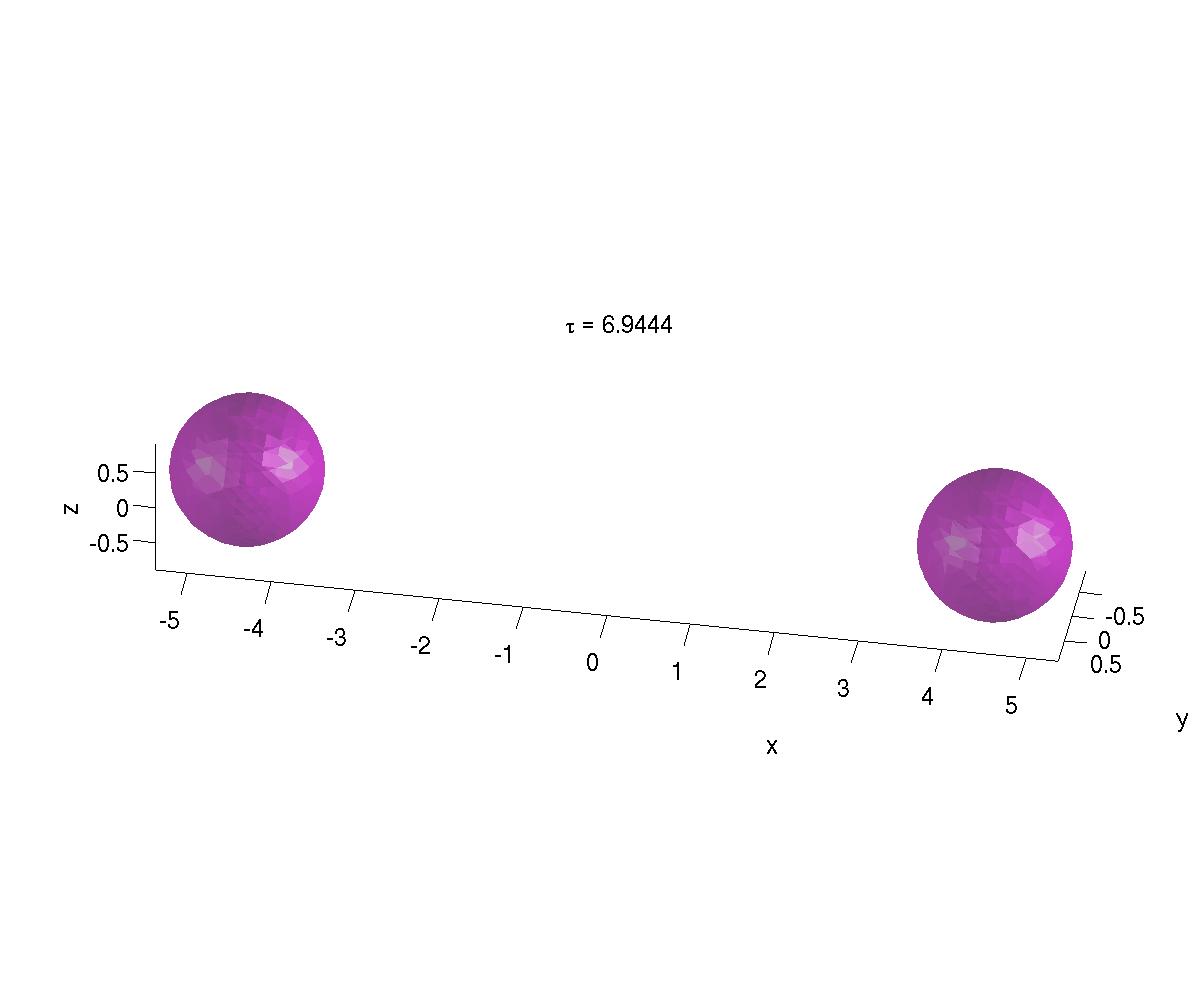}}}
\caption{The global two-monopole of type A with deformation parameter
  $d=h_x$. The figures show isosurfaces of the monopole charge at half
  its maximum value. 
  As the relaxation time progresses, the solution diverges and the
  monopoles eventually escape off to infinity.
  The parameters are chosen as $v=1/4$ and $\lambda=128$. 
  The calculation is made on a $121^3$ cubic lattice. 
  }
\label{fig:m12a1isosurfaces}
\end{center}
\end{figure}

For completeness, we attempt to put the $Q=2$ monopole on a lattice
with the initial condition given by the Ans\"atze \eqref{eq:nhedgehog}
and \eqref{eq:tipoa} with $d=h_x$, i.e.~the deformation parameter
takes the values of a single lattice spacing. We take the profile
function of the initial condition to be
\beq
h = \tanh\left[\kappa r\right], \nonumber
\eeq
where $r$ is the radial coordinate and $\kappa$ is an appropriately
chosen constant. 
In Fig.~\ref{fig:m12a1isosurfaces} is shown a series of snapshots of
the isosurfaces of the monopole charge as the relaxation time
progresses. We observe that the two monopole constituents split 
up, escape the Skyrmion and move off to infinity (at an accelerating
speed). 

\end{document}